\documentclass[fleqn,usenatbib,useAMS]{mnras}
\usepackage{txfonts}

\usepackage[T1]{fontenc}
\usepackage{ae,aecompl}
\usepackage{graphicx}	
\usepackage{amssymb}	

\usepackage{amsbsy}
\usepackage{mathrsfs,bm}
\newcommand{\bcdot}{\ensuremath{%
  \mathchoice%
   {\mskip\thinmuskip\lower0.2ex\hbox{\scalebox{1.5}{$\cdot$}}\mskip\thinmuskip}}%
   {\mskip\thinmuskip\lower0.2ex\hbox{\scalebox{1.5}{$\cdot$}}\mskip\thinmuskip}%
   {\lower0.3ex\hbox{\scalebox{1.2}{$\cdot$}}}%
   {\lower0.3ex\hbox{\scalebox{1.2}{$\cdot$}}}%
}

\newcommand{\ltsima}{$\; \buildrel < \over \sim \;$}
\newcommand{\lsim}{\lower.5ex\hbox{\ltsima}}
\newcommand{\gtsima}{$\; \buildrel > \over \sim \;$}
\newcommand{\gsim}{\lower.5ex\hbox{\gtsima}}
\newcommand {\apgt} {\ {\raise-.5ex\hbox{$\buildrel>\over\sim$}}\ }
\newcommand {\aplt} {\ {\raise-.5ex\hbox{$\buildrel<\over\sim$}}\ } 

\newcommand{\bra}{\langle}
\newcommand{\ket}{\rangle}
\renewcommand{\d}{\rmn{d}}

\newcommand{\eps}{\varepsilon}

\newcommand{\CR}{\rmn{cr}}
\renewcommand{\th}{\rmn{th}}
\newcommand{\inj}{{\rm inj}}

\newcommand{\eff}{\mathrm{eff}}
\newcommand{\e}{\rmn{e}}
\newcommand{\p}{\rmn{p}}
\newcommand{\dps}{\displaystyle}
\newcommand{\B}{{\mathcal B}}
\newcommand{\M}{{\mathcal M}}
\newcommand{\A}{{\mathcal A}}
\newcommand{\s}{\rmn{s}}
\newcommand{\tA}{\tilde{A}}
\newcommand{\ph}[1]{\phantom{#1}}
\newcommand{\msun}{\ensuremath{\rmn{M}_\odot}}

\newcommand{\pp}[2]{\frac{\partial #1}{\partial #2}}
\newcommand{\dd}[2]{\frac{\mathrm{d} #1}{\mathrm{d} #2}}
\renewcommand{\vec}{\ensuremath{\mathbfit}}
\newcommand{\mat}{\ensuremath{\mathbfss}}
\newcommand{\vel}{\ensuremath{\varv}}
\newcommand{\bvel}{\ensuremath{\bm{\varv}}}
\newcommand{\bw}{\ensuremath{\bm{\varw}}}
\newcommand{\bB}{\ensuremath{\mathbfit{B}}}
\newcommand{\bb}{\ensuremath{\mathbfit{b}}}
\newcommand{\com}{\ensuremath{\rmn{c}}}
\newcommand{\bnabla}{\ensuremath{\boldsymbol{\nabla}}}
\newcommand{\btimes}{\ensuremath{\boldsymbol{\times}}}
\newcommand{\bnablaR}{\ensuremath{\boldsymbol{\nabla}_r}}
\newcommand{\bnablaX}{\ensuremath{\boldsymbol{\nabla}_x}}

\title[Simulating cosmic ray physics on a moving mesh]{Simulating cosmic ray physics on a moving mesh}

\author[C. Pfrommer et al.] {C. Pfrommer$^1$\thanks{e-mail:
    christoph.pfrommer@h-its.org (CP)},
  R. Pakmor$^1$, K. Schaal$^{1,2}$, C. M. Simpson$^1$, and V. Springel$^{1,2}$\vspace*{0.2cm}\\
  $^1$Heidelberger Institut f\"{u}r Theoretische Studien,
  Schloss-Wolfsbrunnenweg 35, 69118 Heidelberg, Germany\\
  $^2$Zentrum f\"ur Astronomie der Universit\"at Heidelberg,
  Astronomisches Recheninstitut, M\"{o}nchhofstr. 12-14, 69120
  Heidelberg, Germany\\
}

\begin{document}
\pagerange{\pageref{firstpage}--\pageref{lastpage}} \pubyear{2003}
\maketitle
\label{firstpage}

\begin{abstract}
  We discuss new methods to integrate the cosmic ray (CR) evolution
  equations coupled to magneto-hydrodynamics (MHD) on an unstructured
  moving mesh, as realised in the massively parallel {\sc arepo} code
  for cosmological simulations. We account for diffusive shock
  acceleration of CRs at resolved shocks and at supernova remnants in
  the interstellar medium (ISM), and follow the advective CR transport
  within the magnetised plasma, as well as anisotropic diffusive
  transport of CRs along the local magnetic field.  CR losses are
  included in terms of Coulomb and hadronic interactions with the
  thermal plasma. We demonstrate the accuracy of our formalism for CR
  acceleration at shocks through simulations of plane-parallel shock
  tubes that are compared to newly derived exact solutions of the
  Riemann shock tube problem with CR acceleration. We find that the
  increased compressibility of the post-shock plasma due to the
  produced CRs decreases the shock speed. However, CR acceleration at
  spherically expanding blast waves does not significantly break the
  self-similarity of the Sedov-Taylor solution; the resulting
  modifications can be approximated by a suitably adjusted, but
  constant adiabatic index. In first applications of the new CR
  formalism to simulations of isolated galaxies and cosmic structure
  formation, we find that CRs add an important pressure component to
  the ISM that increases the vertical scale height of disk galaxies,
  and thus reduces the star formation rate. Strong external structure
  formation shocks inject CRs into the gas, but the relative pressure
  of this component decreases towards halo centres as adiabatic
  compression favours the thermal over the CR pressure.
\end{abstract}


\section{Introduction}

Understanding the physics of galaxy formation is arguably one of the
most complicated problems in modern astrophysics. A large body of
theoretical work based on cosmological simulations and semi-analytic
models has demonstrated that so-called feedback processes by stellar
winds and radiation fields, supernovae, and active galactic nuclei
(AGNs) appear to be critical in obtaining realistic galaxy populations
\citep[e.g.][]{2010MNRAS.402.1536S, 2011ApJ...742...76G, Puchwein2013,
  2014MNRAS.445..581H, 2014MNRAS.437.1750M, Henriques2015,
  2014MNRAS.444.1518V, Schaye2015Eagle}. These processes are invoked
to slow down star formation to the small observed rates, to move gas
and metals out of galaxies into the intergalactic medium by means of
galactic winds, to obtain a realistic mix of early- and late-type
galaxies, to quench star formation in elliptical galaxies, and to
balance radiative cooling of low-entropy gas at the centres of galaxy
clusters so that global cluster observables agree with observations in
the X-ray and micro-wave regime via the thermal Sunyaev-Zel'dovich
effect \citep{2012ARA&A..50..353K, 2012ApJ...758...74B,
  2012ApJ...758...75B, 2013ApJ...777..123B, 2014MNRAS.440.3645M,
  2016arXiv160302702M, 2016MNRAS.tmp.1157D}.

While the recent progress of galaxy formation models is remarkable, it
still comes with the caveat that the involved feedback has typically
been modelled empirically and tuned to match observed galaxy scaling
relations, weakening the predictive power of the corresponding
calculations. In particular, feedback in hydrodynamical simulations of
galaxy formation has thus far often been implemented very coarsely,
for example based on explicit subgrid models that aim to represent the
unresolved, multi-phase structure of the ISM with an effective
description that still yields the correct average star formation rate
\citep{2003MNRAS.339..289S, Schaye2008}. The physics behind
galactic winds and outflows remains especially unclear, and is sometimes treated
in a purely phenomenological way where the wind velocity and momentum
flux are prescribed \citep{Oppenheimer2006}.  Similarly, in order to
prevent too many stars from precipitating out of the hot phase of the
intra-cluster medium (ICM), feedback from AGN has often been modelled
by estimating accretion rates with a simple Bondi prescription and
injecting thermal energy as feedback \citep{DiMatteo2005, Springel2005}.

However, rather than just depositing thermal energy by supernovae or
AGNs, a physically more correct solution may involve the formation of
a non-thermal relativistic particle population (i.e., in CRs), created
through the process of diffusive shock acceleration at expanding
supernova remnants \citep[e.g.,][]{2008A&A...481...33J} or in
relativistic jets powered by AGNs
\citep[e.g.,][]{2008MNRAS.387.1403S}. Non-thermal energy is dissipated
on a longer timescale than thermal energy because CR cooling is
generally less efficient than the radiation cooling of a thermal
plasma \citep{2007A&A...473...41E}.  As the non-thermal
energy is not observable through X-ray emission or other thermal
observables, the (temporary) storage of feedback energy in CRs also
avoids problems with the overproduction of these
observables. Moreover, the CR pressure force could accelerate the
ambient ISM and drive powerful galactic outflows and winds.

Another major contender for the physical basis of feedback in galaxies
lies in the momentum and energy deposition of ultra-violet
radiation. Momentum-driven winds can form when radiation pressure acts
efficiently on dust grains and atomic lines in dense gas and imparts
momentum kicks that can expel the gas if it exceeds the escape
velocity. While this mechanism has been argued to provide efficient
feedback during the formation of galaxies, possibly even explaining
strong outflows in starburst galaxies \citep{2005ApJ...618..569M,
  2005ApJ...630..167T}, direct radiation-hydrodynamical simulations of
simplified set-ups of the Rayleigh-Taylor instability
\citep{2012ApJ...760..155K} or full galaxy-scale models
\citep{2015MNRAS.451...34R, 2015ApJ...809..187S} fail to see these
strong winds.  This suggests that radiation feedback is more gentle
and less effective than assumed in some subgrid prescriptions. Also,
the high dust opacities needed for radiation pressure to be efficient
are unlikely to be realised in Milky Way-type galaxies, in particular
at larger galactocentric radii.

On the other hand, CRs and magnetic fields are observed to be in
pressure equilibrium with the turbulence in the midplane of the Milky
Way \citep{1990ApJ...365..544B}. This could be naturally explained if
it is the outcome of a self-regulated feedback loop where CRs and
magnetic fields provide the main wind driving mechanism, as suggested
by a number of theoretical works \citep{1975ApJ...196..107I,
  1991A&A...245...79B, 1996A&A...311..113Z, 1997A&A...321..434P,
  2002A&A...385..216B, 2008ApJ...687..202S, 2008ApJ...674..258E,
  2010ApJ...711...13E, 2010MNRAS.402.2778S, 2012A&A...540A..77D} and
local three-dimensional (3D) simulations of the ISM
\citep{2013ApJ...777L..38H, 2016ApJ...816L..19G}.

In comparison to other wind-driving mechanisms, CRs have a number of
properties that make them advantageous for driving winds: (i) their
pressure drops less quickly upon adiabatic expansion than the thermal
pressure due to their softer equation of state
($P_\CR\propto \rho^{\gamma_\CR}$ with $\gamma_\CR=4/3$), (ii) CRs
cool less efficiently than the thermal gas and can hold on to their
energy for longer time scales, and (iii) the CRs can energise the wind as
it rises from the disk with a rate that depends on the ratio of the
advection-to-streaming speed, thus maintaining the outflows in a warm/hot
state and acting like a `CR battery'.  Polarised radio
observations of radio haloes in edge-on galaxies reveal poloidal field
lines at the interface of the ISM and the galactic halo
\citep[e.g.,][]{2000A&A...364L..36T}, which coincides with the wind
launching site: this argues for a dynamical mechanism that is
responsible for reorienting the preferentially toroidal field in
galactic disks by means of anisotropic CR transport along the magnetic
field. 

If the CR pressure in a disk is super-critical, i.e.~if it is unstable
to a buoyancy instability, CR-loaded gas will start to rise from the
disk and drive a Parker instability \citep{2016ApJ...816....2R},
resulting in a poloidal (`open') field configuration. Pioneering
numerical work has demonstrated that CR-driven winds can expel ISM
from the disk and efficiently regulate star formation in 3D
hydrodynamic simulations of galaxy formation
\citep{2012MNRAS.423.2374U, 2013ApJ...777L..16B, 2014MNRAS.437.3312S,
  2014ApJ...797L..18S, 2016arXiv160204856R}. Unlike the classical
energy- or momentum-driven wind solutions, CR-driven winds impart not
only energy and momentum to the ISM at the wind base, but can also
continuously repower the plasma during its ascent because of dynamic
and thermal coupling of CRs to the plasma.

The ability of CRs to drive winds is intimately tied to their
transport processes. In particular, the fast streaming of CRs along
the magnetic field excites Alfv\'en waves through the streaming
instability \citep{1967ApJ...147..689L,
  1969ApJ...156..445K}. Scattering off of this wave field limits the
CRs' bulk speed in turn. These waves are then damped, effectively
transferring CR energy and momentum to the thermal plasma. Hence CRs
exert a pressure on the thermal plasma by means of scattering off
of Alfv\'en waves. Interestingly, winds driven by CR streaming are
characterised by an increasing mass loading factor $\eta_{\rmn{m}} =
\dot{M}/\dot{M}_*$ of the galactic wind towards smaller galaxies
\citep[where $\dot{M}$ and $\dot{M}_*$ denote the wind mass loss and
  star formation rate, respectively, see][]{2012MNRAS.423.2374U}. This
mass scaling of $\eta_{\rmn{m}}$ may be required to yield a strongly
decreasing star conversion efficiency of baryons towards dwarf
galaxies and thus makes CR physics a prime candidate for the physical
mechanism underlying this conundrum of galaxy formation.

In addition, CRs may hold the key to understanding a similar problem in
giant elliptical galaxies located at the centres of galaxy groups and
clusters. In the absence of heating processes, the hot gaseous
atmospheres of these objects are expected to cool and feed star
formation at rates up to several hundred $\rmn{M}_\odot~\rmn{yr}^{-1}$
\citep[see][ for a review]{2006PhR...427....1P}. Instead, the observed
gas cooling and star formation rates are reduced to levels
substantially below those expected from unimpeded cooling
flows. High-resolution {\em Chandra} and {\em XMM-Newton} observations
of clusters have provided evidence that radiative cooling is offset by
an unidentified heating process, which is associated with the AGN 
jet-inflated radio lobes that are co-local with the cavities seen
in the X-ray maps. The interaction of cooling gas, subsequent star
formation, and nuclear activity seems to be tightly coupled to a
self-regulated feedback loop \citep[for reviews,
see][]{2007ARA&A..45..117M, 2012NJPh...14e5023M}.  While the
energetics associated with AGN feedback are sufficient to balance
radiative cooling, how the available energy is coupled to the cluster
gas is less clear.

Several physical processes have been proposed to mediate the heating, including
dissipation of turbulent energy excited by the rising AGN lobes
\citep{2014Natur.515...85Z}. Alternatively, a net outward flux of streaming CRs
may provide a means to stably heat the cooling cluster plasma by the excitation
of resonant Alfv\'en waves. Once generated, they experience non-linear Landau
damping or decay via a cascading process as a result of strong external
turbulence, and eventually dissipate locally \citep{1991ApJ...377..392L,
  2008MNRAS.384..251G, 2011A&A...527A..99E, 2012ApJ...746...53F,
  2013MNRAS.434.2209W, 2013ApJ...779...10P, 2016arXiv160906321J,
  2016arXiv160906322J}.

This picture is supported by the widespread evidence for non-thermal
activity in galaxy clusters, which also manifests itself on cluster
scales through extended synchrotron emission in the form of radio
haloes and peripheral relics \citep[for a review,
  see][]{2014IJMPD..2330007B}. Despite a large body of work, the
origin of radio haloes and the nature of seed electrons that are
energised in radio relics are still open problems. This is because in
clusters different plausible acceleration mechanisms are operating
that range from diffusive shock acceleration of relativistic
electrons to injection as secondaries in hadronic interactions of
relativistic protons with the dense cluster gas, and even turbulent
re-acceleration through interactions with the compressible cluster
turbulence. One or a combination of these processes could be
responsible for those enigmatic radio phenomena. Cosmological MHD
simulations with CR physics are indispensable for solving these
puzzles by evolving the Fokker-Planck equation for the CR
distribution through cosmic time, following in detail the CR
acceleration and transport into galaxy clusters
\citep{2001ApJ...559...59M, 2001ApJ...562..233M, 2007MNRAS.378..385P,
  2008MNRAS.385.1211P, 2008MNRAS.385.1242P, 2010MNRAS.409..449P,
  2012MNRAS.421.3375V, 2013MNRAS.429.3564D, 2013MNRAS.435.1061P,
  2015arXiv150307870P}.

The above considerations provide ample motivation for outfitting
modern hydrodynamical cosmological simulation codes with a numerically
efficient and accurate treatment of CR physics coupled to MHD. This is
the goal of this paper, which aims to provide a comprehensive
exposition of the most relevant CR physics, and a description of our
numerical implementation of it in the {\sc arepo} moving-mesh
code. We discuss a set of tests of the new code, focusing in
particular on the modifications of strong shocks due to the inlined CR
acceleration. We also present a set of first applications in the form
of isolated disk galaxies and basic cosmological simulations of
structure formation. In a companion paper \citep{Pakmor2016a}, we
provide further details on the technical implementations of our
anisotropic diffusion solver, and in two further companion studies, we
apply our new methodology to wind formation in disk galaxies
\citep{Pakmor2016b} and to the regulation of the ISM by individual
supernova explosions \citep{Simpson2016}.

This paper is organised as follows.  In Section~\ref{sec:equations},
we present the basic system of MHD equations with CRs in physical
coordinates, on a cosmologically expanding background, and describe
our implementation.  In Section~\ref{sec:non-adiabatic}, we detail our
modelling of non-adiabatic processes of CRs that include diffusive
shock acceleration, injection at supernova remnants, and collisional
loss processes such as Coulomb and hadronic interactions with the
ambient plasma.  In Section~\ref{sec:results}, we present our results
on plane parallel shock tubes and a spherically expanding Sedov-Taylor
blast wave with CR shock acceleration. We also discuss MHD simulations
of galaxy formation that include advective CR transport, and initial
results on cosmological simulations that account for CR acceleration
at structure formation shocks.  In Section~\ref{Sec:conclusions}, we
summarise our main findings and conclusions.  Finally, in
Appendix~\ref{sec:CRhydro}, we review the basic considerations
underlying the derivation of CR hydrodynamics, and in
Appendices~\ref{sec:Riemann} and \ref{sec:Riemann+CRs} we detail our
novel derivation of the exact solutions of the Riemann shock-tube
problem when CR acceleration with and without a pre-existing
population of CRs is included.

\section{Basic equations}
\label{sec:equations}

\subsection{Magneto-hydrodynamics with cosmic rays}

The equations of ideal magneto-hydrodynamics of a two-fluid medium composed of
thermal gas and CRs can be written as a system of conservation laws,
\begin{equation}
  \label{eq:conservation}
  \frac{\partial \vec{U}}{\partial t} + \bnabla\bcdot \mat{F}
  = \vec{S} ,
\end{equation}
for a vector of conserved variables $\vec{U}$, a flux function
$\mat{F}\equiv\mat{F}(\vec{U})$, and source terms $\vec{S}$. Those are given in
the local restframe by (see Appendix~\ref{sec:CRhydro})
\begingroup
\renewcommand*{\arraystretch}{1.5}
\begin{eqnarray*}
  \label{eq:matrix}
  \vec{U} = \left( \begin{array}{c} \rho \\ \rho \bvel \\ \eps \\ \eps_\CR \\ \bB \end{array} \right),
  \ \ \ \
  \mat{F} = \left( \begin{array}{c} 
      \rho \bvel \\
      \rho \bvel \bvel^{\rmn{T}} + P\bm{1} - \bB \bB^{\rmn{T}} \\
      (\eps + P) \bvel - \bB \left( \bvel \bcdot \bB \right) \\
      \eps_\CR \bvel + (\eps_\CR+P_\CR)\bvel_{\rmn{st}} - \kappa_\eps \bb \left( \bb \bcdot \bnabla \eps_\CR\right) \\
      \bB \bvel^{\rmn{T}} - \bvel \bB^{\rmn{T}}
    \end{array} \right),
\end{eqnarray*}
\begin{equation}
  \vec{S} = \left( 
    \begin{array}{c} 
      0 \\
      \mathbf{0} \\
      \phantom{-} P_\CR\,\bnabla \bcdot \bvel - \bvel_{\rmn{st}} \bcdot \bnabla P_\CR +\Lambda_\th + \Gamma_\th \\
               -  P_\CR\,\bnabla \bcdot \bvel + \bvel_{\rmn{st}} \bcdot \bnabla P_\CR +\Lambda_\CR + \Gamma_\CR \\
      \mathbf{0}
    \end{array} \right),
\end{equation}
\endgroup
\noindent
where we used the Heaviside-Lorentz system of units. $P$ is the total pressure
due to thermal gas, CRs, and magnetic fields and $\eps$ is the total energy
density excluding CRs,
\begin{eqnarray}
  \label{eq:P_eps}
  P &=& P_\th + P_\CR +\frac{\bB^2}{2}, \\
  \eps &=& \eps_\th + \frac{\rho\bvel^2}{2} + \frac{\bB^2}{2}.
\end{eqnarray}
The local gas density, velocity, and magnetic field strength are given by
$\rho$, $\bvel$ and $\bB$, respectively. The thermal and CR energy densities are
given by $\eps_\th$ and $\eps_\CR$, respectively. The unit vector along the
local magnetic field is denoted by $\bb=\bB/|\bB|$ and $\bm{1}$ is the unit
rank-two tensor. The kinetic energy-weighted spatial diffusion coefficient of CRs is
denoted by $\kappa_\eps$ (see also Appendix~\ref{sec:CRhydro}) and the CR streaming
velocity is given by
\begin{eqnarray}
  \label{eq:vstream}
  \bvel_{\rmn{st}} = -\bvel_{\rmn{A}}\, \rmn{sgn}(\bB\bcdot\bnabla P_\CR)
  = -\frac{\bB}{\sqrt{\rho}}\,
  \frac{\bB\bcdot\bnabla P_\CR}{\left|\bB\bcdot\bnabla P_\CR\right|},
\end{eqnarray}
implying that the CR streaming velocity is oriented along magnetic fields lines
down the CR pressure gradient with a velocity that corresponds in magnitude to
the velocity of Alfv{\'e}n waves, $\bvel_{\rmn{A}}$. Hence, the CR source term
due to the generation of resonant Alfv\'en waves in the CR equation is always a
loss term,
\begin{eqnarray}
  \label{eq:Alfven cooling}
  \bvel_{\rmn{st}}\bcdot\bnabla P_\CR
  = -\frac{1}{\sqrt{\rho}}\,
  \frac{(\bB\bcdot\bnabla P_\CR)^2}{\left|\bB\bcdot\bnabla P_\CR\right|}<0.
\end{eqnarray}
Equivalently, the corresponding term in the thermal energy equation is always a
gain term ($-\bvel_{\rmn{st}}\bcdot\bnabla P_\CR>0$).\footnote{Note that we
  refrain from explicitly integrating the equation for the Alfv\'en wave energy
  that is resonantly generated by the streaming CRs since this energy is quickly
  dissipated.}
Furthermore, there are explicit gain and loss terms ($\Gamma_i$ and $\Lambda_i$
with $i\in\{\rmn{th},~\rmn{\CR}\}$) for the thermal and CR energy density,
respectively. If there are no explicit gain and loss terms, we clearly see that
the total energy (volume integral of $\partial (\eps+\eps_\CR)/\partial t$) is
conserved, i.e., that the source terms of these two equations vanish
identically. To close the system of equations, we assume an equation of state
for the thermal gas as well as for the CRs
\begin{eqnarray}
  \label{eq:EoS}
  P_\th &=& (\gamma_\th-1)\,\eps_\th, \\
  P_\CR &=& (\gamma_\CR-1)\,\eps_\CR,
\end{eqnarray}
with $\gamma_\th=5/3$ and $\gamma_\CR=4/3$ in the relativistic
limit. Additionally, the magnetic field has to fulfil the divergence constraint
$\bnabla \bcdot\bB=0$. To understand the physics of CR transport, we can
rewrite the equation for the CR energy density and obtain
\begin{eqnarray}
  \label{eq:epsCR}
  \frac{\partial \eps_\CR}{\partial t}
  +\bnabla\bcdot\left[ \eps_\CR (\bvel+\bvel_{\rmn{st}})
    - \kappa_\eps \bb \left( \bb \bcdot \bnabla \eps_\CR\right)\right]
  \nonumber\\
  = -P_\CR\,\bnabla\bcdot (\bvel+\bvel_{\rmn{st}}) +\Lambda_\CR + \Gamma_\CR.
\end{eqnarray}
This demonstrates that the spatial transport of CR energy density is a
superposition of advection with the frame propagating at velocity
$\bvel+\bvel_{\rmn{st}}$, as well as anisotropic diffusion along magnetic field
lines with respect to that frame. In addition to the explicit source terms
($\Gamma_\CR$ and $\Lambda_\CR$), CRs experience adiabatic gains and losses
depending on whether the Alfv\'en frame is compressed ($\bnabla\bcdot
(\bvel+\bvel_{\rmn{st}})<0$) or expanded in the laboratory frame.

The physical picture underlying this transport equation can be understood with
the following considerations. CRs with energies around the proton rest mass
energy dominate the CR pressure \citep{2007A&A...473...41E}.  CRs that propagate
faster than the Alfv{\'e}n velocity excite Alfv{\'e}n waves through the CR
streaming instability \citep{1969ApJ...156..445K}. Scattering off of this
self-excited wave field isotropises the CRs' pitch angles, thereby confining the
pressure-carrying CRs almost perfectly to the Alfv{\'e}n wave frame
\citep{2013MNRAS.434.2209W, 2016arXiv160802585W}.\footnote{Note that more
  relativistic CRs with Lorentz factors $\gamma\gtrsim100$ can drift
  super-Alfv{\'e}nically due to the weaker coupling of these CRs to the
  Alfv{\'e}n wave frame. This is because the streaming instability growth rate
  scales with the number of CRs, which is steeply declining with energy for
  usual CR power-law distributions \citep{2013MNRAS.434.2209W,
    2016MNRAS.462.4227R}.}  As a result, those CRs stream along magnetic fields
with the streaming velocity $\bvel_{\rmn{st}}$ (equation~\ref{eq:vstream}).

On the other hand, effective damping processes (such as ion-neutral
interactions) can lead to an incomplete CR confinement. Externally sourced
turbulence can interact with CRs, leading to more spatial confinement of
CRs. Effectively, these effects can be described by a diffusion process with an
energy-weighted spatial diffusion coefficient $\kappa_\eps$. In the absence of
CR diffusion ($\kappa_\eps=0$) and explicit gains and losses
($\Gamma_i=0=\Lambda_i$ for $i\in\{\rmn{th},~\rmn{\CR}\}$), the equation for the
CR energy density (\ref{eq:epsCR}) states that CRs only experience
thermodynamically reversible changes of their population as a result of
adiabatic expansion/compression of the gas as well as adiabatic losses due to CR
streaming. These streaming losses irreversibly heat the thermal gas by
non-linear Landau damping or scattering off of strong external turbulence that
cascade the energy down to the dissipation scale. However, the supply of
mechanical energy, e.g., in form of turbulence can adiabatically add CR energy
through volume-compression work done on the CRs.

CR energy sources ($\Gamma_\CR$) are provided by diffusive shock acceleration at
supernova remnants, at cosmological formation shocks, and in relativistic jets
from active galactic nuclei. Energy gain through second-order Fermi acceleration
by means of momentum space diffusion is another source of CR energy.  CR energy
sinks ($\Lambda_\CR$) are provided by hadronic interactions of CRs with thermal
gas protons, Coulomb and ionisation interactions, which will be detailed below.

\subsection{MHD with cosmic rays in comoving coordinates}

In the previous section, all spatial derivatives are taken with respect to
physical coordinates $\vec{r}$. In an expanding universe, it is convenient to
introduce spatial coordinates $\vec{x}$ comoving with the cosmological expansion
so that only gas motions with respect to that comoving rest frame need to be
computed. We parameterise the global expansion of the universe with the
time-dependent scale factor $a(t)$ that obeys Friedmann's equations. To this
end, we define a set of `comoving' variables (denoted with a subscript `c') for
velocity, mass density, magnetic field, thermal and CR pressure:
\noindent
\resizebox{\hsize}{!}{
\begin{minipage}[t]{0.5\hsize}
\begin{eqnarray*}
\vec{r}&=& a\vec{x},\\
\rho &=& \rho_\com a^{-3},\\
P_\th &=& P_{\th,\com} a^{-3},\\
\eps_\th &=& \eps_{\th,\com} a^{-3},
\end{eqnarray*}
\end{minipage}\hfill
\begin{minipage}[t]{0.5\hsize}
\begin{eqnarray}
\vec{u} &=& \bvel - \dot{a} \vec{x},\\
\bB &=& \bB_\com a^{-2},\\
P_\CR &=& P_{\CR,\com} a^{-4},\\
\eps_\CR &=& \eps_{\CR,\com} a^{-4}.
\end{eqnarray}
\vspace{0.1cm}
\end{minipage}}
Here, $\bvel=\dot{\vec{r}}$ is the physical velocity and $\vec{u}=
a\dot{\vec{x}}$ is the peculiar velocity. With the exception of the magnetic
field and CR pressure, these transformations are standard definitions. The
adopted definitions for $\bB_\com$ and $P_{\CR,\com}$ ensure that there are no
cosmological source terms in the induction equation and the equation for the CR
energy density. The latter property is a consequence of the assumption of a constant
$\gamma_\CR$. More realistic models of CRs that go beyond this
simplification would have to account for a cosmological source term. The
transformation to comoving coordinates implies the following transformation of
derivatives,
\begin{equation}
  \bnabla \equiv \bnablaR \to \frac{1}{a}\bnablaX
  \quad\mbox{and}\quad
  \frac{\partial}{\partial t} \to \frac{\partial}{\partial t} - H\vec{x}\bcdot\bnablaX,
\end{equation}
where $H\equiv H(a)=\dot{a}/a$ is the Hubble function, the transformed time derivatives are
defined at constant comoving position $\vec{x}$, and $\bnablaX$ is the vector of
spatial derivatives with respect to $\vec{x}$. Adopting our replacement rules, we find
for the comoving evolution equations
\begin{equation}
  \frac{\partial \vec{U}_{\rmn{c}}}{\partial t} + \frac{1}{a} \bnablaX\bcdot \mat{F}_{\rmn{c}}
  = \vec{S}_{\rmn{c}} ,
\end{equation}
for a vector of conserved comoving variables $\vec{U}_{\rmn{c}}$, a comoving
flux function $\mat{F}_{\rmn{c}}\equiv\mat{F}_{\rmn{c}}(\vec{U}_{\rmn{c}})$, and
comoving source terms $\vec{S}_{\rmn{c}}$. Those are given in the local
restframe by
\begingroup
\renewcommand*{\arraystretch}{1.5}
\begin{eqnarray*}
  \vec{U}_\com = \left(
    \begin{array}{c} 
      \rho_\com \\ \rho_\com \vec{u} \\ \eps_\com \\ \eps_{\CR,\com} \\ \bB_\com 
    \end{array} \right),
  {\ }
  \mat{F}_\com = \left(
  \begin{array}{c} 
      \rho_\com \vec{u} \\
      \rho_\com \vec{u} \vec{u}^{\rmn{T}} + P_\com\bm{1} - a^{-1}\bB_\com \bB_\com^{\rmn{T}} \\
      (\eps_\com + P_\com) \vec{u} - \frac{\dps\bB_\com}{\dps a} \left( \vec{u} \bcdot \bB_\com\right)
      + \frac{\dps\bvel_{\rmn{s},\com}}{\dps a^{3/2}}P_{\CR,\com} \\
      \eps_{\CR,\com} \vec{u}_{\rmn{A}}
      - a^{-1} \kappa_\eps \bb\left( \bb \bcdot \bnablaX \eps_{\CR,\com}\right) \\
      \bB_\com \vec{u}^{\rmn{T}} - \vec{u} \bB_\com^{\rmn{T}}
  \end{array} \right),
  \label{eq:flux}
\end{eqnarray*}
\begin{equation}
  \vec{S}_\com = \left(
    \begin{array}{c} 
      0 \\
      - H\rho_\com \vec{u} \\
      \frac{\dps P_{\CR,\com}}{\dps a^2}\bnablaX \bcdot \vec{u}_{\rmn{A}}
      + \Gamma_{\th,\com} + \Lambda_{\th,\com}
      - H\left(2\eps_\com-\frac{\dps\bB_\com^2}{\dps2a}\right)\\
      -\frac{\dps P_{\CR,\com}}{\dps a}\,\bnablaX\bcdot \vec{u}_{\rmn{A}}
      + \Gamma_{\CR,\com} + \Lambda_{\CR,\com}
      +H\eps_{\CR,\com} (4-3\gamma_\CR)   \\
      \mathbf{0}
    \end{array} \right).
\end{equation}
\endgroup
\noindent
Here, we introduced the total comoving pressure ($P_\com$) and the total comoving
energy density ($\eps_\com$) (excluding the CR energy density),
\begin{eqnarray}
  \label{eq:P_com}
  P_\com &=& P_{\th,\com} + \frac{1}{a}\left(P_{\CR,\com} + \frac{\bB_\com^2}{2}\right)
  = P_{\th,\com} + \tilde{P}_{\CR,\com} + \frac{\tilde{\bB}_\com^2}{2}, \\
  \eps_\com &=& \eps_{\th,\com} + \frac{\rho_\com}{2}\vec{u}^2 + \frac{\bB_\com^2}{2a}.
\end{eqnarray}
In practice, the {\sc arepo} code employs a redefinition of the CR pressure,
$\tilde{P}_{\CR,\com} = P_{\CR,\com}/a$, and magnetic field, $\tilde{\bB}_\com =
\bB_\com/\sqrt{a}$, to facilitate the notation of the total pressure.  We further
define the comoving gain and loss terms for the thermal and CR energy
densities, respectively,
\begin{eqnarray}
  \Gamma_\th = a^{-3} \Gamma_{\th,\com},
  &\qquad&
  \Lambda_\th = a^{-3}\Lambda_{\th,\com}, \\
  \Gamma_\CR = a^{-4} \Gamma_{\CR,\com},
  &\qquad&
  \Lambda_\CR = a^{-4}\Lambda_{\CR,\com}.
\end{eqnarray}
The peculiar velocity of the frame comoving with forward Afv{\'e}n waves that
are excited by streaming CRs is given by
\begin{equation}
  \vec{u}_{\rmn{A}} = \vec{u}+a^{-1/2}\bvel_{\rmn{st},\com},
  \quad\mbox{with}\quad
  \bvel_{\rmn{st},\com} =
  -\rmn{sgn}(\bB_\com\bcdot\bnabla P_{\CR,\com})\, \frac{\bB_\com}{\sqrt{\rho_\com}}.
\end{equation}
The cosmological source terms in the energy and momentum equations can
be absorbed through a redefinition of velocity and total energy density via
\begin{eqnarray}
  \label{eq:redef}
  \bw &=& a\vec{u},\\
  \tilde\eps_\com &=& a^2 \eps_\com =
  a^5\left(\eps_\th + \frac{\rho}{2}\vec{u}^2 + \bB^2\right).
\end{eqnarray}
Substituting these variables into the cosmological MHD equations with
CRs enables us to derive a set of equations in which the new variable
$\bw$ is used for the conservative variables while the fluxes are
formulated with $\vec{u}$,
\begin{equation}
  \frac{\partial \bm{\tilde{U}}_{\rmn{c}}}{\partial t}
  + \frac{1}{a} \bnablaX\bcdot \bm{\tilde{\mat{F}}}_{\rmn{c}}
  = \bm{\tilde{S}}_{\rmn{c}} ,
\end{equation}
where we redefined the conservative variables $\bm{\tilde{U}}_{\rmn{c}}$, fluxes
$\tilde\mat{F}_{\rmn{c}}\equiv\tilde\mat{F}_{\rmn{c}}(\bm{\tilde{U}}_{\rmn{c}})$,
and source terms $\bm{\tilde{S}}_\com$ via
\begingroup \renewcommand*{\arraystretch}{1.5}
\begin{eqnarray*}
  \bm{\tilde{U}}_\com = \left( \!\!
    \begin{array}{c} 
      \rho_\com \\ \rho_\com \bw \\ \tilde\eps_\com \\ \eps_{\CR,\com} \\ \bB_\com 
    \end{array} \!\!\right),
  \
  \bm{\tilde{\mat{F}}}_\com
  = \left(\!\!
  \begin{array}{c} 
      \rho_\com \vec{u} \\
      a\left(\rho_\com \vec{u} \vec{u}^{\rmn{T}} + P_\com\bm{1} - a^{-1}\bB_\com \bB_\com^{\rmn{T}}\right) \\
      a^2\left[(\eps_\com + P_\com) \vec{u} - \frac{\dps\bB_\com}{\dps a} \left( \vec{u} \bcdot \bB_\com \right)
        + \frac{\dps\bvel_{\rmn{st},\com}}{\dps a^{3/2}}P_{\CR,\com} \right] \\
      \eps_{\CR,\com} \vec{u}_{\rmn{A}}
      - a^{-1} \kappa_\eps \bb\left( \bb \bcdot \bnablaX \eps_{\CR,\com}\right) \\
      \bB_\com \vec{u}^{\rmn{T}} - \vec{u} \bB_\com^{\rmn{T}}
    \end{array} \!\!\right),
\end{eqnarray*}
\begin{equation}
  \bm{\tilde{S}}_\com = \left(
    \begin{array}{c} 
      0 \\
      \mathbf{0} \\
      P_{\CR,\com}\,\bnablaX \bcdot \vec{u}_{\rmn{A}} 
      +a^2\left(\Gamma_{\th,\com} + \Lambda_{\th,\com}\right)
      + \frac{\dps a H}{\dps2}\,\bB_\com^2\\
      -a^{-1} P_{\CR,\com}\,\bnablaX\bcdot \vec{u}_{\rmn{A}}
      + \Gamma_{\CR,\com} + \Lambda_{\CR,\com} \\
      \mathbf{0}
    \end{array} \right),
\end{equation}
\endgroup
\noindent
and we have assumed $\gamma_\CR=4/3$. This rescaling procedure
minimises the appearance of cosmological source terms in the energy equation
to a single term.

\subsection{Implementation}

To solve this coupled system of hyperbolic conservation laws in practice, we
discretize quantities on a moving unstructured mesh defined by the Voronoi
tessellation of a set of discrete points as realised in the {\sc arepo} code
\citep{2010MNRAS.401..791S}. This numerical technique is known to cure
numerical inaccuracies of smoothed particle hydrodynamics as well as adaptive
mesh-refinement techniques \citep[e.g.,][]{2010MNRAS.401..791S,
  2012MNRAS.423.2558B, 2012MNRAS.424.2999S}.  

We reconstruct the (comoving) primitive variables in the rest frame of a cell
interface (with normal vector $\vec{n}$), using an improved second-order
hydrodynamic scheme with Green-Gauss gradient estimates and a Runge-Kutta time
integration \citep{2016MNRAS.455.1134P}.  We calculate the fluxes across the
moving interface from the reconstructed primitive variables using the HLLD
Riemann solver \citep{2005JCoPh.208..315M} as previously described
\citep{2011MNRAS.418.1392P}. While the shear-Alfv{\'e}n mode remains unaffected
by a CR component, the magnitude of the fast and slow magneto-acoustic modes are
modified according to
\begin{eqnarray}
  \label{eq:cf,s}
  c_{\rmn{f,s}} \!&=&\! \left[
  \frac{\gamma_{\rmn{eff}} P + |\bB|^2 \pm
    \sqrt{\left(\gamma_{\rmn{eff}} P + |\bB|^2\right)^2
      - 4 \gamma_{\rmn{eff}} P B_n^2} }{2\rho}\right]^{1/2}\\
  \gamma_{\rmn{eff}} P \!&=&\! \gamma_\th P_\th + \gamma_\CR P_\CR,
\end{eqnarray}
where $B_n = \bB \bcdot \vec{n}$ (and equivalently for the cosmological
analogue). We use the Powell scheme for divergence control
\citep{1999JCoPh.154..284P} to evolve the (cosmological) MHD equations on our
unstructured mesh \citep{2013MNRAS.432..176P}.

To solve the CR energy equation, we passively advect $\eps_{\CR}$ (or
$\eps_{\CR,\com}$ in cosmological simulations) on the modified Courant timestep
$\Delta t_{\rmn{Cour}}=f_{\rmn{Cour}} \Delta x/(\vel_{\rmn{m}}+c_{\rmn{f,max}})$
with
\begin{eqnarray}
  \label{eq:Courant}
  c_{\rmn{f,max}} \!&=&\! \left(
  \frac{\gamma_{\rmn{eff}} P + |\bB|^2}{\rho}\right)^{1/2},
\end{eqnarray}
where $\vel_{\rmn{m}}=|\bvel_{\rmn{m}}|$ is the gas velocity in the frame of the
mesh generating point, $\Delta x$ is the cell radius, and $f_{\rmn{Cour}}$ is
the Courant factor. Note that this represents a conservative choice for the
signal speed as $c_{\rmn{f,max}}$ is the maximum speed of the fast mode, which it
acquires for propagating perpendicularly to the mean magnetic field.

The adiabatic source term $P_\CR \bnabla\bcdot\bvel$ (or its comoving analogue)
is calculated by employing Gauss' divergence theorem in every Voronoi cell and
exchanging the corresponding fluxes across the interfaces to the neighbouring
cells.  Non-adiabatic CR source terms and active CR transport processes are
treated in an operator-split fashion after evolving the homogeneous system by
one time step. First, we account for sources of CR energy such as acceleration
at resolved shocks or via a subgrid-scale model of injection at supernova
remnants.  We subsequently follow anisotropic CR diffusion.  Finally, we cool
the CR population by accounting for Coulomb and hadronic interactions.  An
implementation for CR streaming is left to future work.

Also note that we only provide the general formalism and simulations of
advective CR transport in this paper. We defer a detailed exposition of a
numerical algorithm that is capable of following anisotropic CR transport on an
unstructured mesh to a companion paper \citep{Pakmor2016a}.

\section{Non-adiabatic processes}
\label{sec:non-adiabatic}

In this section, we describe implementations for non-adiabatic processes that
provide sources ($\Gamma_\CR$) and sinks ($\Lambda_\CR$) to the CR energy
equation.  The methods described here are for the acceleration of CRs either at
shock fronts or injection via a subgrid treatment for supernova remnants and are
for the cooling of CRs through Coulomb and hadronic processes.

\subsection{CR shock acceleration}

Diffusive shock acceleration is a universal process at collisionless shocks,
which enables a small fraction of particles that impinge on the shock to gain
substantially more energy than the average particle through multiple shock
crossings. Provided that the shock propagates almost along the background
magnetic field (quasi-parallel shock geometry with respect to the shock normal),
it reforms quasi-periodically on ion cyclotron scales. Ions that enter the shock
region when the discontinuity is the steepest are specularly reflected by the
strong electrostatic shock potential and can be injected into the process of
diffusive shock acceleration \citep{2015ApJ...798L..28C}. In order to model the
CR shock acceleration process in our hydro-dynamic simulations, we need to (i)
find and characterise shocks within our simulation volume and (ii) model the CR
acceleration with an efficiency that depends on shock Mach number and magnetic
obliquity, i.e., the angle enclosed by the shock normal and upstream magnetic
field orientation.

\subsubsection{Shock finding}
\label{sec:shock_finding}

We employ the shock finding method developed by \citet{2015MNRAS.446.3992S} and
generalise it to include pre-existing as well as freshly accelerated CRs. Note
that we restrict ourselves to the case without magnetic fields -- an inclusion of
those would add two more degrees of freedom to the system so that the resulting
solution allows for two types of compressible shocks (slow-mode and fast-mode
shocks) that are in principle able to accelerate particles. Moreover, efficient
CR shock acceleration can amplify magnetic fields resonantly and non-resonantly
\citep{2014ApJ...794...46C}, further modifying the conservation laws.

A composite of thermal gas and CRs cannot be described by the equation
of state of an ideal gas with an effective adiabatic index that is
constant across the shock surface. Hence, the Rankine-Hugoniot shock
jump conditions are necessarily modified from the the case of
a single component fluid. Employing the continuity of mass and
momentum across the shock in its rest frame
(equation~(\ref{eq:generalRH}) with $\varv_\rmn{s}=0$), we can derive
a closed form for the Mach number,
\begin{equation}
  \label{eq:Mach}
  \mathcal{M}_1^2 = \frac{\varv_1^2}{c_1^2}
  =\left(\frac{P_2}{P_1}-1\right)
  \frac{x_\rmn{s}}{\gamma_{\eff}(x_\rmn{s} - 1)}.
\end{equation}
Up- and downstream quantities are denoted by subscripts 1 and 2, respectively,
$x_\rmn{s} = \rho_2/\rho_1$ is the density jump at the shock, $P_i = P_{\th\,i}+
P_{\CR\,i}$ is the total pressure with $i\in\{1,2\}$, and $\varv_1$ denotes the
velocity component parallel to the shock normal that is impinging on the shock
from the upstream. The effective sound speed, $c_1$, and adiabatic index,
$\gamma_{\eff}$, are given by
\begin{eqnarray}
  \label{eq:cs}
  c_1^2&=&\frac{\gamma_\CR P_{\CR1} + \gamma_\th P_{\th1}}{\rho_1}
  =\frac{\gamma_{\eff} P_1}{\rho_1},\mbox{ and}\\
  \gamma_{\eff} &\equiv& \left.\dd{\ln (P_\CR+P_\th)}{\ln\rho}\right|_s
  = \frac{\gamma_\CR P_{\CR1} + \gamma_\th P_{\th1}}{P_1}.
\end{eqnarray}
Here the derivative is taken at constant generalised entropy~$s$.
Moreover, it proves beneficial to generalise the expression for the
thermodynamic temperature by defining a pseudo temperature,
\begin{equation}
  \label{eq:temp}
  k \tilde{T} = \frac{P}{n} = \frac{\mu_{\rmn{mw}} m_\p (P_\th + P_\CR)}{\rho},
\end{equation}
where $n$ is the gas number density, $m_\p$ is the proton rest mass, and $\mu_{\rmn{mw}}$ is the
mean molecular weight.

We summarise the main steps of the generalised shock finding algorithm. First,
we determine the direction of shock propagation in each Voronoi cell by
identifying it with the unlimited gradient of the pseudo-temperature,
\begin{equation}
  \label{eq:direction}
  \vec{d}_\rmn{s} = -\frac{\bnabla \tilde{T}}{\left|\bnabla \tilde{T}\right|}.
\end{equation}
The shock finding algorithm identifies a {\em shock zone} by applying the
following {\em local cell-based} criteria to (i) find converging flows, (ii)
filter spurious shocks such as tangential discontinuities and contacts, and
(iii) providing a safeguard against labelling numerical noise as physical
shocks:
\begin{enumerate}
\item $\bnabla\bcdot\bvel < 0,$\\[-0.5em]
\item $\bnabla\tilde{T}\bcdot \bnabla\rho > 0,$\\[-0.5em]
\item $\tilde{\M}_1 > \tilde{\M}_{\rmn{min}}.$
\end{enumerate}
We find that equation~(\ref{eq:Mach}) is numerically not robust for spurious
shocks or very weak shocks in combination with numerical noise. Thus, we employ
the following, mathematically equivalent formulation to estimate the Mach number
for the third criterion, using the density and pressure of neighbouring cells
along the direction of shock propagation,
\begin{eqnarray*}
  \tilde{\M}_1^2 = \left\{\!
    \begin{array}{ll} 
      \frac{\dps 1}{\dps2\gamma}\left[(\gamma+1)\frac{\dps P_2}{\dps P_1}+\gamma-1\right],
      & \!\!\!\mbox{for }\gamma_1\approx\gamma_2\equiv\gamma,\\[1em]
      \frac{\dps1}{\dps\gamma_{\eff}}\,
      \frac{\dps(y_\rmn{t}-1)\mathcal{C}}
           {\dps\mathcal{C} - 
                [(\gamma_1+1)+(\gamma_1-1)y_\rmn{t}](\gamma_2-1)},
      & \!\!\!\mbox{for } \gamma_1\neq\gamma_2,
    \end{array} \right.
  \label{eq:tildeM}
\end{eqnarray*}
where
$\mathcal{C}\equiv[(\gamma_2+1)y_\rmn{t}+\gamma_2-1](\gamma_1-1)$. Here,
$y_\rmn{t} = P_2/P_1$ denotes the total pressure jump across the
interface of neighbouring cells and $\gamma_i = P_i/\eps_i + 1$, where
$P_i = P_{\th\,i}+ P_{\CR\,i}$ and
$\eps_i = \eps_{\th\,i}+ \eps_{\CR\,i}$ denote the total pressure and
energy densities in the regions $i\in\{1,2\}$, respectively. The first
line is a simplification provided $\gamma_1=\gamma_2$, and we use this
expression if $2 |\gamma_1-\gamma_2| / (\gamma_1+\gamma_2) < 0.01$.
The tilde symbol indicates that $\tilde{\M}_1$ is only a lower limit
to the true Mach number if the shock is broadened over more than the
neighbouring cells. For concreteness, we chose a minimum Mach number
$\tilde{\M}_{\rmn{min}}=1.3$. We note that the combination of
adiabatic indexes $\gamma_1$, $\gamma_2$, and $\gamma_{\eff}$ of the
previous equation are related by the shock adiabat and are {\em not}
independent variables. As a result, the Mach number has to obey the
consistency relation (which derives from the requirement that the
pressure jump is a real quantity),
\begin{equation}
  \tilde{\M}_1^2 > \frac{1}{\gamma_{\eff}}
  \left[\frac{\gamma_2^2-\gamma_1}{\gamma_1-1}+
    \sqrt{\left(\frac{\gamma_2^2-\gamma_1}{\gamma_1-1}\right)^2-\gamma_2^2}\right].
\end{equation}

We tag cells that are inside the shock zone and show a maximum compression value
along the direction of shock propagation $\vec{d}_\rmn{s}$ as {\em shock surface
  cells}. In walking along the ray that is oriented along $\vec{d}_\rmn{s}$ from
the shock surface to the `true' post-shock state, we collect the total energy
$E_{\rmn{tot},i} = E_{\th,i} + E_{\CR,i}$ of each of those cells $i$, which
represent the numerically broadened post-shock states. After reaching the
post-shock state (which represents the first cell outside the shock zone), we
reverse the direction of the ray and pass through the entire shock zone, which
has a typical thickness of 3 to 4 cells, to determine the pre-shock energy
$E_{\rmn{tot},1}$. This will be needed later on for modelling CR shock
acceleration. The Mach number for a shock surface cell is calculated with
equation (\ref{eq:Mach}) across the shock zone by using the pre- and post-shock
states in the cells directly adjacent to the shock zone. Finally, we account for
overlapping shock zones which can be present in colliding shocks \citep[for
  details, please see][]{2015MNRAS.446.3992S}.

\subsubsection{Diffusive shock acceleration of CRs}

A shock dissipates kinetic energy into thermal and non-thermal energy
with a corresponding dissipated energy density that is equal to the total
post-shock energy density corrected for the adiabatically compressed
pre-shock energy density,
\begin{eqnarray}
  \label{eq:e_diss}
  \eps_{\rmn{diss}} =
   \eps_{\th2} + \eps_{\CR2} - \eps_{\th,1} x_\rmn{s}^{\gamma_\th}
  - \eps_{\CR,1} x_\rmn{s}^{\gamma_\CR},
\end{eqnarray}
where $\eps_{\CR2}$ attains contributions from the adiabatically
compressed pre-existing CRs and the freshly accelerated CRs. The
dissipated energy flux at a shock ($f_{\rmn{diss}}$ in units of energy
per unit area and per unit time) is given by the dissipated energy
density times the post-shock velocity ($\varv_2$),
\begin{eqnarray}
  \label{eq:fth}
  f_{\rmn{diss}} &=& \eps_{\rmn{diss}} \varv_2 =
  \eps_{\rmn{diss}} \frac{\M_1 c_1}{x_\rmn{s}},
\mbox{ so that}\\
  \label{eq:dotEdiss}
  \dot{E}_{\rmn{diss}} &=& f_{\rmn{diss}} A_{\rmn{shock}}
\end{eqnarray}
is the shock-dissipated energy per unit time and $A_{\rmn{shock}}$ the area of
the shock surface within a shocked cell.

Diffusive shock acceleration is able to convert a fraction $\zeta$ of the
shock-dissipated energy to CRs,
\begin{equation}
  \label{eq:Einj}
  \Delta E_\CR = \zeta(\M_1, \theta)\, E_\rmn{diss},
\end{equation}
where $E_\rmn{diss} = \dot{E}_{\rmn{diss}} \Delta t$ in our simulations with a
discretized time integration scheme. As indicated, the CR acceleration
efficiency depends on the shock strength as given by the Mach number $\M_1$, the
magnetic obliquity $\theta$, and perhaps the plasma beta parameter.  In practice,
we define the magnetic obliquity as the angle between the shock normal and
the magnetic field direction in the shock surface cell.  We adopt an acceleration
efficiency that is determined from recent hybrid particle-in-cell (PIC)
simulations with kinetic protons and a charge neutralising electron fluid
\citep{2014ApJ...783...91C}. Provided the magnetic obliquity
$\theta\lesssim45^\circ$ (quasi-parallel geometry) and $\M_1 \geq
\M_\rmn{crit}$, the CR acceleration efficiency is $\zeta\approx
0.1$.\footnote{The acceleration efficiency in those PIC simulations is defined
  to be the energy density fraction contained in non-thermal particles and was
  determined in the post-shock rest frame, which by definition has no kinetic
  contribution to the energy. This coincides with our definition provided we
  consider strong shocks that have a negligible pre-shock thermal energy
  density.} Depending on the exact physics that determines CR injection into the
diffusive shock acceleration process, we typically use
$\M_\rmn{crit}\approx3$.\footnote{We defer a detailed study of how the magnetic
  obliquity dependence of the acceleration efficiency influences the MHD
  solution to future work.}

We inject CR energy into the shock surface cells (labelled with a subscript $s$)
as well as into the numerically broadened post-shock cells $i$ (including the
`true' post-shock cell, i.e., the first cell outside the `shock zone')
according to the following prescription,
\begin{eqnarray}
  \label{eq:DelEinj}
  \Delta E_{\rmn{tot}} &=& \sum_{j=s}^{s+n}
  \left(E_{\rmn{tot},j} - E_{\rmn{tot},1}\right),\\
  \Delta E_{\CR,i} &=& \zeta(\M_1, \theta)\, E_{\rmn{diss}}\,
  \frac{E_{\rmn{tot},i} - E_{\rmn{tot},1}}{\Delta E_{\rmn{tot}}}.
\end{eqnarray}
Here, $E_{\rmn{tot},1}$ is the pre-shock energy, $E_{\rmn{tot},i}=
(\eps_{\th,i}+\eps_{\CR,i}) V_i$, and $V_i$ is the volume of a Voronoi cell $i$.
This sum typically extends over $n=2$ to 3 cells. This prescription neglects the
adiabatically compressed thermal and CR energies over the shock, which are
insignificant for strong shocks that efficiently accelerate CRs. We found that
this prescription yields robust results in a uniform medium as well as in cases
where the shock propagates into a steeply stratified atmosphere.

\subsection{CR injection at supernova remnants}
\label{sec:SNe}

Supernova explosions drive strong shocks into the ISM, 
which are believed to be one of the most important sources of
the galactic CRs up to PeV energies. In our global (cosmological)
galaxy formation models, we cannot afford to resolve individual supernova
remnants \citep[unlike in our companion paper,][]{Simpson2016}. Hence, we
need to resort to a sub-resolution treatment of star formation and its
regulation by supernova remnants. To compute radiative cooling in the
galaxy formation simulations presented here, we assume an optically
thin gas in collisional ionisation equilibrium and include heating by
a photo-ionising, time-dependent, spatially uniform ultraviolet
background. Star formation and thermal supernovae feedback are
modelled using the hybrid multiphase model for the interstellar medium
of \citet{2003MNRAS.339..289S}. In this model, star formation is
treated stochastically and the star formation rate is correlated to
the free-fall time of self-gravitating sheets of gas, according to the
observed Kennicutt law \citeyearpar{1998ApJ...498..541K}. 

In the original model, the energy released by supernovae heats the
ambient hot phase of the ISM and evaporates clouds in supernova
remnants. These effects establish a tightly self-regulated regime for
star formation in the ISM. In practice, the ISM
is described by a sub-resolution model, which assumes the ISM to be
pressurised by star formation feedback so that there exists a
coexistence between the hot phase and embedded cold clouds in pressure
equilibrium. This is parametrized by a stiff effective equation of
state above a critical density that sets the star formation threshold,
$\rho_0 \simeq 4 \times 10^{-25}\,\rmn{g~cm}^{-3}$, above which the model
interpolates between the hot and cold phases.

In order to model CR injection\footnote{Note that we use the terminology `CR
  injection' for a one-time CR energy gain whereas we adopt the phrase `CR
  acceleration' for a continuous acceleration process at resolved shocks.} at
supernova remnants, we assume that an energy fraction $\zeta_{\rmn{SN}}=0.1$ of
the supernova energy can be transferred to a CR population
\citep{2012SSRv..173..369H, 2012A&A...538A..81M, 2013Sci...339..807A}. So far,
we only model CR injection at core-collapse supernovae and leave the modelling
of type Ia supernovae for future work.  The total energy injection rate by
supernovae for a given star formation rate density $\dot\rho_\star$ depends on
the initial mass function (IMF).  Assuming a Kroupa IMF
\citep{2001MNRAS.322..231K} and that stars above a mass of $8\,\rmn{M}_\odot$
explode as supernovae, we find a mass fraction of 0.128 of stars that end up in
supernovae, and about one supernova per $100\,\rmn{M}_\odot$ of stellar mass
formed.

With a canonical kinetic energy release of $10^{51}$~erg per
supernova, this translates to an energy injection rate per unit volume
of $\epsilon_{\rmn{SN}}\dot\rho_\star$ with
$\epsilon_{\rmn{SN}}=10^{49}\, \rmn{erg}\,\rmn{M}_\odot^{-1}$. We
model the CR energy injection per timestep of a gas cell as,
\begin{equation}
  \label{eq:Eacc}
  \Delta E_\CR = \zeta_{\rmn{SN}}\epsilon_{\rmn{SN}}\dot{m}_\star \Delta t,
\end{equation}
where $\dot{m}_\star=m_\star\dot{\rho}_\star/\rho$ is the star formation rate of
the mesh cell. The uncertainties in any of the parameters are parametrized with
$\zeta_{\rmn{SN}}$, which controls the amount of CR energy that is generated by
supernovae. We inject this energy into the local environment of every newly
created star particle using a spherical top hat filter that contains the closest
32 mesh cells of the star. The remaining supernova energy is assumed to be
transferred to subgrid-scale turbulence that dissipates and heats the
ambient gas, thereby establishing a tightly self-regulated regime of the ISM.

\subsection{Collisional loss processes of CRs}
\label{sec:loss}

CR particles propagating through a plasma will gradually dissipate their kinetic
energy, and transfer it to the surrounding thermal plasma through individual
electron scatterings in the Coulomb field of the CR particle as well as via
small momentum transfers through excitations of quantised plasma
oscillations. We refer to the sum of both effects as Coulomb
losses. Additionally, energetic CRs collide inelastically with ambient thermal
gas to produce pions, causing a catastrophic energy loss for the CRs.  This
process we refer to as hadronic losses.  The rate of these energy loss
mechanisms depends both on the physical properties of the ambient medium and on
the detailed momentum spectrum of the CR population.

Hadronic interactions only affect the high-momentum regime with a rate that is
to good approximation independent of momentum so that the spectral shapes
remains largely invariant. In contrast, an accurate determination of the Coulomb
loss rate of a CR population would require a full dynamical modelling of the CR
momentum spectrum. Particles with low momenta are most strongly affected by
Coulomb interactions with a rate that increases towards lower momenta. As a
consequence, this induces a spectral break of the CR momentum distribution:
while the high-momentum tail remains invariant, the low-momentum CR particles
transfer their energy effectively to the thermal gas, and eventually get
thermalized.  Since we are not following the momentum distribution in this work,
we will construct an equilibrium momentum distribution that balances CR
injection on the one side and CR losses on the other side. The CR cooling and
thermal heating rates are then derived from this equilibrium distribution, which
is valid on timescales long compared to impulsive changes in the injection
process.

\subsubsection{Coulomb losses}
\label{sec:Coulomb}

The kinetic energy of a proton with dimensionless momentum
$p=P_\p/(m_\p c)$ is given by
\begin{equation}
\label{eq:Tp}
E_\p(p) = \left(\sqrt{1+p^2} -1\right)\, m_\p\,c^2.
\end{equation}
The kinetic energy loss of a proton with $\gamma\ll m_\p/m_\e$ due to Coulomb
interactions in a plasma is \citep{1972Physica....58..379G}
\begin{equation}
\label{eq:Coul}
- \left( \dd{E_\p(p)}{t} \right)_{\rm Coul} = \frac{4 \upi
 \, e^4 \,n_{\rm e}}{m_{\rm e}  \, \beta \, c}
 \left[ \ln \left( \frac{2 m_{\rm e} c^2\, \beta p}{\hbar
     \omega_{\rm pl}} \right) - \frac{\beta^2}{2} \right]
 \equiv \frac{A_{\rmn{Coul}}}{\beta}.
\end{equation}
Here, $\omega_{\rm pl} = \sqrt{4 \upi e^2 n_{\rm e} /m_{\rm e}}$ is the plasma
frequency, $n_\e$ is the number density of free electrons, and $\beta \equiv
\vel_\p/c = p/\sqrt{1+p^2}$ is the dimensionless CR velocity. The energy loss
due to Coulomb interactions in a neutral gas can also be approximately
calculated with this formula, provided $n_\e$ is taken to be the total electron
number density (free plus bound). Note that atomic charge shielding effects
lower the actual Coulomb loss rate so that a more accurate description of
ionisation losses is desirable \citep{2007A&A...473...41E}. To obtain an
approximate analytical expression for the CR equilibrium distribution, we
replace the term $\beta p$ in the Coulomb logarithm with its mean value for the
spectrum, which can be written as $\langle \beta\,p \rangle =
3\,P_\CR/(m_\p\,c^2\,n_\CR)$.  We also define a CR cooling timescale due to
  Coulomb cooling as
\begin{equation}
  \label{eq:tauC}
  \tau_{\rmn{Coul}}\equiv
  \frac{\eps_\CR}{\left|\,\d\eps_\CR/\d t\,\right|_{\rmn{Coul}}}.
\end{equation}

\subsubsection{Hadronic losses}
\label{sec:hadronic}

CR protons interact inelastically with nuclei of the surrounding thermal gas and
produce mainly pions, provided their momentum exceeds the kinematic threshold
$p_\rmn{thr} m_\p c = 0.78 \mbox{ GeV}/c$ for the hadronic
reaction. The neutral pions decay after a mean lifetime of $9\times
10^{-17}\mbox{ s}$ into $\gamma$-rays, and the charged pions decay into
secondary electrons/positrons and neutrinos:
\begin{eqnarray}
  \pi^\pm &\rightarrow& \mu^\pm + \nu_{\mu}/\bar{\nu}_{\mu} \rightarrow
  e^\pm + \nu_{e}/\bar{\nu}_{e} + \nu_{\mu} + \bar{\nu}_{\mu}\nonumber\\
  \pi^0 &\rightarrow& 2 \gamma \,.\nonumber
\end{eqnarray}

The presence of CRs can be uniquely identified by the characteristic spectral
signature in $\gamma$-rays, the pion-decay bump, which is centred on half the
pion's rest mass in the differential spectrum, as well as by the neutrino
emission. The luminosity of secondary electrons/positrons produced in hadronic
interactions amounts to a fraction of 1/6 of that of the total pion luminosity
(using isospin symmetry, there is a branching ratio of 2/3 to produce charged
pions and the mean energies of the produced secondaries in the laboratory frame
is $\bra E_{e^\pm}\ket=\bra E_{\pi^\pm}\ket/4$). Those secondaries cool via
synchrotron emission and Compton scattering off photons of any radiation field
(cosmic microwave background or star light) but the associated emission
signatures are not unique due to the presence of other possible relativistic
electron distributions.

The hadronic energy loss of a CR proton is independent of the partitioning among
the pions during hadronic interactions and given by
\begin{eqnarray}
  \label{eq:had}
  -\left(\dd{E_\p}{t}\right)_{\rmn{hadr}}
  &=& n_\rmn{N} \, \sigma_\rmn{pp} K_\p \, m_\p c^3
  (\gamma-1) \, \theta(p - p_\rmn{thr})\nonumber\\
  &\equiv&A_{\rmn{hadr}}\,(\gamma-1) \, \theta(p - p_\rmn{thr}).
\end{eqnarray}
Here, $\sigma_\rmn{pp}\approx44.2$~mbarn is the total pion cross section
\citep[][ assuming that the one-dimensional CR distribution function has a
  momentum spectral index $\alpha=2.2$]{2004A&A...413...17P}, $K_\p \approx 1/2$
denotes the inelasticity of the hadronic reaction in the limiting regime
\citep{1994A&A...286..983M}, and $n_\rmn{N} = n_\e/(1 - X_\rmn{He}/2)$ is the
target nucleon density in the ICM, assuming primordial element composition with
$X_\rmn{He} = 0.24$.  By analogy with Coulomb interactions, we define a CR
  cooling timescale owing to hadronic interactions as
\begin{equation}
  \label{eq:tauC}
  \tau_{\rmn{hadr}}\equiv
  \frac{\eps_\CR}{\left|\,\d\eps_\CR/\d t\,\right|_{\rmn{hadr}}}.
\end{equation}

\subsubsection{Equilibrium distribution}
\label{sec:equilibrium}

To estimate the total loss rate due to Coulomb and hadronic reactions, we will
derive an equilibrium spectrum which balances continuous CR injection and the
described energy losses \citep[following][]{2007A&A...473...41E}. Assuming a
homogeneous environment and an isotropic CR distribution in momentum space, the
effective one-dimensional CR spectrum $f_\p^{(1)} (p, t) = 4\upi p^2 f_\p(p, t)$ is
governed by the Fokker-Planck equation,
\begin{equation}
  \label{eq:equilibrium}
  \pp{f_\p^{(1)} (p,t)}{t} + \pp{}{p} \left[\dot{p}(p,t) f_\p^{(1)} (p,t)\right]=
  Q^{(1)}(p) - \frac{f_\p^{(1)} (p,t)}{\tau_{\rmn{loss}}(p)},
\end{equation}
where the momentum loss rates due to Coulomb and hadronic interactions are
\begin{equation}
  \label{eq:loss}
  \dot{p}(p) = \left[
    \left( \dd{E_\p(p)}{t} \right)_{\rmn{Coul}} +
    \left(\dd{E_\p}{t}\right)_{\rmn{hadr}}\right]
  \left(\dd{E_\p(p)}{p}\right)^{-1} .
\end{equation}

We approximate hadronic losses as continuously occurring rather than
catastrophically, assume a sufficiently extended environment so that we can
ignore escape from the system ($\tau_{\rmn{loss}}\to \infty$), and search for
steady state solutions ($\partial f^{(1)}/\partial t=0$). We assume that the CR
injection spectrum is a power-law in momentum,
\begin{equation}
  \label{eq:inj}
  Q^{(1)}(p) = Q_{\rmn{inj}} p^{-\alpha_{\rmn{inj}}}\,\theta(p-p_{\rmn{l}}),
\end{equation}
where $\theta(x)$ is the Heaviside distribution, which is unity for positive
arguments and zero otherwise, and $p_{\rmn{l}}$ is the low-momentum cutoff.
Using the Jacobian transformation $\d E_\p(p)/\d p = m_\p c^2
\beta(p)$, we find the asymptotic steady state spectrum by assuming negligible
hadronic and Coulomb losses in the low- and high-momentum regimes, respectively,
\begin{equation}
  f_\p^{(1)}(p)= 
  \frac{Q_{\rmn{inj}}p^{-\alpha_{\rmn{inj}}}}{(\alpha_{\rmn{inj}}-1)\,A_{\rmn{Coul}}}\times
  \left\{
  \begin{array}{ll}
        p^3,    & p_{\rmn{l}} \ll p\ll p_*,\\
        p_*^3,  & p\gg p_*.\\
    \end{array}
    \right.
\end{equation}
The intersection momentum $p_*$ depends on the ratio of the Coulomb-to-hadronic
loss rates
\begin{equation}
  p_*=\sqrt[3]{\frac{A_{\rmn{Coul}}}{A_{\rmn{hadr}}}}
  =\sqrt[3]{\frac{4 \upi \, e^4 \,n_\e}{m_\e m_
    p c^4 \sigma_{\rmn{pp}} K_\p n_{\rmn{N}}}
  \ln \left( \frac{2 m_{\rm e} c^2 \bra \beta p\ket}{\hbar
     \omega_{\rm pl}} \right)}\approx1.1,
\end{equation}
where we have dropped the small term due to momentum transfers through
excitations of quantised plasma oscillations in the Coulomb cooling rate. The
numerical value of $p_*\approx1.1$ is accurate to 10 per cent for electron
densities in the range $[10^{-6},10^3]\,\rmn{cm}^{-3}$
\citep{2007A&A...473...41E}.  

We provide an analytic approximation to the steady
state equilibrium CR spectrum and introduce a matched asymptotic solution,
\begin{equation}
  f_\p^{(1)}(p)= 
  \frac{Q_{\rmn{inj}}p^{-\alpha_{\rmn{inj}}}\,\theta(p-p_{\rmn{l}})}
       {(\alpha_{\rmn{inj}}-1)\,A_{\rmn{Coul}}(p_*^{-3}+p^{-3})}.
\end{equation}
Using this approximate steady state spectrum, we derive the
Coulomb loss rate of a CR population,
\begin{eqnarray}
  \Lambda_{\rmn{Coul}}
  &=& \int_{p_{\rmn{l}}}^\infty f_\p^{(1)}(p)\left(\dd{E_\p}{t}\right)_{\rmn{Coul}}\d p\\
  &=& -2.78\times 10^{-16}
  \left(\frac{n_\e}{\rmn{cm}^{-3}}\right)
  \left(\frac{\eps_\CR}{\rmn{erg~cm}^{-3}}\right)\rmn{erg~s}^{-1}\rmn{cm}^{-3},
  \nonumber
\end{eqnarray}
where we expressed the normalisation of the injection spectrum $Q_{\rmn{inj}}$
in terms of the CR energy density $\eps_\CR$ and we adopted $p_{\rmn{l}}=0.1$
and $\alpha_{\rmn{inj}}=2.2$, which is characteristic of the CR injection
spectrum at supernova remnants or jets of active galactic nuclei. The loss rate
$\Lambda_{\rmn{Coul}}$ depends somewhat on the value of $p_{\rmn{l}}$, but only
weakly as the low-momentum part attains a weak momentum scaling,
$p^{3-\alpha_{\rmn{inj}}}$. It does, however, depend on the precise value of
$\alpha_{\rmn{inj}}$ and varies in the range $\Lambda_{\rmn{Coul}}= -1.35 \times
10^{-16}$ to $-12.8 \times 10^{-16}\,\rmn{erg~s}^{-1}\rmn{cm}^{-3}$ for
$\alpha_{\rmn{inj}}=2.1$ to $2.7$.  Similarly, the hadronic loss rate of an
equilibrium CR population is
\begin{eqnarray}
  \Lambda_{\rmn{hadr}}
  &=& \int_{p_{\rmn{l}}}^\infty f_\p^{(1)}(p)\left(\dd{E_\p}{t}\right)_{\rmn{hadr}}\d p\\
  &=& -7.44\times 10^{-16}
  \left(\frac{n_\e}{\rmn{cm}^{-3}}\right)
  \left(\frac{\eps_\CR}{\rmn{erg~cm}^{-3}}\right)\rmn{erg~s}^{-1}\rmn{cm}^{-3}.
  \nonumber
\end{eqnarray}
The hadronic loss rate depends less sensitively on the precise value of
$\alpha_{\rmn{inj}}$; it varies as $\Lambda_{\rmn{hadr}}= -8.1\times 10^{-16}$
to $-5.6 \times 10^{-16}\,\rmn{erg~s}^{-1}\rmn{cm}^{-3}$ for
$\alpha_{\rmn{inj}}=2.1$ to $2.7$. While the Coulomb loss rate is sub-dominant
for hard injection indexes $\alpha_{\rmn{inj}}\lesssim2.4$, it dominates over
hadronic losses for softer injection spectra.

The total CR energy-loss rate due to Coulomb and hadronic interactions is
$\Lambda_\CR = \Lambda_{\rmn{hadr}} + \Lambda_{\rmn{Coul}} = - \lambda_\CR n_\e
\eps_\CR$ where $\lambda_\CR = 1.022\times 10^{-15}~\rmn{cm}^3~\rmn{s}^{-1}$ is
the rate coefficient for collisional energy loss of the CRs. This allows us to
implicitly follow the associated energy loss of CRs according to the solution
\begin{equation}
  E_\CR(t)=E_{\CR}(t_0)\e^{-\lambda_\CR n_\e t}.
\end{equation}
The CR energy lost in Coulomb interactions is entirely thermalized, thereby
heating the surrounding plasma of the ISM or intracluster medium. In contrast,
most of the hadronic energy losses escape in $\gamma$ rays and neutrinos as the
interaction regions are optically thin to these hadronic decay
products. 

\begin{figure*}
\resizebox{0.325\hsize}{!}{\includegraphics{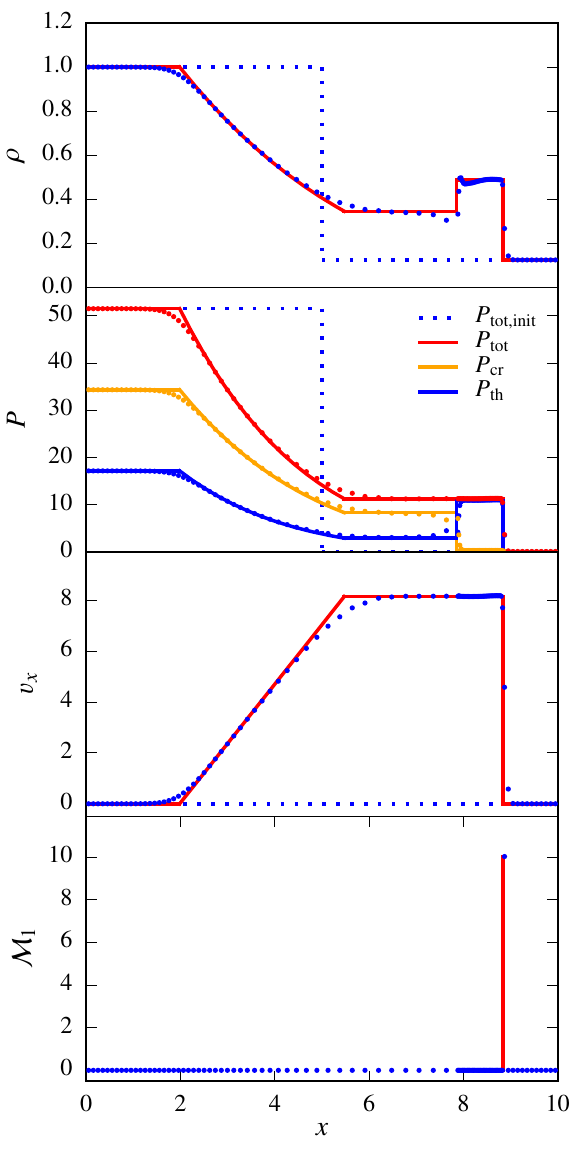}}
\resizebox{0.325\hsize}{!}{\includegraphics{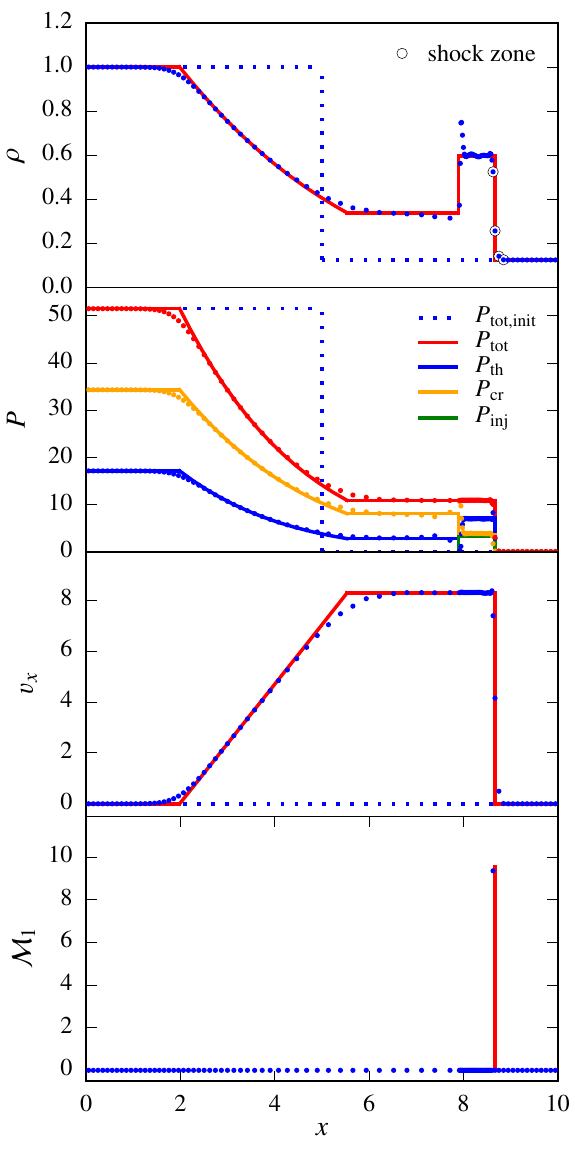}}
\resizebox{0.325\hsize}{!}{\includegraphics{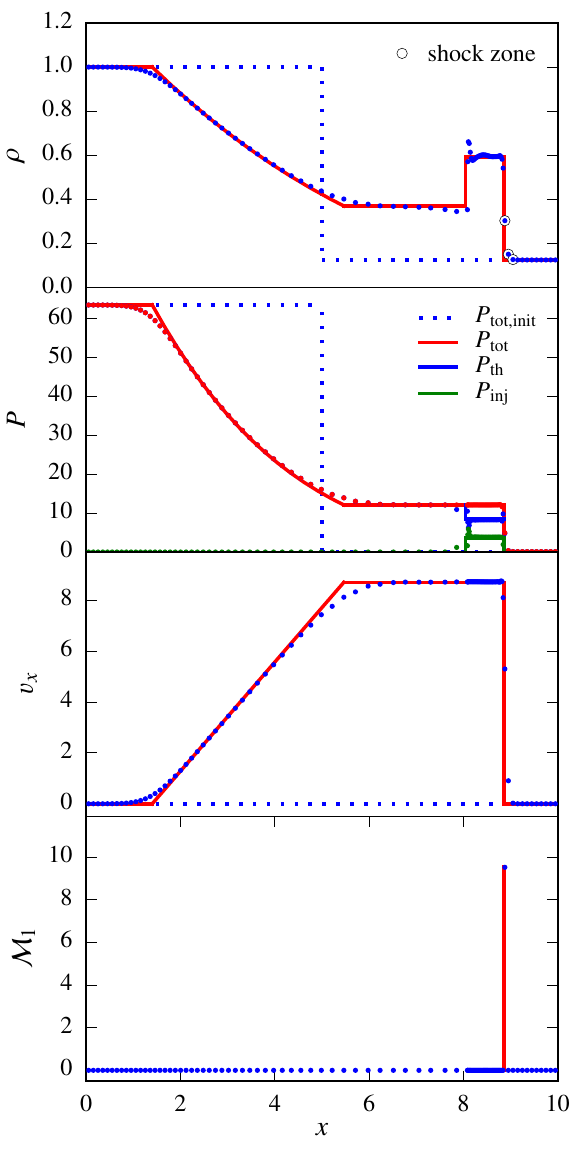}}
\caption{CR shock-tube tests of our moving mesh code. Shown are one-dimensional
  simulations with 100 cells at time $t=0.35$. In each of the three columns, we
  show from the top to bottom: mass density $\rho$, pressure $P$, velocity
  $\vel_x$, and pre-shock Mach number $\M_1$ of our simulations (points) and the
  exact solution (solid). The initial state of the Riemann problem is also shown
  (blue dotted).  Left-hand panels: the shock propagates through a composite of
  CRs and thermal gas without taking into account CR acceleration.  Initially,
  the CR pressure is twice the thermal pressure in the left-half space and equal
  to the thermal pressure in the right-half space. Upon encountering the shock,
  the CRs are only adiabatically compressed.  Middle panels: Same initial
  conditions as before, but now also accounting for CR shock acceleration
  ($P_{\rm inj}$ shown in green). Shown is also the shock zone (open circles)
  and the shock surface (cell with $\M_1\ne 0$). As a result of CR acceleration,
  the post-shock gas is more compressible and denser. Thus, the shock does not
  propagate as fast as in the case without CR acceleration.  Right-hand panels:
  this shock tube shows CR-shock acceleration in a thermal gas without
  pre-existing CRs. }
\label{fig:shock_tube}
\end{figure*}

As explained in Sect.~\ref{sec:hadronic}, on average a fraction of 1/6 of the
hadronically dissipated luminosity ends up in secondary
electrons/positrons. While the highly energetic electrons ($\gamma_\e\gtrsim
10^3$) cool via synchrotron emission as well as Compton scattering, most of the
mildly relativistic electrons ($\gamma_\e\lesssim 200$) will thermalize this
energy via Coulomb collisions \citep{1979ApJ...227..364R,
  1999ApJ...520..529S}. The secondary electrons generated at the kinematic
threshold of the hadronic reaction have a mean energy of $\bra
E_{\e^\pm}\ket=m_{\pi^\pm}c^2/4= 35\,\rmn{MeV}= 68\,m_\e
c^2$.  Hence, we assume that the majority of the energy of
secondary electrons/positrons is used for heating the surrounding plasma by
Coulomb interactions. As a result, the collisional heating rate due to Coulomb
and hadronic interactions is given by $\Gamma_\th = - \Lambda_{\rmn{Coul}} -
\Lambda_{\rmn{hadr}}/6 = \lambda_\th n_\e \eps_\CR$, where $\lambda_\th = 4.02\times
10^{-16}~\rmn{cm}^3~\rmn{s}^{-1}$ is the rate coefficient for collisional
heating of the ambient plasma by CRs. We obtain a gain of thermal energy of
\begin{equation}
  \Delta E_\th=E_{\CR}(1-\e^{-\lambda_\th n_\e t}).
\end{equation}
due to these collisional CR heating processes.

\subsubsection{Limitations}
\label{sec:limitations}

What are the conditions of validity for our approach of computing the CR loss
rate from the equilibrium distribution? Clearly, this is an excellent
description if CR injection balances the non-adiabatic cooling processes.  As we
will show in the following, this equilibrium distribution also provides a good
approximation to the case of a freely cooling CR population in the absence of
injection at late times.  While Coulomb cooling increases the spectral break,
catastrophic CR losses due to hadronic interactions lowers the break. As a
result, the CR population evolves towards an attractor solution that resembles
the equilibrium distribution and exhibits an identical spectral break, but shows
a low-momentum spectrum with a modified power-law index \citep[see figures 5 and
  6 in][]{2007A&A...473...41E}.

In general, a freely cooling CR spectrum has to be integrated numerically, but
we can analytically demonstrate this effect for a simplified case of a CR
power-law distribution,
\begin{equation}
  \label{eq:fp}
  f(p)=Cp^{-\alpha}\theta(p-q).
\end{equation}
Non-adiabatic cooling processes lead to a change of the energy density, $\d
\eps_\CR = \left(\tau_{\rmn{hadr}}^{-1} + \tau_{\rmn{Coul}}^{-1}\right)\,
\eps_\CR\,\d t$ and a change of the number density $\d n_\CR = \eps_\CR \d t /
[\tau_{\rmn{Coul}}E_\p(q)]$ because hadronic interactions conserve the number of
CRs. The implied change in the normalisation $C$ and low-momentum cutoff $q$ can
be straightforwardly calculated by means of a Jacobian transformation. Hence, at
late times Coulomb and hadronic interactions reach a balance with fixed cutoff
$q_{\rmn{fix}}$, which is given as a solution of the equation
\citep{2008A&A...481...33J}
\begin{equation}
  \label{eq:qfix}
  \frac{T_\CR(\alpha,q_{\rmn{fix}})}{E_\p(q_{\rmn{fix}})}
  = 1 + \frac{\tau_{\rmn{Coul}}(q_{\rmn{fix}})}{\tau_{\rmn{hadr}}(q_{\rmn{fix}})},
\end{equation}
where $T_\CR = \eps_\CR/n_\CR$ is the average kinetic energy of a CR
population. The normalisation decreases with time as
\begin{equation}
  \label{eq:dCdt}
  \frac{\d\ln C}{\d t}
  = \left[\tau_{\rmn{hadr}}(q)\,
    \left(1 - \frac{E_\p(q)}{T_\CR(\alpha,q)}\right)\right]^{-1}
  \to \frac{\tau_{\rmn{hadr}} + \tau_{\rmn{Coul}}}
  {\tau_{\rmn{hadr}} \tau_{\rmn{Coul}}},
\end{equation}
for $q\to q_{\rmn{fix}}$. This shows that a freely cooling CR distribution has
an invariant spectral shape at late times and decays with a constant rate that
is given by the right-hand side of equation~(\ref{eq:dCdt}).

However, this approach is potentially problematic if the real CR spectrum is not
in equilibrium.  This is particularly true after a fresh CR injection event, in
which case Coulomb losses should efficiently thermalize the low-momentum part of
the CR distribution. Instead, Coulomb cooling of our equilibrium spectrum only
removes CRs at low (late-time) cooling rates.  The net result is an initially
substantially slower Coulomb cooling than a full (non-equilibrium) solution with
a time-dependent spectrum would provide. We defer a detailed study, which
follows the CR spectrum in time and space, to future work.

\section{Results}
\label{sec:results}

All simulations and application runs discussed in the following section were
performed with the moving mesh set-up of {\sc arepo} using standard parameters
for mesh regularisation \citep{2012MNRAS.425.3024V, 2016MNRAS.455.1134P}.  We
also repeated some of these simulations with a spatially fixed mesh and report
any significant differences to the moving mesh case.

\subsection{Shock tubes}

To assess the validity of our shock acceleration algorithm and to gain
confidence in our numerical CR implementation, we perform a sequence
of shock-tube simulations in one, two, and three dimensions, with a
range of shock strengths, and with various resolutions. In
Fig.~\ref{fig:shock_tube}, we present three types of problem set-ups:
a shock tube that encounters a pre-existing population of CRs, with
and without taking into account CR acceleration, as well as a shock
propagating in thermal gas with inlined CR acceleration. Note that
since this problem is scale-free, the solution can be scaled to any
astrophysical problem at hand.

Solutions of the Riemann shock-tube problem exist in the case of a
single polytropic fluid \citep{1948sfsw.book.....C} as well as for a
two-component fluid composed of CRs and thermal gas
\citep{2006MNRAS.367..113P}, in which the CRs are adiabatically
compressed over the shock. So far, an exact solution that also
accounts for CR shock acceleration has not been reported in the
literature to our knowledge. To close this gap, we derive such an
exact solution of a shock tube with CR shock acceleration in the case
of a shock propagating in purely thermal gas
(Appendix~\ref{sec:Riemann}) as well as for the case of a shock
propagating in a composite of CRs and thermal gas
(Appendix~\ref{sec:Riemann+CRs}).

Figure~\ref{fig:shock_tube} shows the resulting shock-tube setup for our
one-dimensional simulations with 100 mesh cells that are initially equally
spaced. We choose our initial conditions such that a shock with Mach number of
$\M_1=10$ forms in the absence of CR acceleration. Any deviation of that is an
immediate consequence of CR acceleration. The detailed simulation set-up is
reported in Table~\ref{tab:shock tubes}.  To examine our CR shock-injection
algorithm at the extreme, we choose a high value for the CR acceleration
efficiency of $\zeta=0.5$ and do not account for non-adiabatic CR loss
processes.

\begin{table*}
\caption{Initial set-up of our shock tubes and resulting shock strengths.}
\begin{tabular}{l cc cc cc cc}
\hline
\hline
& $\rho_5^{(1)}$ & $\rho_1$ & $P_5$ & $P_1$ & $X_{\CR5}$ & $X_{\CR1}$ &
$x_\rmn{s}^{(2)}$  & $\M_1$\\ 
\hline
th     & 1 & 0.125 & 63.499 & 0.1 & 0 & 0 & 3.88 &  10.00 \\
th+inj & 1 & 0.125 & 63.499 & 0.1 & 0 & 0 & 4.74 & ~~9.56 \\
CR     & 1 & 0.125 & 51.516 & 0.1 & 2 & 1 & 3.90 &  10.00 \\
CR+inj & 1 & 0.125 & 51.516 & 0.1 & 2 & 1 & 4.78 & ~~9.56 \\
\hline
\end{tabular}
\begin{quote}
\ \\
(1) The left-half space in the initial conditions ($x<5$) is denoted with a
  subscript 5 while the right-half space ($x>5$) has a subscript 1.\\
(2) The density jump at the shock is given by $x_\rmn{s}=\rho_2/\rho_1$.
\end{quote}
\label{tab:shock tubes}
\end{table*} 

Overall, there is excellent agreement of our simulations with the exact
solutions. This agreement implies an excellent performance of the approximate
HLLD Riemann solver for the two-fluid problem of thermal gas and CRs that our
exact solution adopts. We note that the averaged hydrodynamic values in our
two- and three-dimensional shock tube simulations match the analytic solutions
similarly well. To understand how our CR injection algorithm works in practice,
we also show the shock zone (Section~\ref{sec:shock_finding}). The shock
surface, i.e., the cell within the shock zone that has the highest velocity
divergence, characterised by $\M_1\ne 0$, is almost always situated at the edge
of the shock zone. This result holds also for weaker shock waves. It implies
that our algorithm injects CRs mostly into two cells, the shock surface and the
first cell outside the shock zone, in the direction towards the post-shock
region. Injecting and accelerating CRs from the thermal pool lowers the
effective adiabatic index of the post-shock gas, making it more compressible and
thus denser. As a result of this and as a consequence of mass conservation, the
shock does not propagate as fast as in the case without CR acceleration.

The only noticeable difference to the exact solution is the high-density blip in
the first two cells past the contact discontinuity. This comes about because in
the first few time steps after the start of the simulation when the shock has
not yet fully developed and the post-shock regime is about to form, our
algorithm injects too much CR energy because the estimated pressure jump is
initially too large. While this causes an increased compressibility in
comparison to the exact solution, the algorithm recovers as soon as the shock
and post-shock regime have formed and then performs correctly.

We note that our second-order hydrodynamic scheme with a {\em fixed mesh}
exhibits chequerboard instabilities within the rarefaction wave at the position
of the high-pressure state of the initial conditions. These are known numerical
artefacts that result from the large expansion velocity relative to the fixed
mesh and cause the approximate Riemann finder to return erroneously the
characteristics of the shock state instead of the intermediate state
representing the rarefaction fan.\footnote{We caution that a better
  reconstruction scheme for the fixed-mesh (that avoids the pairwise decoupling)
  should also be able to correct for these inaccuracies.} If run with the moving
mesh, these chequerboard instabilities vanish identically because the lower
velocity relative to the moving mesh enables the Riemann solver to correctly
identify the intermediate states.

\subsection{Sedov-Taylor blast wave}

\begin{figure*}
\includegraphics{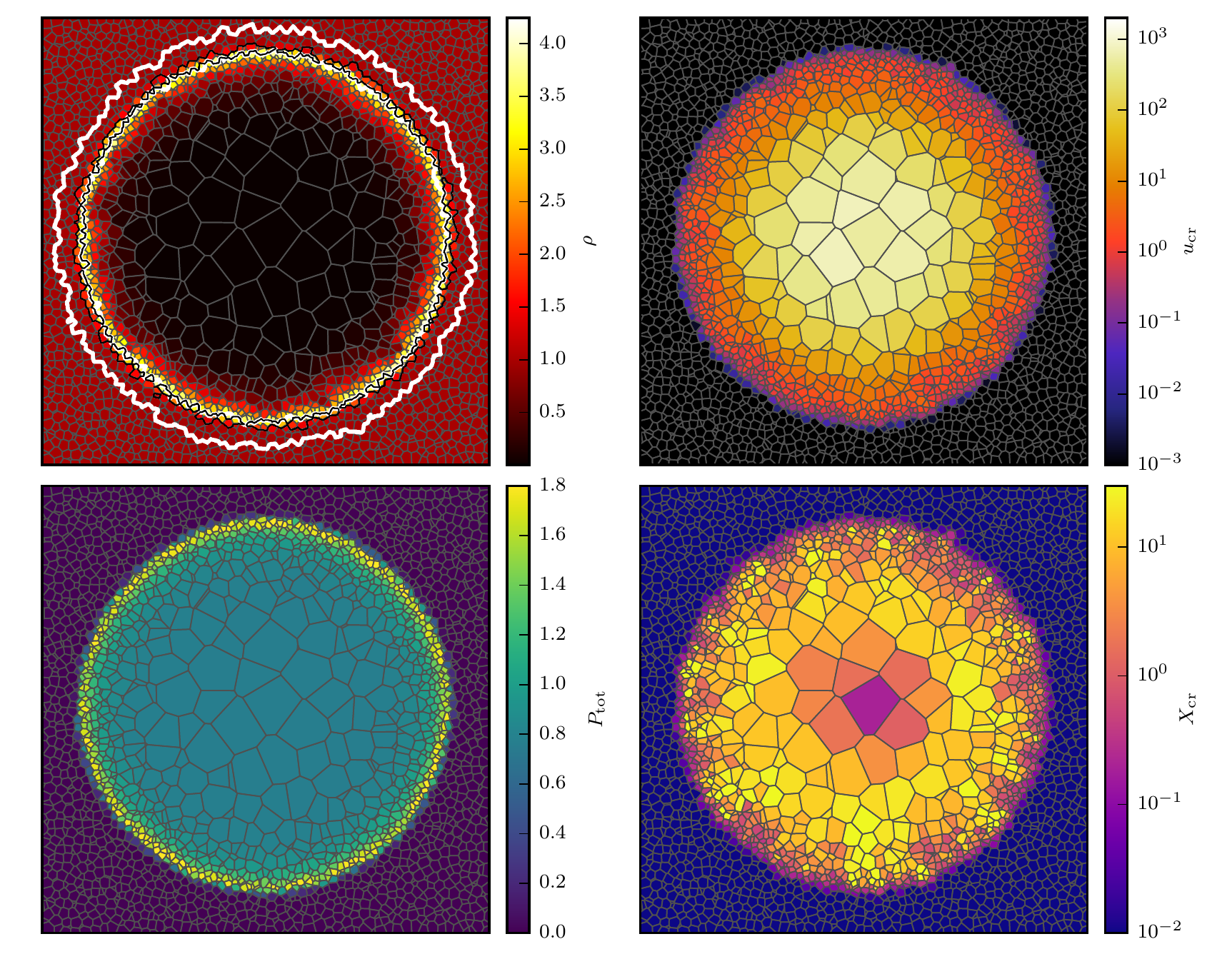}
\caption{Sedov-Taylor blast wave with CR acceleration. We present cross-sections
  through the centre of the 3D simulation volume of our CR shock acceleration
  simulation with $50^3$ cells at time $t=0.1$. Shown are the mass density
  $\rho$ (top left), the CR energy per unit mass $u_\CR$ (top right), the total
  pressure $P_{\rm tot}=P_\th+P_\CR$ (bottom left), and the relative CR
  pressure $X_\CR=P_\CR/P_\th$ (bottom right). In the density cross-section, we also
  visualise the shock zone (bounded by two thick white contour lines) as well as
  the shock surface (thin black contour).}
\label{fig:sedov}
\end{figure*}

\begin{figure*}
\includegraphics{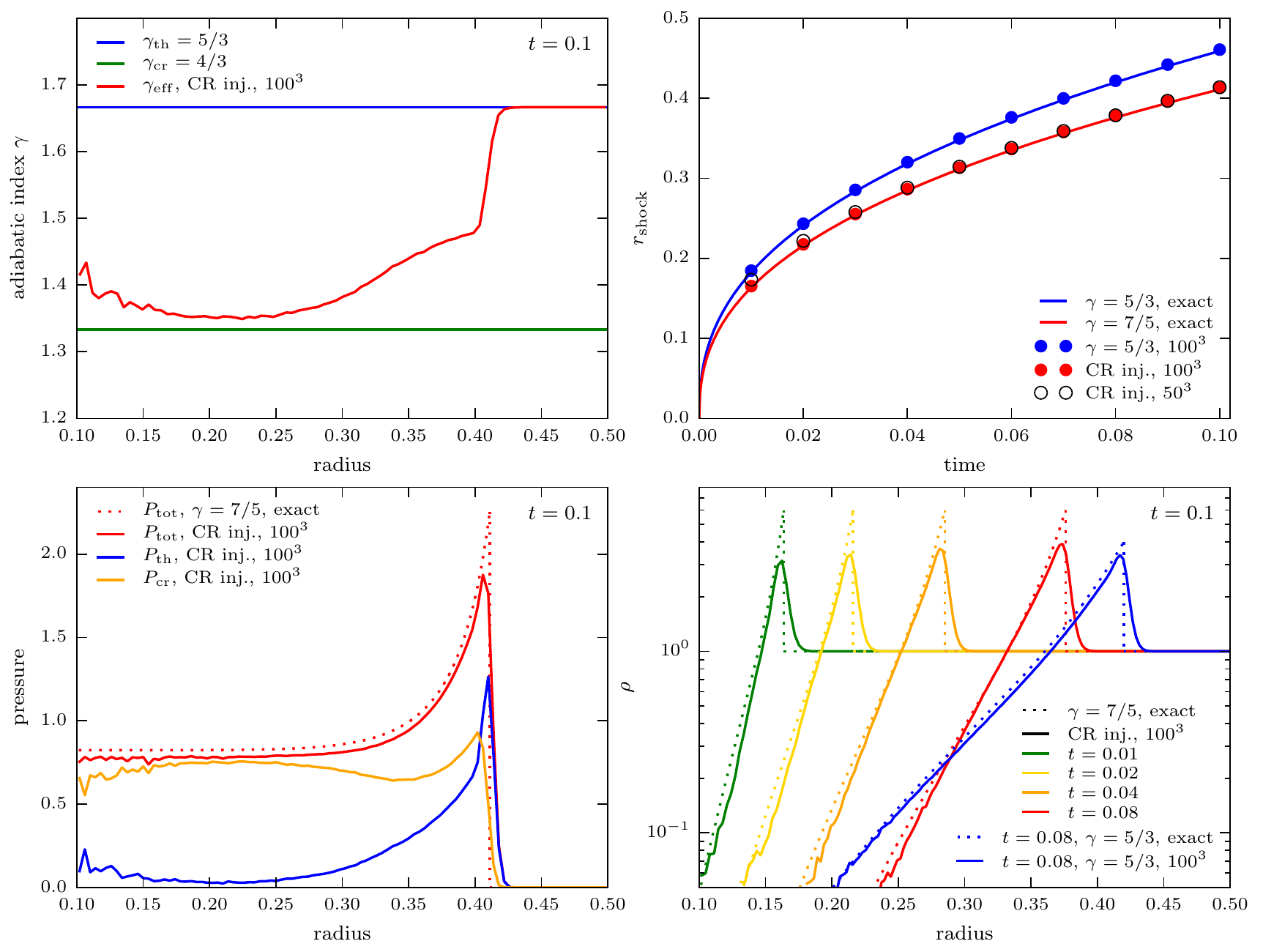}
\caption{3D Sedov-Taylor blast wave simulations with CR shock acceleration. All
  radial profiles show azimuthally averaged quantities and are centred on the
  centre-of-mass of the initial cell into which the energy was injected. Top
  left: we show the effective adiabatic index $\gamma_\eff$ of our $100^3$-cell
  simulation, which transitions from the thermal gas value of $\gamma=5/3$
  outside the shock to a value that is almost fully CR-dominated well inside the
  blast wave. Top right: time evolution of the azimuthally averaged shock radius
  of our simulations (data points, determined from the shock surface property)
  and the exact solution (solid line). Except for the innermost data point at
  $t=0.01$, the $100^3$-cell simulation has well converged to the $50^3$-cell
  run.  Bottom left: we compare the simulated total pressure (solid red) to the
  exact solution for a polytropic gas with $\gamma=7/5$ (dotted red) and show
  the contributions from the partial pressures of the thermal (blue) and CR
  constituents (orange). Bottom right: time sequence of the density profiles of
  our simulation with CR acceleration (solid, green-to-red) and the exact
  solution for a polytropic gas with $\gamma=7/5$ (dotted, green-to-red). Also
  shown is the density profile of a simulation without CR acceleration (solid blue)
  and the exact solution (dotted blue), which has advanced further in comparison
  to the case with CR acceleration.}
\label{fig:sedov analysis}
\end{figure*}

We are interested in how CR shock acceleration changes the expansion
behaviour of spherical shocks and whether it breaks self-similarity of
the exact solution in the case without CR acceleration. To this end,
we perform a sequence of simulations of the Sedov-Taylor problem,
i.e.~a point explosion leading to an energy-driven, spherically
symmetric strong shock that expands into a cold external medium with
negligible pressure. The curved geometry of the shock surface in this
problem is well suited to explore the behaviour of our CR shock
acceleration algorithm with a fully unstructured 3D Voronoi mesh free
of any preferred directions in the initial conditions.

The initial Voronoi mesh is generated by distributing mesh-generating points
randomly in the unit box $(x, y, z) \in [0, 1]^3$. We relax the mesh via Lloyd's
algorithm \citeyearpar{Lloyd} to obtain a glass-like configuration. The initial
conditions are as follows: we fill the box with a uniform density gas of
$\rho_0=1$ and pressure $P_0=10^{-4}$, the initial velocities are identically
zero and we adopt an thermal adiabatic index of $\gamma=5/3$.  We inject an
initial thermal energy of $E_0=1$ into the central mesh cell. To explore
numerical convergence, we repeat the simulations at various resolutions
corresponding to $25^3$, $50^3$, and $100^3$ cells.  In these simulations, we
follow hydrodynamics without self-gravity and -- in case of CR-shock
acceleration -- assume a large CR acceleration efficiency of $\zeta=0.5$ in
order to examine our algorithm under extreme conditions. We do not account for
non-adiabatic CR loss processes.
 
Figure~\ref{fig:sedov} shows cross-sections through the centre of the
3D volume of our simulation with CR shock acceleration at $t=0.1$. We
chose the low-resolution simulation with $50^3$ cells so that features
of the CR distribution can be related to individual mesh cells, which
are also shown in the images. The original glass-like distribution of
mesh cells is still visible in the corners of the maps.  We note that
the cross-section of a 3D Voronoi mesh is in general no longer a
Voronoi tessellation.  We address the details of the CR shock
acceleration algorithm in the density cross-section
(Fig.~\ref{fig:sedov}, top left). The area in between the two white
contour lines shows the {\em shock zone}, i.e. it is characterised by
converging flows for which spurious shocks such as tangential
discontinuities and contacts have been filtered out. The velocity
convergence along the direction of shock propagation is maximised in
the cell labelled as {\em shock surface}. The collection of
shock-surface cells is enclosed by the black contour lines. As before
in the case of the (strong) shock tube simulations, the shock zone
extends typically over three to four cells along the shock normal and
the inner boundary that connects to the post-shock region often
coincides with the shock surface cells. This implies that we typically
inject CRs into two (or rarely three) cells along the {\em direction
  of shock propagation} $\vec{d}_\rmn{s}$ as inferred by the local
orientation of the temperature gradient.

The shock velocity decreases steeply with time in the Sedov solution. Thus, most
of the specific energy is dissipated at early times. Because we funnel a fixed
fraction of the dissipated energy into CRs in our model, the CR energy per unit
mass, $u_\CR$, decreases also steeply with radius (Fig.~\ref{fig:sedov}, top
right). In contrast, the total pressure peaks at the shock surface and plateaus
to a constant value inside the blast wave (Fig.~\ref{fig:sedov}, bottom left),
indicating that the density decrease due to adiabatic expansion is exactly
countered by the increase in specific energy. Most interestingly, the relative
CR pressure, $X_\CR=P_\CR/P_\th$, increases steeply with increasing distance from
the shock (Fig.~\ref{fig:sedov}, bottom right). This is because adiabatic
expansion of a composite of CR and thermal gas favours the CR pressure over the
thermal pressure due to the softer equation of state of CRs. 

For our adopted
value for the CR acceleration efficiency $\zeta=0.5$, we obtain a CR pressure
ratio at the shock of $X_{\CR0}=(\gamma_\CR-1)\eps_{\CR0} /
[(\gamma_\th-1)\eps_{\th0}] = \zeta/[2\,(1-\zeta)]= 0.5$. Adiabatic expansion
over a density expansion factor $\delta$ yields
\begin{equation}
  X_\CR =  \frac{P_{\CR0}\,\delta^{\gamma_\CR}}{P_{\th0}\,\delta^{\gamma_\th}}
  = X_{\CR0}\, \delta^{-1/3}\approx 5\, \left(\frac{\delta}{10^{-3}}\right)^{-1/3},
\end{equation}
in agreement with the simulated average values. However, the cross-section of
$X_\CR$ shows a substantial scatter in this quantity at any given radius. A
close inspection reveals radial `fingers' of CR over-pressured pockets close to
the shock surface.  This can be traced back to the scatter in the temperature
gradient that defines the opposite {\em direction of shock propagation}. This
causes individual `shock rays' that point opposite to the direction of shock
propagation to sometimes converge in a single post-shock cell. As a result, we
inject CR energy (and accordingly remove thermal energy) twice in those cells
and consequently skip CR injection in adjacent cells if those are missed by a
`shock ray'. While this may represent a weakness of our CR injection
algorithm, we emphasise that physical CR diffusion (which is neglected here for
clarity) will smooth out these irregularities on short timescales.\footnote{The
  fact that this scatter in $X_\CR$ can be maintained by the code, is a
  manifestation of the low numerical diffusivity of our moving mesh technique.}

In order to quantify these visual impressions, we would like to
compare radial profiles of azimuthally averaged quantities to exact
solutions. Unfortunately, a solution of the Sedov problem for a
two-component fluid which accounts for CR acceleration at the
expanding shock does not exist. In order to check whether such a
simulation with CR acceleration obeys a self-similar behaviour, we
need to derive a constant adiabatic index that best describes the
solution. To this end, we show the radial profile of the effective
adiabatic index $\gamma_\eff$ (equation~\ref{eq:cs}) of our
$100^3$-cell simulation at $t=0.1$ (Fig.~\ref{fig:sedov analysis}, top
left). It transitions from the thermal gas value of $\gamma=5/3$
outside the shock to a value that is almost fully CR dominated well
inside the blast wave, because the adiabatic expansion favours the CR
pressure over the thermal gas pressure (Fig.~\ref{fig:sedov analysis},
bottom left). Overall, a value of $\gamma=7/5$ characterises the
average behaviour of $\gamma_\eff$ well, as can be validated by
comparing the total simulated pressure profile to the exact profile of
a polytropic fluid of $\gamma=7/5$.

In the case of a single polytropic fluid, the shock radius of a 3D explosion
evolves as
\begin{equation}
r_{\rmn{shock}}(t) = \left(\frac{E_0}{\alpha \rho_0}\right)^{1/5} t^{2/5},
\end{equation}
where $\alpha = (0.49,0.851)$ for $\gamma=(5/3,7/5)$ according to Sedov
\citeyearpar{1959sdmm.book.....S}. In Figure~\ref{fig:sedov analysis} (top
right) we show the time evolution of the azimuthally averaged shock radius of
our simulation without (blue points) and with CR acceleration (red points). In
agreement with our shock tube simulations, the blast wave that accelerates CRs
propagates at a slower rate in comparison to the case without CR
acceleration. This comes about because of the softer effective equation of state
of the composite fluid that allows higher post-shock densities and thus cannot
advance as fast. We demonstrate that our $100^3$-cell simulation has converged
to the $50^3$-cell run, with the exception of the innermost data point at
$t=0.01$. These measured shock positions compare well to the exact values for
$\gamma=5/3$ and $7/5$, respectively. This shows that a constant value for
$\gamma_\eff$ captures the overall expansion behaviour of the blast wave despite
the fact that the adiabatic index is changing as a function of radius. This can
be explained because in our model, a fixed fraction of the dissipated energy is
injected into CRs so that the CR pressure ratio experiences the same adiabatic
expansion for the same dilution factor.

This can be substantiated further by comparing the density profiles of
our simulation with CR acceleration (solid, green-to-red) to the
self-similar analytical solution for a polytropic gas with
$\gamma=7/5$ (dotted, green-to-red) at different times. The shapes of
the density profiles stay approximately self-similar and show similar
deviations from the analytical solution as the density profile of a
simulation without CR acceleration (solid blue) and the exact solution
(dotted blue). However, the CR acceleration run differs by a larger
factor from the exact solution at the shock in comparison to the
polytropic run with $\gamma=5/3$. This is because our approximation of
$\gamma=7/5$ clearly breaks down at the shock, where we obtain an
effective adiabatic index, $\gamma_\eff$, and shock compression ratio,
$x_\rmn{s}$, of
\begin{eqnarray}
  \gamma_\eff &=& \frac{\gamma_\CR X_{\CR0} + \gamma_\th}{X_{\CR0} + 1}=
  \frac{14}{9}=1.\bar{5},
  ~\mbox{and}\\
  \label{eq:xs}
  x_{\rmn{s}} &=& \frac{\gamma_\eff+1}{\gamma_\eff-1}=\frac{23}{5}=4.6.
\end{eqnarray}
The expression for $x_\rmn{s}$ in equation~(\ref{eq:xs}) is valid in the limit of a
strong shock, and the numerical values are obtained by adopting the assumed CR
acceleration efficiency of $\zeta=0.5$. We conclude that the azimuthally
averaged post-shock density falls short of the theoretically expected value by
about $15\%$, which is a similar deficiency factor as for the polytropic run
with $\gamma=5/3$.

\subsection{Isolated models of galaxy formation}

We now assess the impact of CR pressure feedback on the formation and evolution
of isolated disk galaxies in dark matter haloes that range in mass from
$10^{10}$ to $10^{12}\msun$. We would like to understand how exactly CRs
accelerated by supernova remnants are able to regulate star formation by means
of their pressure feedback.  Moreover, we are interested how the structure of
the ISM changes as a consequence of CR feedback and whether this has any direct
consequences on magnetic dynamo amplification mechanisms.\footnote{Here, we
  report values for the magnetic field strength in cgs units.} We are in
particular interested how the strength of CR pressure feedback depends on halo
mass, because the global star conversion efficiency needs to be a strong
function of halo mass in order to explain the shallow faint end of the galaxy
luminosity function. While strong galactic outflows in form of winds are
potentially responsible for the majority of this mass-dependent regulation of
star formation, we will study here how much of this reduction of star formation
is directly related to the additional non-thermal pressure support of CRs.

We model the ISM by an effective pressurised equation of state and follow
radiative cooling and star formation using the approach by
\citet{2003MNRAS.339..289S}. We employ ideal MHD using cell-centred magnetic
fields and the \citet{1999JCoPh.154..284P} scheme for divergence control
\citep{2011MNRAS.418.1392P, 2013MNRAS.432..176P}.  In these simulations, we
account for CR injection at supernovae with a CR energy injection efficiency of
$\zeta_{\rmn{SN}}=0.1$, follow advective CR transport, and account for adiabatic
changes of the CR energy as well as Coulomb and hadronic CR cooling as detailed
in Sects.~\ref{sec:SNe} and \ref{sec:loss}.\footnote{Here and in
  Sect.~\ref{sec:cosmo}, we use the collisional heating rate due to Coulomb
  interactions only, where $\Gamma_\th = - \Lambda_{\rmn{Coul}} =
  \tilde\lambda_\th n_\e \eps_\CR$ and $\tilde\lambda_\th =2.78\times
  10^{-16}~\rmn{cm}^3~\rmn{s}^{-1}$.}  Note that in these simulations, we use
the sub-resolution model of CR injection at supernovae (Sect.~\ref{sec:SNe}) and
do not employ our explicit shock finding method and associated CR
acceleration. Also, we neglect active CR transport in the form of anisotropic
diffusion and streaming, which is studied in detail in a companion paper
\citep{Pakmor2016b} and only consider advective CR transport with the
gas. Because in this approximation, CRs are tied to the gas, they cannot diffuse
ahead of the gas and, as a result of this, their pressure gradient cannot
impulsively start to dominate the force balance, which is a necessary condition
for accelerating the gas in order to launch a powerful galactic wind.

The initial conditions are given by a prescribed dark matter potential that
results from a density distribution motivated from cosmological simulations
\citep{1997ApJ...490..493N}. The density profile is characterised by a
concentration parameter, which we keep fixed at a value of $c_{200}=12$ across
our halo mass range. Hence, the haloes are scaled versions of each other which
would evolve in a self-similar way if we only considered gravity and ideal
\mbox{(magneto-)}hydrodynamics. However, the physics of cooling, star formation
and CRs breaks this scale-invariance and any quantitative differences among the
haloes can be directly related to additional scales introduced by these
processes.

We adopt a hydrostatic gas distribution that is initially in equilibrium within
the halo. We assume that the halo carries a small amount of angular momentum,
parametrized by a spin parameter $\lambda = 0.05$, which is close to the median
found in cosmological simulations.  In all cases, we use a baryon mass fraction
of $\Omega_\rmn{b}/\Omega_\rmn{m} = 0.155$. The magnetic field is initialised
through a uniform homogeneous seed field along the $x$-axis with an initial
strength of $10^{-10}\mathrm{G}$, and there are no CRs in the initial setup.

\begin{figure*}
\includegraphics{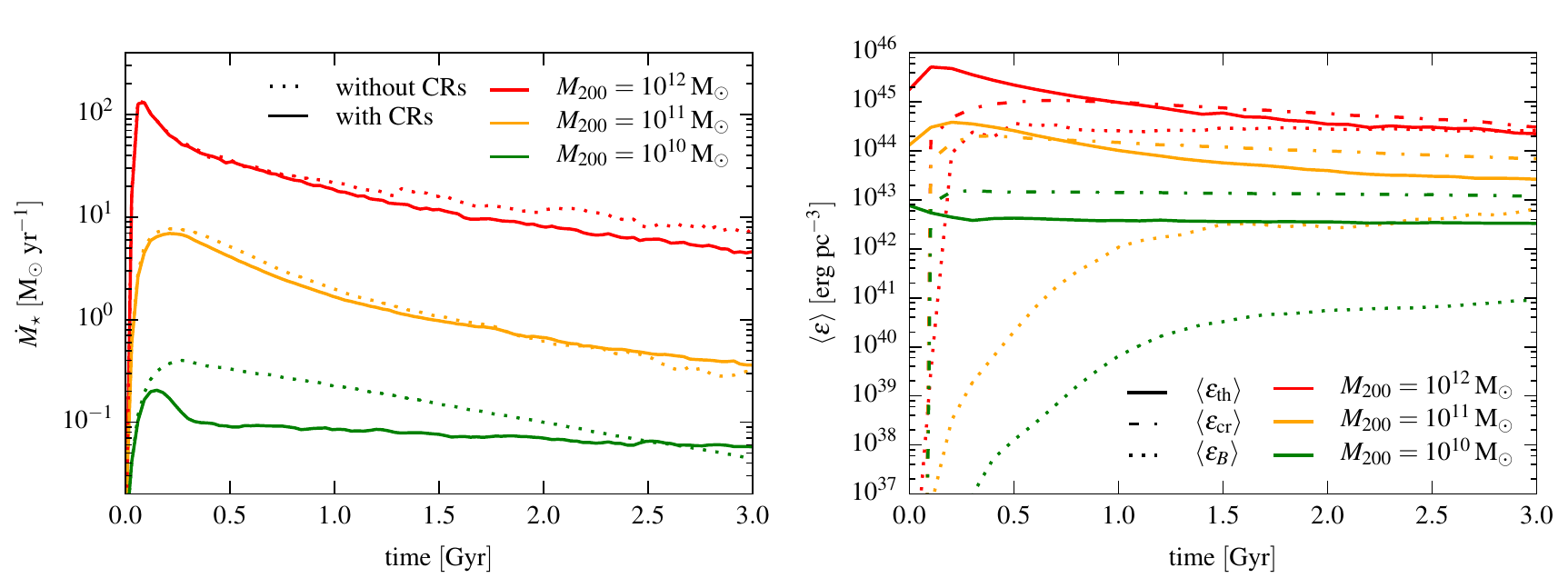}
\caption{Time evolution of the star formation rate (left-hand panel) and the
  average energy densities in a disk of radius 10~kpc and height 1~kpc that is
  centred on the mid-plane (right-hand panel). Different halo masses with
  $10^{10},~10^{11}$, and $10^{10}\,\msun$ are colour coded. Simulations with
  advective CR feedback (solid lines) suppress star formation more strongly in
  smaller galaxies in comparison to simulations without CRs (dotted lines).  The
  right panel shows the evolution of the thermal energy density, CR energy
  density, and magnetic energy density, respectively, in our MHD simulations
  with CR feedback. }
\label{fig:disk_evolution}
\end{figure*}

\begin{figure*}
\includegraphics{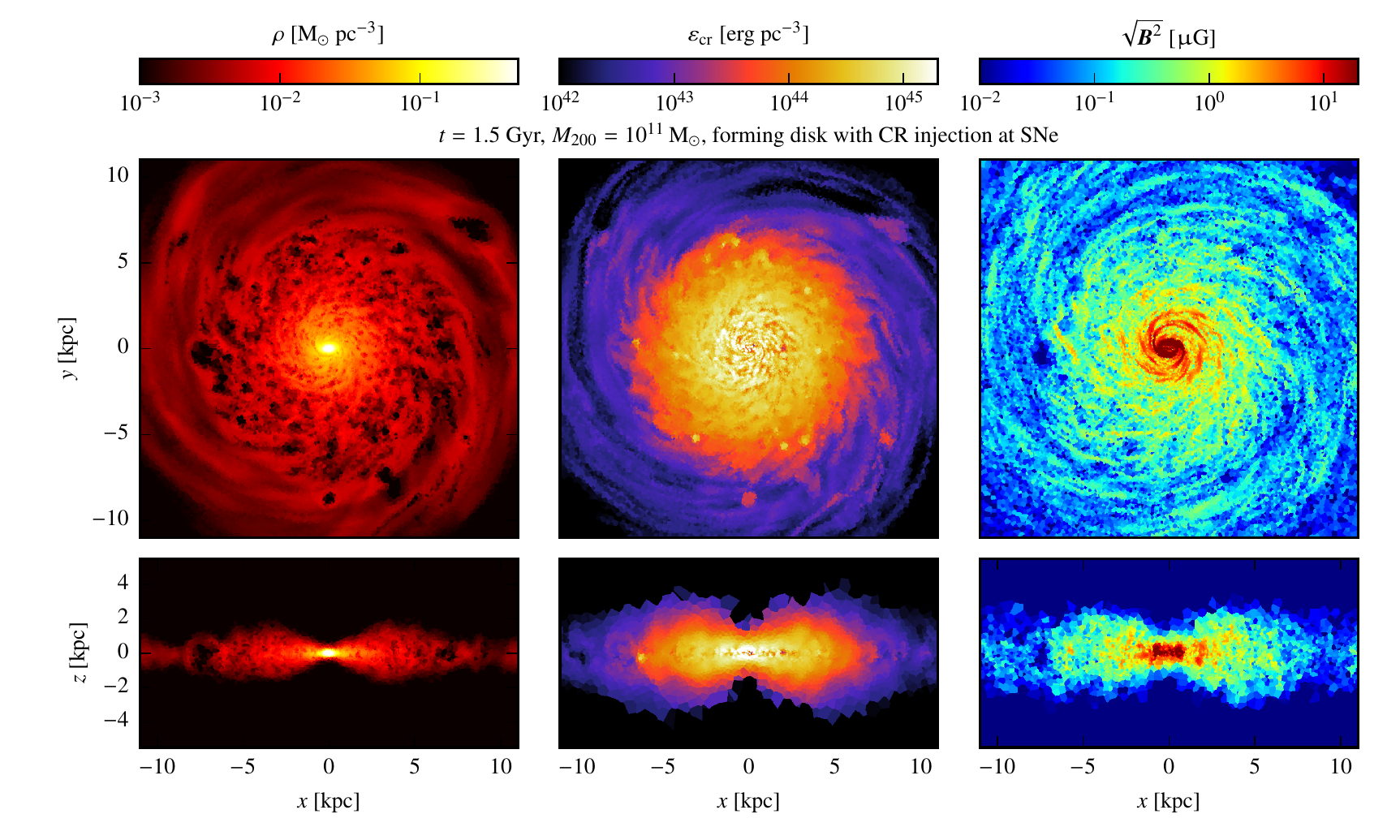}
\includegraphics{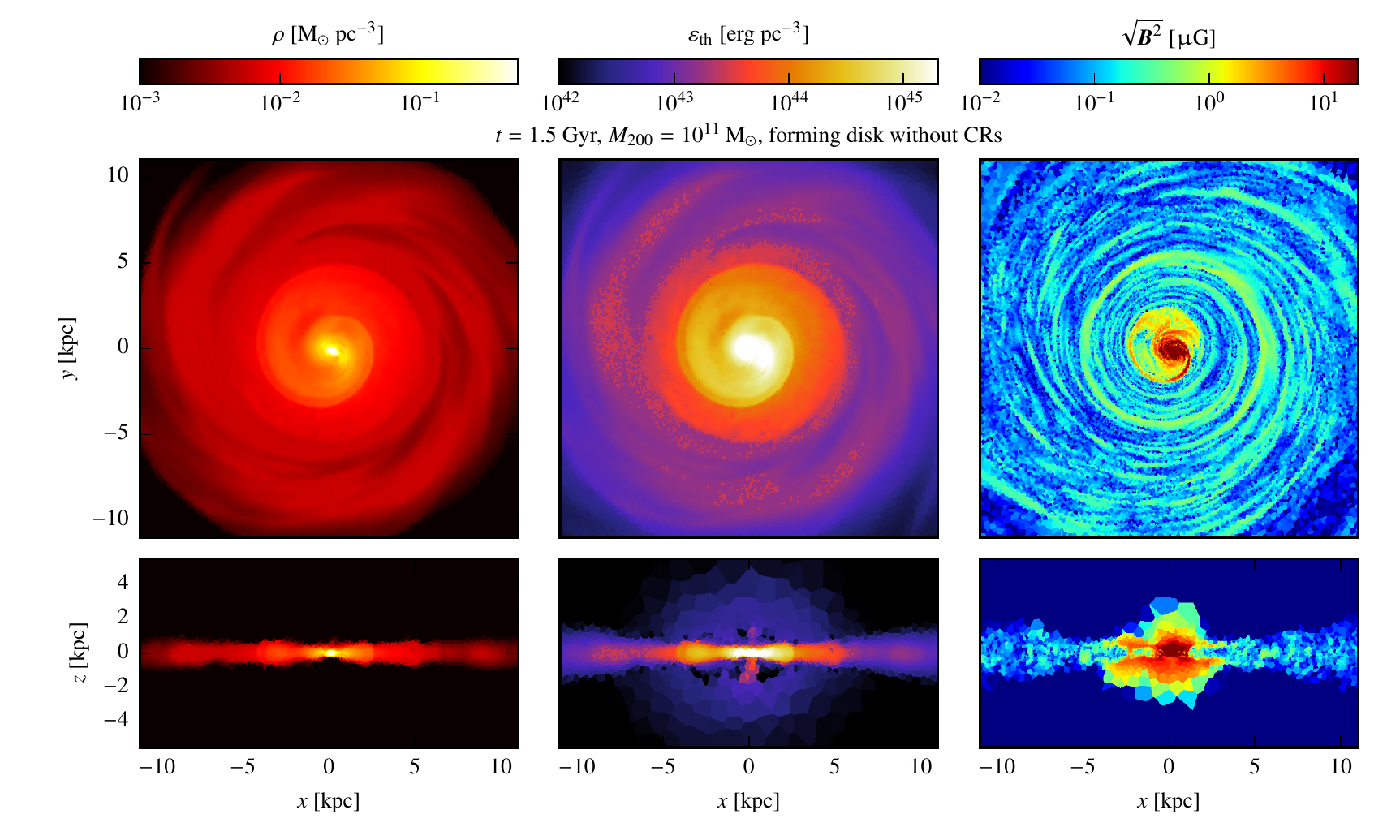}
\caption{Properties of the gas disk in our $10^{11}\,\msun$ halo after 1.5~Gyr
  in MHD simulations where we inject CRs with our supernova remnant model and
  follow their advection with the gas (top six panels) and simulations without
  CRs (bottom six panels). We show cross-sections of gas properties in the
  mid-plane of the disk (face-on views) and vertical cut-planes through the
  centre (edge-on views) of the gas density (left-hand panels), CR and thermal
  energy density (middle panels, top and bottom, respectively), and the magnetic
  field strength (right-hand panels).}
\label{fig:disk_1e11}
\end{figure*}

\begin{figure*}
\includegraphics{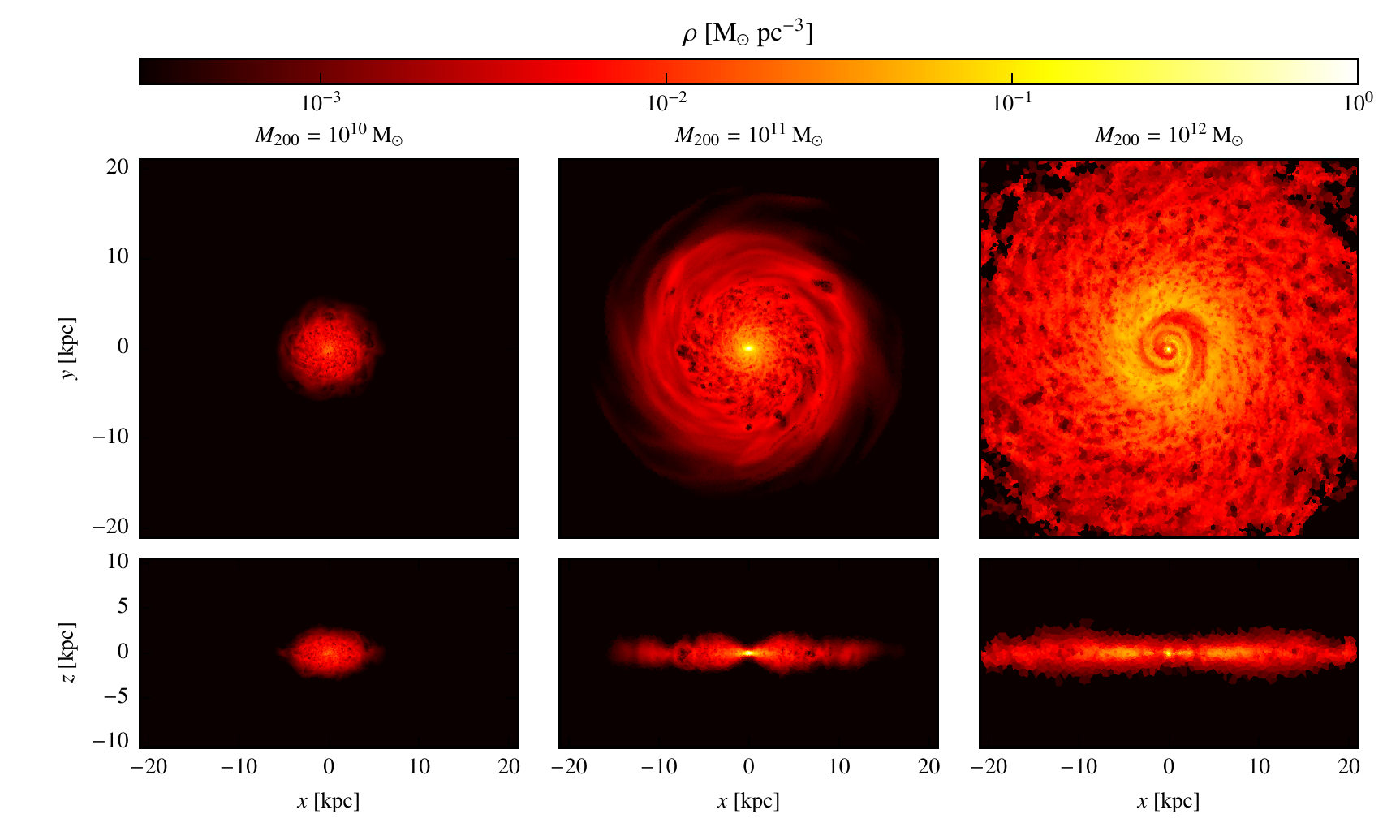}
\includegraphics{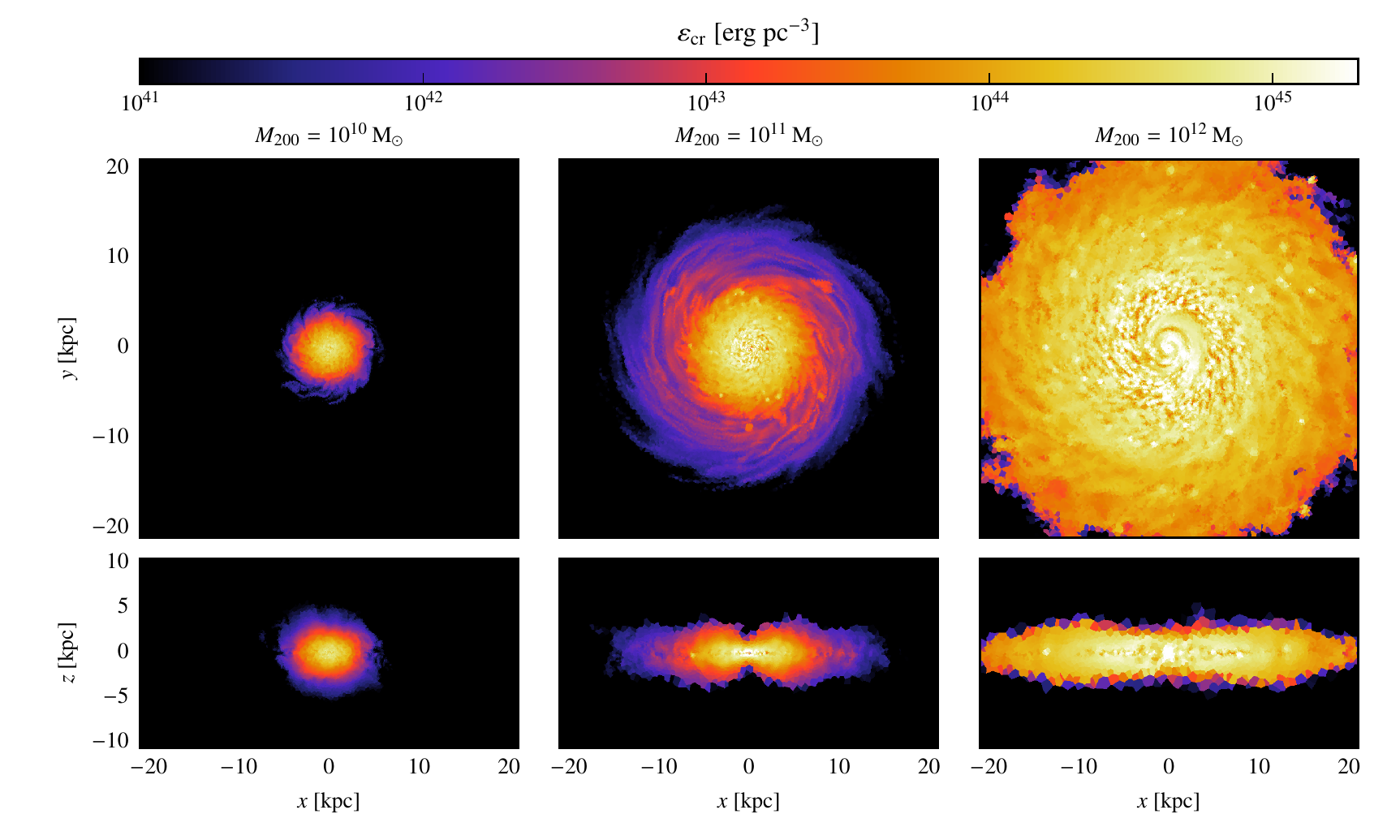}
\caption{Properties of the gas disk after 1.5~Gyr in MHD simulations where we
  inject CRs with our supernova remnant model and follow their advection with
  the gas. We present the gas density (top six panels) and CR energy density
  (bottom six panels), and show cross-sections in the mid-plane of the disk
  (face-on views) and vertical cut-planes through the centre (edge-on
  views). From the left to right, we compare different halo masses of
  $10^{10},~10^{11}$, and $10^{12}\,\msun$.}
\label{fig:disk_halo_masses}
\end{figure*}

\begin{figure*}
\includegraphics{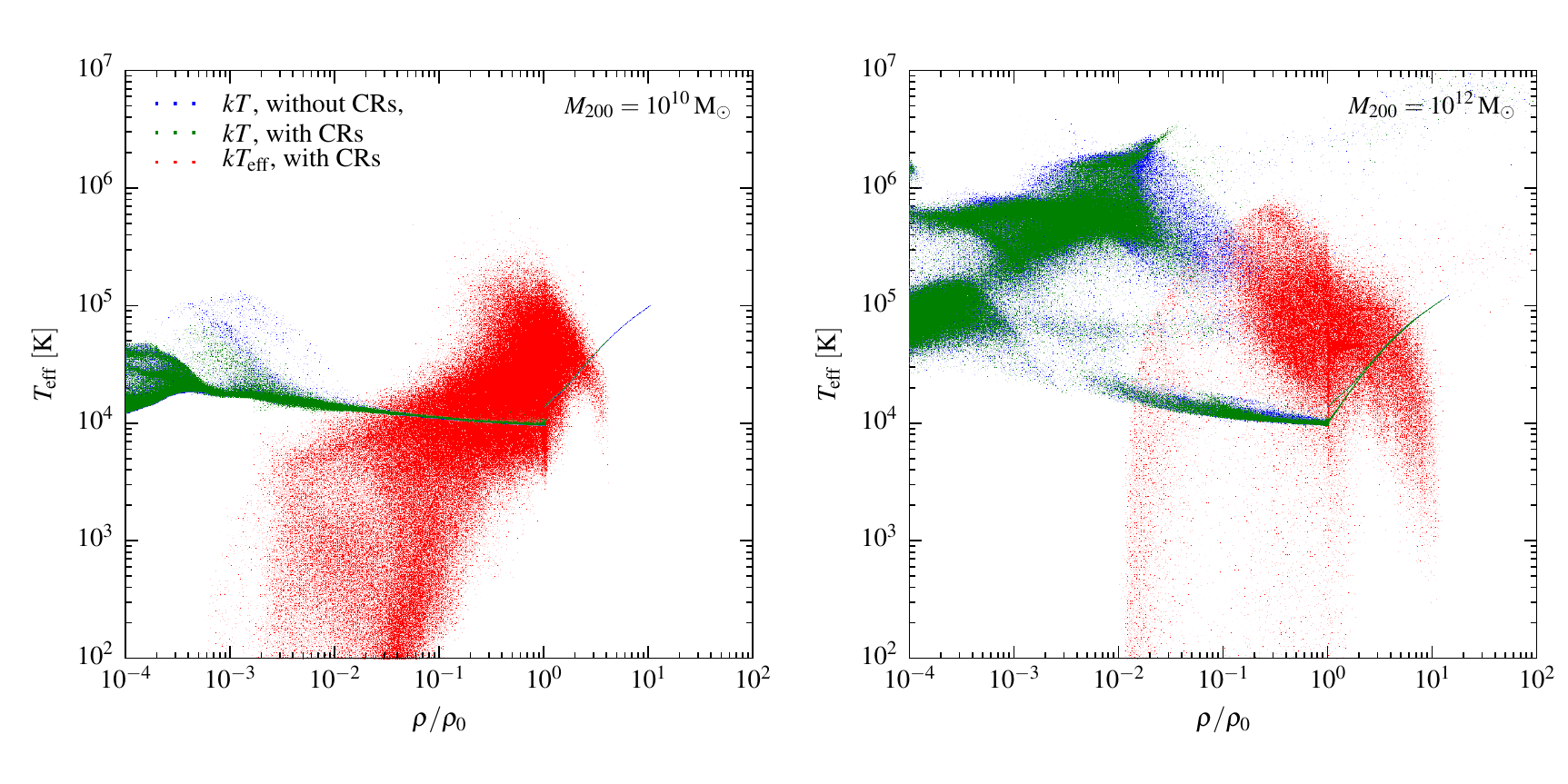}
\caption{Distribution of the gas in the temperature-density plane for our halo
  of mass $10^{10}\,\msun$ (left-hand panel) and $10^{12}\,\msun$ (right-hand
  panel). All densities are scaled to the star formation threshold $\rho_0 =
  4.05\times 10^{-25}\,\rmn{g~cm}^{-3}$. We compare the thermal gas properties
  ($T$ and $\rho$) in our simulation without CR physics (blue points) to the
  simulation with advective CR physics (green points). To assess the impact of
  CR pressure feedback, we also show the CR pseudo temperature
  $T_{\rmn{eff}}=P_\CR \mu_{\rmn{mw}} m_\p \rho^{-1}$ versus gas density
  (red). We show the systems at time $t = 2.0$~Gyr after the start of the
  evolution.}
\label{fig:disk_phase_space}
\end{figure*}

In the initial conditions, we have $10^6$ gas cells inside the virial radius,
each of which with a target mass of $1.55 \times 10^3\,\msun\times M_{10}$, where
$M_{10}=M_{200}/(10^{10}\,\msun)$. This target gas mass also corresponds to the
typical mass of a stellar macro-particle representing a stellar population. We
enforce that the mass of all cells is within a factor of two of the target mass
by explicitly refining and de-refining the mesh cells. We additionally require
that adjacent cells differ in volume by less than a factor of $10$ and refine
the larger cell if this condition is violated.

When we evolve one of these haloes forward in time, radiative cooling diminishes
the pressure support of the gas in the centre, which then collapses while
conserving its specific angular momentum, thus settling into a rotationally
supported cold disk. In the disk, the gas is compressed by self-gravity to
sufficiently high densities that star formation ensues. In our model, CRs are
injected into the ambient ISM surrounding stellar macro-particles, providing the
gas with additional non-thermal pressure.  Because the energy stored in CRs is
subject to different dissipative loss processes in comparison to the thermal
gas, their additional pressure support alters the radiative cooling of
galaxies. This reduces the overall cooling efficiency of gas in haloes, directly
resulting in a reduction of the condensated phase of cold gas and quenches
subsequent star formation (see Fig.~\ref{fig:disk_evolution}, left-hand panel).
In agreement with findings by \citet{2008A&A...481...33J}, during the first
2~Gyr of the simulation this suppression due to CR pressure feedback is larger
for more shallow gravitational potential wells, which are provided by the stars
and the dark matter halo, and hence the smaller galaxies are more strongly
quenched. However, this trend is reversed at late times because the galaxy in
the model without CRs has exhausted its available gas reservoir.

The star formation rate peaks around $0.1-0.2$~Gyr and then declines
exponentially, so that most of the stars are formed in the first Gyr. As the
first stars are forming, the CR energy density in the disk quickly reaches
equilibrium with the thermal energy density and dominates the internal energy
budget soon thereafter (see Figure~\ref{fig:disk_evolution}, right
panel). Smaller galaxies show a shorter time scale for energy equilibration and
the CR energy dominance, $\bra\eps_\CR\ket/\bra\eps_\th\ket$, is also larger in
small galaxies because of the shallower potential wells which amplify the impact
of CR pressure feedback.

Figure~\ref{fig:disk_1e11} shows the disk at time $t=1.5$~Gyr after the start of
the evolution and compares the simulations with and without advective CR
transport (top and bottom panels, respectively).  At this time, the gas density
strongly peaks in the centre of the disk, where most of the stars are formed and
thus most of the CRs are injected there, which is reflected in the distribution
of CR energy density. A visual difference of the density distributions between
the two types of simulations is apparent. While the simulation without CR
feedback shows the smoothed density and thermal energy of the ISM as predicted
by the subgrid-scale model of the pressurised effective equation of state, the
ISM is very clumpy in the run with CR feedback. Every low-density cavity
corresponds to the location of a star forming region (represented by a stellar
macro-particle) and was evacuated by the Sedov-Taylor blast wave that has formed
as a result of the CR energy deposition of the collection of supernovae with an
energy that corresponds to the stellar mass formed (Sect.~\ref{sec:SNe}). This
becomes evident from the tight spatial correlation of the density cavities and
peaks in the CR energy density (top left and middle panel of
Fig.~\ref{fig:disk_1e11}). The cavity sizes grow larger with galacto-centric
radius because of the lower ambient gas densities that the blast waves encounter
there. We note that the cavity morphologies are not smoothed due to numerical
effects thanks to the very low numerical diffusivity of the moving mesh
technique that enables mesh cells to co-rotate with the rotating
disk. Eventually the cavities are sheared and dispersed by differential rotation
on time scales that increase with galacto-centric radius, leaving the
supernovae-blown cavities at galactic outskirts intact for longer times.

While CR streaming and diffusion are believed to smooth out the inhomogeneous CR
distribution to some extent, we envision that some of the results here should
also carry over to situations where more realistic spatially varying diffusion
coefficients are employed. Upstream of supernova remnant shocks, high-energy CRs
drive non-resonant hybrid plasma instabilities \citep{2004MNRAS.353..550B},
which generate strong electromagnetic turbulence that causes the CRs to
experience on average one scattering event per gyro revolution. The resulting
motion of CRs can be described by a diffusive transport that is characterised by
a Bohm diffusion coefficient of $\kappa_{\rmn{B}}\approx p \beta m_\p c^2/(3e B)
\approx 3\times 10^{21}\,(p m_\p c^2/\rmn{GeV})\, (B/\umu\rmn{G})^{-1}\,
\rmn{cm}^2\, \rmn{s}^{-1}$ \citep{2006NatPh...2..614S}.  If the CRs propagate
far into the upstream of the shock or if the supernova remnant shock wave has
sufficiently slowed down and entered the snowplough phase so that the CRs that
have been trapped in the interior of the remnant can escape the supernova
environment, the CR-wave scattering rate starts to decline. As a result, the CR
mean free path increases so that the effective diffusion coefficient -- as long
as the mode of CR propagation can reliably be described by a diffusion process
-- approaches the average galactic value of $3\times10^{28}\,\rmn{cm}^2\,\rmn{s}^{-1}$.

In addition to the increased clumpiness of the ISM, the CR feedback simulation
shows a larger density scale height in comparison to the MHD simulation without
CRs (left-hand panels of Fig.~\ref{fig:disk_1e11}). This increase is due to the
additional CR pressure that is subject to different dissipative loss processes
in comparison to the thermal gas, and has a cooling time well above the
radiative loss time of thermal gas. In Fig.~\ref{fig:disk_halo_masses}, we
compare the gas density and CR energy density after $1.5$~Gyr in our CR
simulations for different halo masses of $10^{10},~10^{11}$, and
$10^{12}\,\msun$. The aspect ratio of the gas disk (disk length-to-height)
decreases for smaller disks to almost unity for the $10^{10}\,\msun$ halo as a
result of the relatively larger impact of CR pressure feedback in smaller
haloes. This is in qualitative agreement with the results of smoothed particle
hydrodynamics simulations by \citet{2008A&A...481...33J}, although the extents
of the disks in our moving mesh simulations are nearly twice of that of their
simulations owing to improvements of the hydrodynamical method here.  Comparing
the distribution of CR energy density across the mass sequence reveals a more
homogeneous distribution in the larger halo of mass $10^{12}\,\msun$: star
formation proceeds here throughout the entire disk rather than being only
concentrated towards the centre as in the smaller haloes, and hence most of the
CRs are injected more homogeneously into the ISM.

In the initial phase of the galaxy assembly (within the first $200$~Myr), the
magnetic field strength is exponentially amplified on small length and time
scales. This is consistent with a turbulent small-scale dynamo that operates to
an average field strength of about $10\,\mathrm{\umu G}$ for our Milky Way-type
halo of $10^{12}\,\msun$ and to $10^{-2}\,\mathrm{\umu G}$ for our less massive
galaxies (see right panel in Fig.~\ref{fig:disk_evolution}). After this initial
assembly phase the disk has finished forming and the dominating differential
rotation of the gas in the disk stretches the magnetic flux tubes so that the
coherence scale can grow. In the Milky Way-type galaxy, the field strength has
saturated and only grows in scale while the magnetic field continues to grow
exponentially in the smaller haloes, but on much longer time scales, suggesting
that the dominant dynamo amplification mechanism has changed. Interestingly, the
morphology of the magnetic field is also substantially modified by CR feedback
(right panels of Fig.~\ref{fig:disk_1e11}). While the field structure is very
regular in the case without CRs, it attains a chaotic small-scale structure
which is superimposed on the dominant azimuthal component. Apparently, the
additional turbulent velocity field as a result of the supernova explosions has
intensified the dynamo action which further amplified the field.

We complement our discussion on the exact mechanisms of CR pressure feedback in
regulating star formation by considering the distribution of the gas in the
temperature-density plane for our small and massive galaxy
(Fig.~\ref{fig:disk_phase_space}). We compare the thermal gas properties ($T$
and $\rho$) in our simulation without CR physics (blue points) to the simulation
with CR pressure feedback (green points). As the cooling gas is infalling onto
the disk, it is accelerated by the gravitational potential and shocks to the
virial temperature of the halo. It continues to loose energy through radiative
cooling and slowly moves onto the nearly isothermal branch of the ISM along
which it moves to higher densities as it cools further up to the critical
density of the star formation threshold. At this point, the ISM is parametrized
by a stiff effective equation of state, which interpolates between the hot and
cold phases and provides an effective pressure to the ISM. 

Clearly, there are regions in the $T$-$\rho$ plane, which are avoided by the
thermal gas in the run with CR injection (visible by the dominant blue
colour). This is due to CR pressure feedback as can be directly assessed by
considering the CR pseudo temperature $T_{\rmn{eff}}=P_\CR \mu_{\rmn{mw}} m_\p
\rho^{-1}$ (red points), which dominates over the thermal temperature at the
density of interest, i.e., where the thermal gas has been pushed out in
comparison to the simulation without CRs. We note that the normalisation of the
CR pseudo temperature depends on the hadronic and Coulomb cooling rate of CRs. A
temporarily increased cooling rate as a result of fresh CR injection would have
lowered the CR pseudo temperature in comparison to our approach that is based on
a CR equilibrium distribution (see Section~\ref{sec:limitations}). This explains
differences to the Milky-Way type galaxy simulations by
\citet{2008A&A...481...33J} that follow a simplified CR spectrum with a single
power-law spectrum and momentum cutoff, which provides a temporal resolution of
the Coulomb cooling rate. As a result, they find the CR pressure to be
insufficient to significantly affect massive galaxies.\footnote{However, in
  order not to overestimate the CR cooling by artificially lowering the
  energy-weighted momentum during a supernova injection event, the formalism by
  \citet{2008A&A...481...33J} only injects CRs above a specific momentum, which
  guarantees that the spectral component can only grow. This implies a lower
  effective injection efficiency in comparison to a case that follows a
  multi-component CR distribution. } This demonstrates the need of future work
to model the spectral CR distribution in space and time to accurately follow
Coulomb cooling processes.

\subsection{Cosmological simulations}
\label{sec:cosmo}

\begin{figure*}
\includegraphics{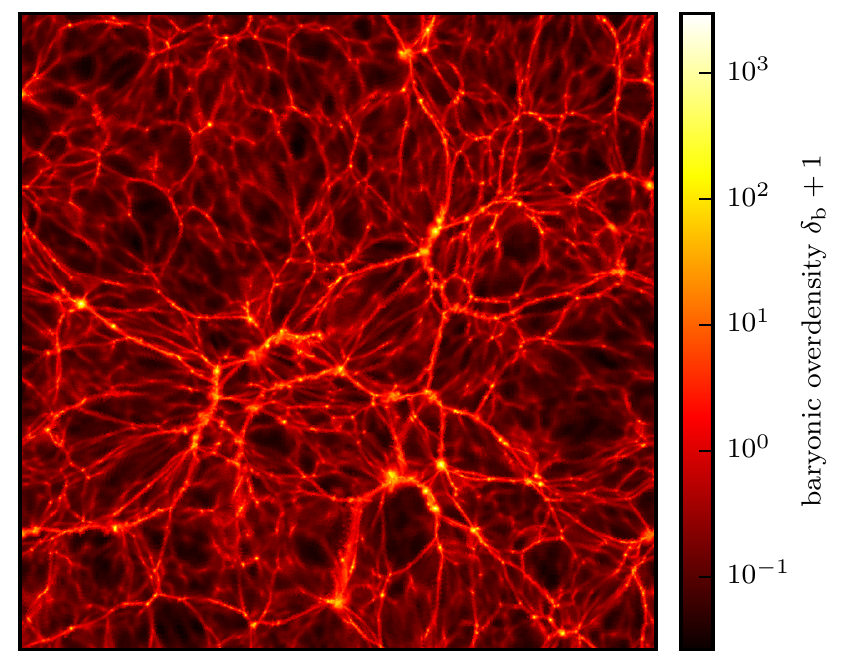}
\includegraphics{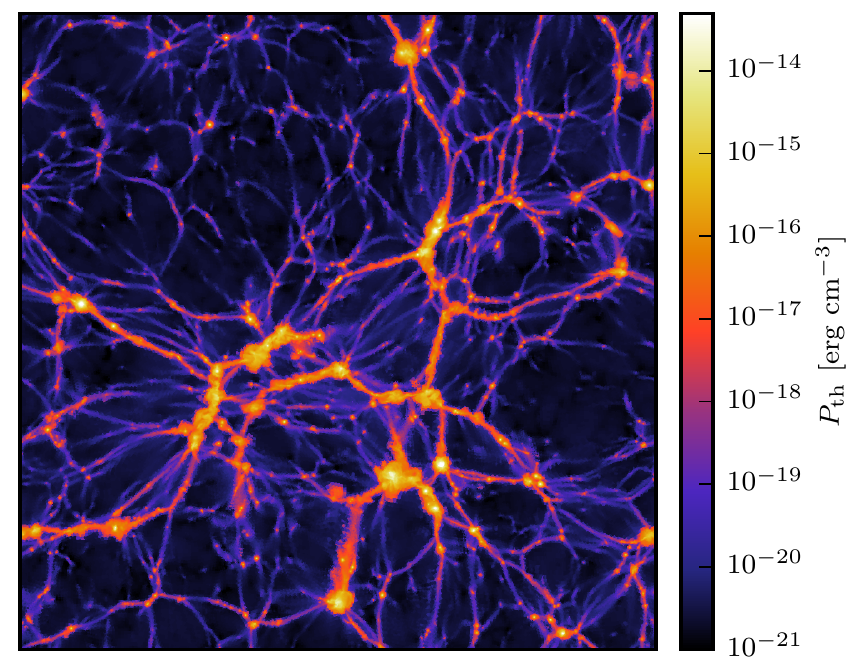}\\
\includegraphics{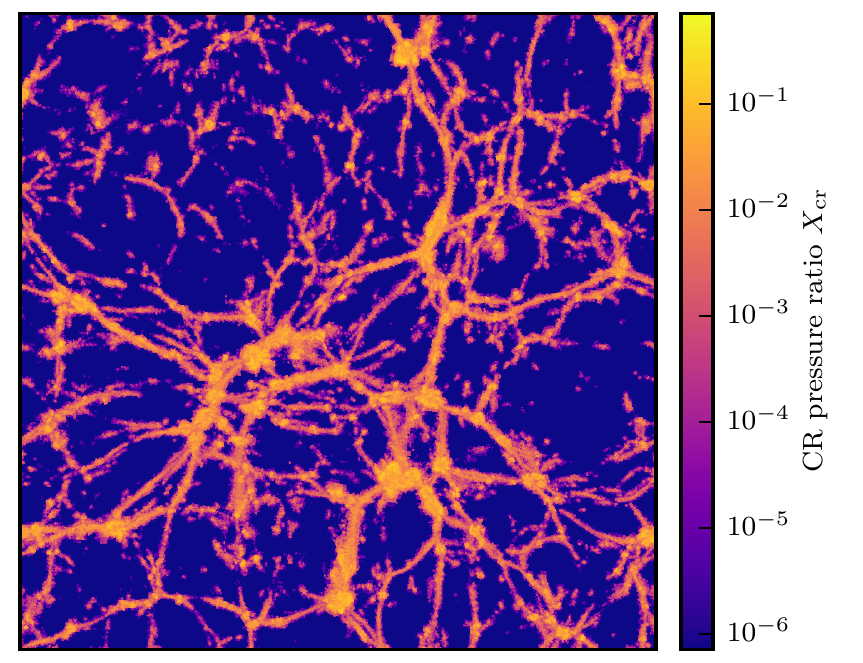}
\includegraphics{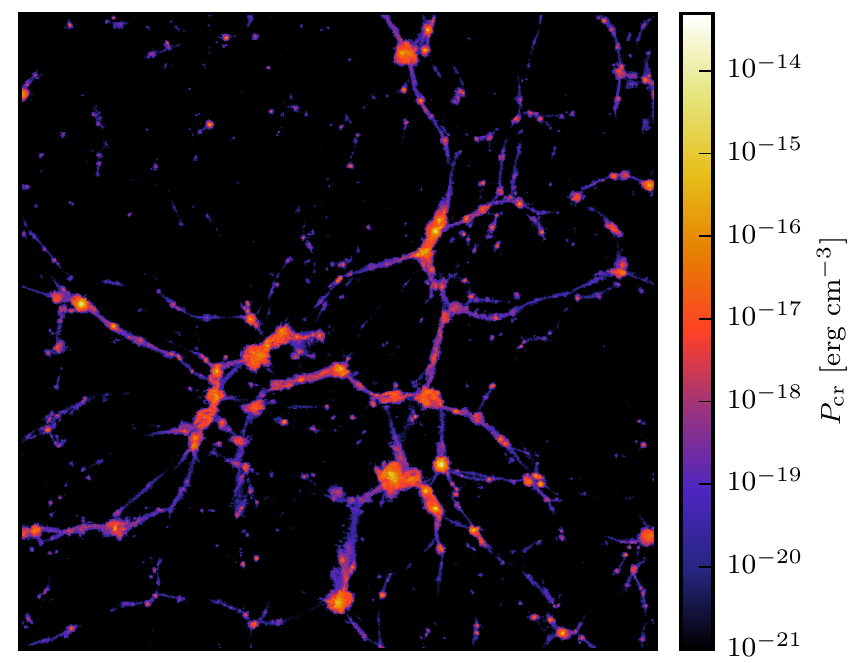}\\
\includegraphics{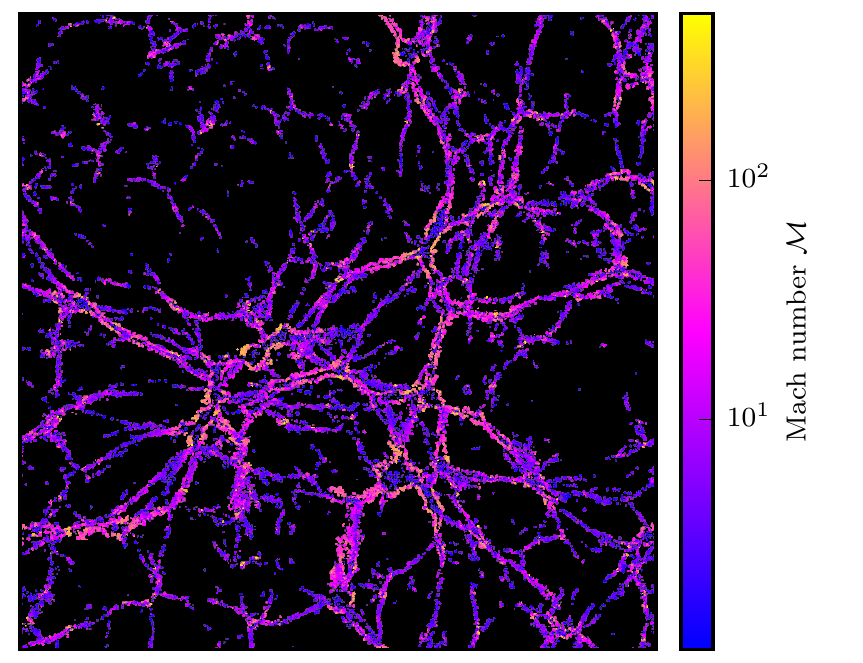}
\includegraphics{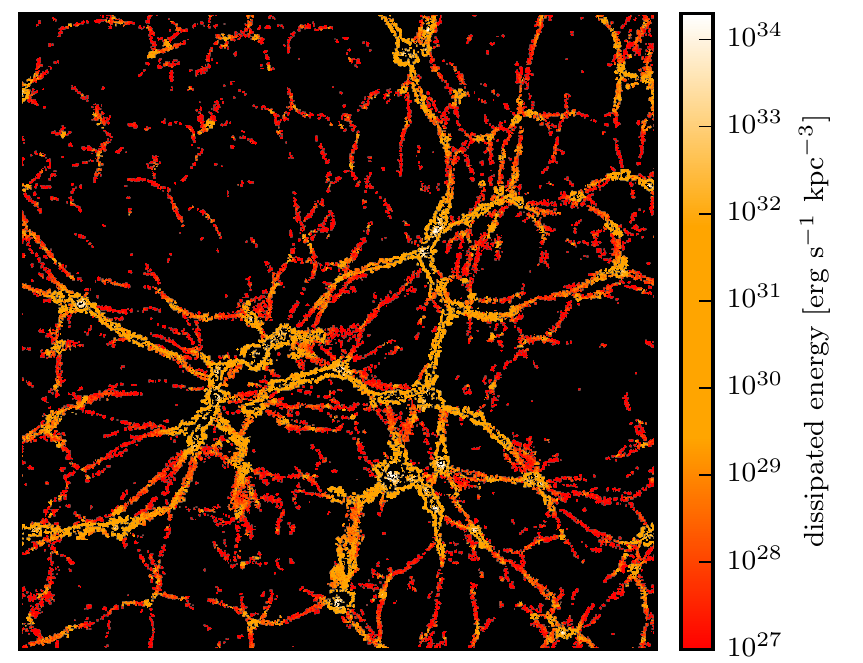}\\
\caption{Visualisation of a non-radiative cosmological MHD simulation at
  redshift $z=0$ which accounts for CR acceleration at structure formation
  shocks.  Top panels: projections of the volume-weighted baryonic overdensity
  and the thermal pressure. The width and the height of the plots correspond to
  the full box size ($100\, h^{-1}\,{\rm Mpc}$). All projections exhibit a depth of
  300~kpc. Middle panels: projection of the CR pressure ratio
  $X_\CR=P_\CR/P_\th$ and the volume-weighted CR pressure. Bottom panels:
  projections of the Mach number of structure formation shocks weighted with the energy
  dissipation (left-hand panel) and energy dissipation rate density (right-hand
  panel). }
\label{fig:cosmo}
\end{figure*}

\begin{figure}
\includegraphics{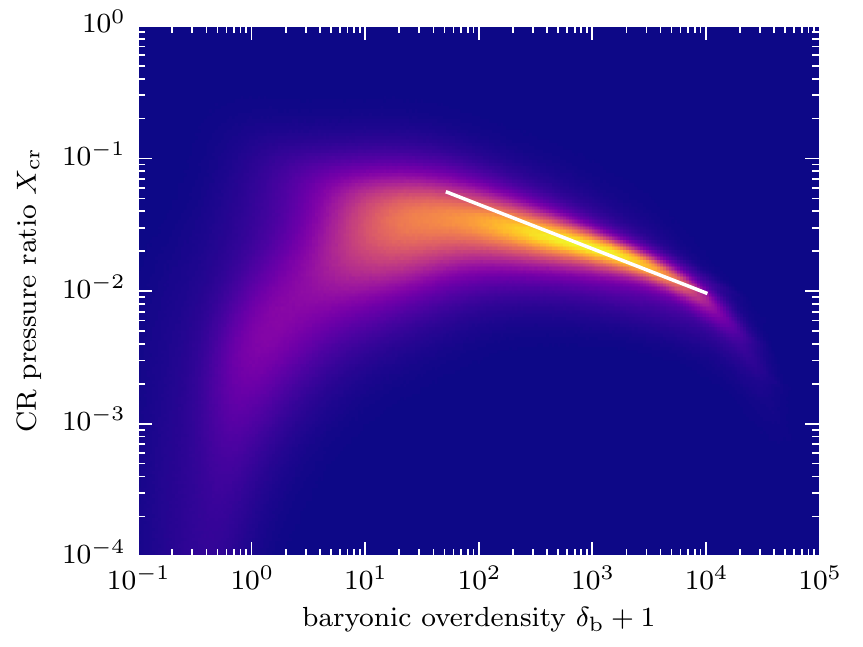}
\caption{Distribution of the CR pressure ratio $X_\CR=P_\CR/P_\th$ as
  a function of baryonic overdensity, $\delta_\rmn{b}$, at $z=0$. The
  colours show a linear scaling of the phase space density. CRs are
  accelerated at strong external formation shocks that form at
  overdensities of $\delta_\rmn{b}\sim5$ to $50$. During the collapse
  into haloes, the composite of CRs and thermal gas experiences
  adiabatic compression, which favours the thermal pressure over that
  provided by CRs and causes the CR pressure ratio to drop as
  $X_\CR\propto \rho^{-1/3}$ (white line).}
\label{fig:phase_space}
\end{figure}

\begin{figure*}
\includegraphics{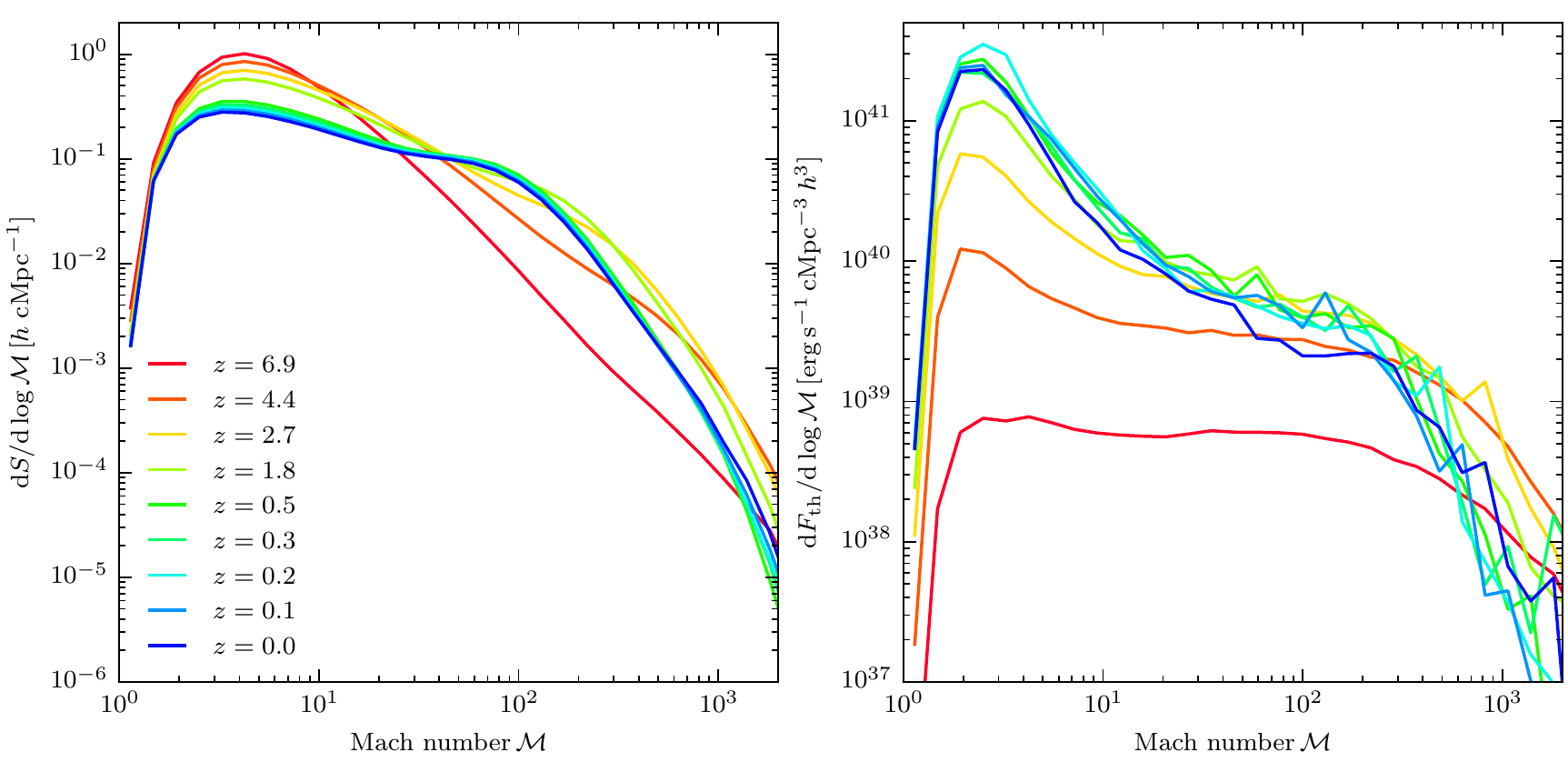}
\caption{Left-hand panel: we show the differential shock surface area per unit
  volume as a function of Mach number for different redshifts. Right-hand
  panel: distribution of the dissipated energy density per time, which accounts
  for the total of generated thermal and CR energy at the shock. While there is
  more energy dissipated at late times in weak internal shocks, the shock
  surface per volume decreases for weak shocks.}
\label{fig:mach_distribution}
\end{figure*}

In order to study the impact of diffusive shock acceleration on cosmological
structure formation shocks as well as to examine the generalised shock finder in
combination with the CR shock acceleration algorithm in a non-trivial
realisation of shock morphologies, we perform here non-radiative cosmological
MHD simulations that do not include radiative cooling or star formation. We
focus on analysing the spatial distribution of the CR distribution in relation
to the cosmological structure formation shocks, and on Mach number statistics.

We adopt a standard cold dark matter model with a cosmological constant
($\Lambda$CDM), as recently inferred by the \citet{2016A&A...594A..13P}. The
cosmological parameters of our model are as follows: $\Omega_\rmn{m} =
\Omega_\rmn{dm} + \Omega_\rmn{b} = 0.3089$, $\Omega_\rmn{b} = 0.0864$,
$\Omega_\Lambda = 0.6911$, $h = 0.6774$, $n_\rmn{s} = 0.9667$, and $\sigma_8 =
0.8159$. Here, $\Omega_\rmn{m}$ denotes the total matter density in units of the
critical density for geometrical closure, $\rho_\rmn{crit} = 3 H_0^2 / (8 \upi
G)$.  $\Omega_\rmn{b}$ and $\Omega_\Lambda$ specify the density of baryons and
the cosmological constant at the present day, respectively. The present day
Hubble constant is parametrized as $H_0 = 100\,h \mbox{ km s}^{-1}
\mbox{Mpc}^{-1}$, $n_\rmn{s}$ denotes the spectral index of the primordial
power-spectrum, and $\sigma_8$ is the {\em rms} linear mass fluctuation within a
sphere of radius $8\,h^{-1}$Mpc extrapolated to $z=0$.

Initially, our simulation employed $2\times 512^3$ gas and dark matter
resolution elements, which were regularly distributed within a
periodic box of comoving size $100\,h^{-1}$Mpc. As a result, the dark
matter particles have masses of $4.6 \times 10^8\,h^{-1}\msun$ and the
gas cells have a target mass of $1.79 \times 10^8\,h^{-1}\msun$.
Using explicit refinement and de-refinement we ensure that the mass of
all cells remains within a factor of two of the target mass. The
gravitational force softening was of a spline form with a
Plummer-equivalent softening length of $6.5\,h^{-1}$kpc comoving.

Initial conditions were created by perturbing the homogeneous particle
distribution with a realisation of a Gaussian random field characterised by the
$\Lambda$CDM linear power spectrum. The displacement field in Fourier space was
laid down using the Zel'dovich approximation, and the amplitude of each random
phase mode was drawn from a Rayleigh distribution. For the initial redshift we
chose $z_\rmn{init}=127$ which translates to an initial temperature of the gas
of $T_\rmn{init}=244.8$~K. The adiabatic index of the gas is set to $\gamma_\th
= 5/3$.

In these simulations, we account for CR acceleration at resolved cosmological
structure formation shocks and adopt a realistic value for the acceleration
efficiency of $\zeta=0.1$ at all shocks exceeding a pre-shock Mach number of
$\M_1=3$, and zero otherwise. For simplicity, we assume that the acceleration
efficiency does not depend on magnetic shock obliquity. We adopt an
ultra-relativistic CR population which translates to a CR adiabatic index of
$\gamma_\CR = 4/3$.  As a result, the CR pressure ratio in the immediate
post-shock regime amounts to $X_\CR = \zeta/[2\,(1-\zeta)] =0.0\bar{5}$,
assuming a cold pre-shock gas with negligible pressure support characteristic of
external accretion shocks. Once injected, we follow advective CR transport,
account for adiabatic changes of the CR energy as well as Coulomb and hadronic
CR cooling. The magnetic field is initialised as a uniform homogeneous
seed field along the $x$-axis with an initial comoving strength of
$10^{-15}\mathrm{G}$. There are no CRs in the initial setup.

Figure~\ref{fig:cosmo} shows a visualisation of the simulation volume at $z=0$
that reveals several quantities of relevance for CR acceleration at structure
formation shocks as well as the successive CR transport. The lower panels show
strong external shocks with Mach numbers exceeding $\M_1\sim 100$ that occur
when the cosmic fluid accretes into cosmic filaments and super-cluster
regions. Interior to these structures, for the most part weak shocks are
visible. Still, most of the energy is dissipated inside collapsed structures due
to the higher pre-shock density and the increased shock velocities there. In
contrast, little energy is dissipated in cosmic voids.  These results are in
excellent agreement with previous work that uses non-radiative physics
\citep{2003ApJ...593..599R, 2006MNRAS.367..113P, 2007MNRAS.378..385P,
  2008ApJ...689.1063S, 2009MNRAS.395.1333V, 2011MNRAS.418..960V,
  2014ApJ...785..133H, 2015MNRAS.446.3992S}, while the addition of radiative
physics introduces new populations of shocks as a result of non-gravitational
energy release \citep{2007MNRAS.378..385P, 2008MNRAS.385.1211P,
  2007ApJ...669..729K, 2013MNRAS.428.1643P, 2016MNRAS.461.4441S}.

The thermal pressure distribution traces the cosmic web as revealed by the
overdensity map (Fig.~\ref{fig:cosmo}, top panels). However, the pressure shows
an increased contrast in comparison to the density distribution because of the
multiplication with the temperature field that drops steeply outside the
location of cosmological formation shocks. At collisionless shocks, not all the
free energy is thermalized but a fraction is funnelled to CRs, provided the
shock strength exceeds a critical Mach number. Hence the CR pressure represents
a biased distribution of the thermal pressure distribution, with groups and
super-cluster regions being prominently visible (Fig.~\ref{fig:cosmo}, middle
panels). While this compares qualitatively well with previous work
\citep{2008MNRAS.385.1211P,2012MNRAS.421.3375V}, we defer a detailed comparison
study to future work.  The relative CR pressure traces the overall morphology of
the cosmic web well, and attains values that typically range up to
$X_\CR=0.0\bar{5}$. As in the case of the Sedov explosion problem, individual
mesh cells can have CR pressure ratios that scatter above this value if the
`shock rays' delineating the direction of shock propagation converge in
post-shock cells. The low-density web in cosmic voids remains almost invisible
and exhibits CR pressure ratios less than $10^{-4}$. This is because the shocks
responsible for forming these structures are weak and dissipate little energy
during their assembly.

A close inspection of the $X_\CR$ projection reveals that this quantity does not
peak toward the densest centres, but at the location of the strong external
formation shocks, which are the primary sources of CR acceleration in cosmic
structures. This can be understood by looking at the distribution of the CR
pressure ratio $X_\CR=P_\CR/P_\th$ as a function of baryonic overdensity
(Fig.~\ref{fig:phase_space}) in combination with spherical collapse theory.  As
an overdensity collapses under the influence of its own self-gravity, its
expansion rate drops below the Hubble expansion and its overdensity starts to
increase (while its physical density continues to decrease with the Hubble
expansion). As the overdensity reaches a value of $\delta_\rmn{b}=5.55$ (in the
spherical collapse model), the expanding shell of pristine cosmic gas turns
around and collapses onto filaments and super-cluster regions. As a result,
cosmological formation shocks form at the location where the ram pressure of the
collapsing gas, $\rho\vel^2$, balances the total pressure,
$P_\rmn{tot}=P_\th+P_\CR$, of the previously collapsed shells of gas and the
cold pristine gas is shock-heated for the first time in cosmic history. In our
simulations, these shocks are characterised by overdensities of
$\delta_\rmn{b}\sim5$ to $50$, and the CR pressure ratio in the shock-heated gas
amounts to $X_\CR \lesssim0.0{5}$, as expected for our injection efficiency and
cold pre-shock gas.  During the continued collapse into haloes, the composite of
CRs and thermal gas is adiabatically compressed, which favours the thermal
pressure over the CR pressure and hence causes the CR pressure ratio to drop as
$X_\CR\propto \rho^{-1/3}$. Finally, at the highest densities, the Coulomb and
hadronic cooling time scales are shorter than the Hubble time and CRs start to
loose pressure support at overdensities of $\delta_\rmn{b}\gtrsim10^4$.  The
density dependence of $X_\CR$ nicely resembles the smoothed particle
hydrodynamics results by \citet{2007MNRAS.378..385P} while the decline of
$X_\CR$ as a result of adiabatic compression in the adaptive-mesh refinement
simulations by \citet{2014MNRAS.439.2662V} is virtually absent and converges
only very weakly with increasing resolution.

Figure~\ref{fig:mach_distribution} quantifies the shock distribution
and the associated energy dissipation in our simulation. The left-hand
panel shows the differential shock surface area per unit volume as a
function of Mach number for different redshifts that are equally
spaced in look-back time. Note that in our analysis we neither account
for radiative cooling nor for reionisation.  We find that the
cumulative area of shocks is dominated by weak shocks with Mach
numbers $\M_1=3$ to $4$ and decreases towards lower redshift, while
the surface area increases for strong shocks with $\M_1\gtrsim20$. The
trend of the shock area with redshift is reversed for weak shocks if
the pre-shock gas is assumed to be photo-heated to $10^4$~K, in which
case we reproduce the result found by
\citet{2015MNRAS.446.3992S}. Essentially, such an analysis increases
the Jeans mass in voids above the masses of most of the haloes so that
shocks would not have formed in voids in such a realisation of the
universe. Instead, we show here the unmodified shock distribution,
which was responsible for the CR acceleration our simulations. At low
redshift, the accretion from previously unshocked gas onto hot
filaments and cluster outskirts forms shocks with Mach numbers of
$\M_1\sim100$, which decisively changes the slope of the Mach number
distribution and introduces a shoulder around $\M_1\sim100$ at late
times.

The right-hand panel of Fig.~\ref{fig:mach_distribution} shows the
distribution of dissipated energy density per time, which includes
the generated thermal and CR energy at the shock. {\em Most importantly,
  CR shock acceleration does not change the overall appearance of the
  shock distribution, which is shaped by the distribution of
  gravitational potentials in space and potential depths.}
Theoretically, we would expect the Mach numbers to be slightly weaker
in the case with CR acceleration but this difference is not visible on
the large logarithmic scale in Fig.~\ref{fig:mach_distribution} (it is
only a 5\% effect at $\M_1=10$ for an extreme CR injection efficiency
of $\zeta=0.5$).  The total dissipated energy increases with time
until $z=0.5$ and slightly decreases thereafter. The increase of the
shock-dissipated energy with time is due to the increasing densities
in collapsed structures and increasing shock velocities as deeper
potential wells are forming, since the kinetic energy flux through the
shock scales as $F_\rmn{kin}\propto\rho\vel_\rmn{s}^3$.  At low
redshifts, this effect saturates because of the self-similar density
profiles of collapsed haloes in non-radiative simulations and,
furthermore, dark energy slows structure growth and dilutes the
pre-shock gas inside voids, which leads to a drop of the thermal
energy flux for high Mach numbers \citep{2006MNRAS.367..113P}.

To test for numerical convergence, we additionally simulated the same
cosmological model by degrading the particle resolution in steps of
8. We find convergence in the Mach number distributions at late times
$z<3$,  indicating that the bulk of the relevant shock structures
are well resolved.

\section{Summary and Conclusions}  \label{Sec:conclusions}

In this paper, we have detailed a new versatile approach for treating CR physics
coupled to MHD in a hydrodynamical code with an unstructured moving mesh, as
realised in the massively parallel {\sc arepo} code. This enables us to perform
self-consistent MHD-CR simulations in a cosmological framework. We model the CR
distribution as a second fluid with an adiabatic index (usually taken to be
$\gamma_\CR=4/3$).  Our implementation accounts for diffusive shock acceleration
of CRs at resolved shocks in the computational domain, and additionally from
supernova remnants that are individually not resolved in simulations of galaxy
formation, but that follow star formation and feedback in a sub-resolution
framework.

So far, our approach follows the advective transport of CRs with the magnetised
plasma as well as the anisotropic diffusive transport along the orientation of
the local magnetic field lines. For the latter, we employ a gradient-limited,
conservative, semi-implicit scheme for anisotropic CR diffusion that supports
local time stepping, as described in a companion paper \citep{Pakmor2016a}. We
account for the most important CR loss processes in the form of Coulomb and
hadronic interactions with the thermal plasma by adopting an equilibrium CR
distribution that results from balance between injection and dissipation
processes. Furthermore, we also model CR energy losses as a result of the
generation of Alfv\'en waves by the CR streaming instability. This novel element
enables us to study problems associated with CR acceleration at supernova
remnants, and to understand the dynamical impact of CRs on galaxy formation and
the evolution of galaxy clusters.

We have validated our new numerical methods in a number of different problem
set-ups, including CR acceleration at planar and spherically expanding shocks,
isolated galaxy formation, and CR acceleration and transport in non-radiative
cosmological simulations.

\begin{itemize}

\item {\bf Riemann shock tube.}  We demonstrate the accuracy of our CR
  implementation in a Riemann shock tube problem with a pre-existing CR
  population that modifies the dynamics, but -- in the absence of CR shock
  acceleration -- is only adiabatically compressed across the shock jump.  To
  model diffusive shock acceleration at shocks in our simulation, we find and
  characterise them on-the-fly and accelerate CRs according to a sub-resolution
  prescription from PIC plasma simulations. Our shock tube simulations compare
  very well to newly derived, exact solutions of the Riemann shock-tube problem
  with CR acceleration (as shown in the Appendices~\ref{sec:Riemann} and
  \ref{sec:Riemann+CRs}). As expected, injecting and accelerating CRs from the
  thermal distribution lowers the effective adiabatic index of the post-shock
  gas, increases its compressibility and causes it to become denser. Because of
  mass conservation and of the higher post-shock gas density, the shock does
  then not propagate as fast as in the case without CR acceleration.

\item {\bf Sedov-Taylor blast wave.}  A slower propagating shock in the case of
  diffusive shock acceleration of CRs is also seen for a spherically expanding
  shock as a result of a point explosion, corresponding to the well-known
  Sedov-Taylor problem. We observe a strongly varying effective adiabatic index
  $\gamma_{\rmn{eff}}$ of the two-fluid medium as a result of CR shock
  acceleration: the value of $\gamma_{\rmn{eff}}$ decreases from its canonical
  value of $5/3$ in the external medium to a lower value that depends on the CR
  shock-acceleration efficiency because the freshly accelerated CRs exhibit a
  softer equation of state. Importantly, $\gamma_{\rmn{eff}}$ drops further to a
  value close to that of the intrinsic CR population of $\gamma_\CR\approx 4/3$
  because of the successive adiabatic expansion in the interior of the blast
  wave that favours the CR pressure over the thermal pressure. Despite this
  strong variation of the adiabatic index, the self-similarity of the solution
  is almost conserved. In particular, the evolution of the expanding shock
  radius as well as the detailed pressure and density profiles can be well
  approximated by a constant but modified adiabatic index which we find to be
  close to $7/5$ for our optimistic CR acceleration efficiency of
  $\zeta=0.5$.

\item {\bf Galaxy formation.}  To explore the impact of CR physics on MHD
  simulations of galaxy formation, we model CR injection at supernovae, advective
  CR transport, and Coulomb and hadronic CR interactions with the ambient gas. In
  line with previous findings that employed smoothed particle hydrodynamics, we
  find that CR pressure feedback suppresses star formation more strongly in
  smaller galaxies in comparison to simulations without CRs. During the first
  starburst, the CR energy density in the disk quickly reaches equilibrium with
  the thermal energy density and dominates the internal energy budget soon
  thereafter. The slowly cooling non-thermal pressure reservoir provided by CRs
  causes the disks to be more expanded in the vertical direction, providing
  additional dynamical stability to a disk that would otherwise be unstable to
  gravitational collapse.

  The local injection of CR energy at supernovae modifies the multi-phase
  structure of the ISM, which exhibits an amorphous clumpy structure when CR
  physics is included.  Without CRs, this multiphase structure is mostly
  suppressed by the stiff effective equation of state above of the adopted
  subgrid model for star formation and its regulation. We envision that such
  a multiphase structure as a result of local CR feedback will also be
  maintained when we additionally account for CR streaming and diffusion,
  and employ a temporarily and spatially varying diffusion coefficient. The
  latter is expected to increase with distance from the location of
  supernova remnants and reach the average Galactic value on large scales.

  Initially, all our galaxy models with halo masses ranging from $10^{10}$
  to $10^{12}\,\rmn{M}_\odot$ exhibit a rapid turbulent dynamo for the
  magnetic field that eventually transitions to a slower amplification
  process. This also increases the magnetic coherence scale of the field as
  it is wound up by differential rotation of the disk. The field structure
  is very regular and quiet in the simulations without CRs, reflecting the
  pressurised equation of state of the ISM. In contrast, in our CR
  simulations, it attains a chaotic small-scale component which is
  superimposed on the dominant azimuthal structure. We find that the
  additional turbulent velocity field is a result of supernova explosions
  that sustain an additional dynamo action, amplifying the field to observed
  strengths exceeding $10~\umu$G in the centres of the disks.

\item {\bf Cosmological simulations.}  To study the impact of diffusive shock
  acceleration on cosmological structure formation shocks, we have simulated a
  representative volume of the universe with the currently favoured $\Lambda$CDM
  cosmology, following non-radiative MHD. By employing our on-the-fly shock
  finder, we model CR acceleration at formation shocks that exceed pre-shock
  Mach numbers $\M_1=3$, and subsequently followed their advective transport as
  well as their Coulomb and hadronic interactions with the ambient gas. We
    find that CRs do not modify the shock statistics, which is shaped by the
  distribution of gravitational potentials in space and potential depths.  CRs
  are mostly accelerated at strong external formation shocks that form at
  overdensities of $\delta_\rmn{b}\sim5$ to $50$ as the pristine cosmic fluid
  collapses onto filaments and sheets. During the collapse into haloes, the
  composite of CRs and thermal gas experiences adiabatic compression, which
  increases both pressure components. However, this favours the thermal pressure
  over that provided by CRs and causes the CR pressure ratio to drop with
  density as $X_\CR\propto \rho^{-1/3}$. At the highest densities, the Coulomb
  and hadronic cooling time scales are shorter than the Hubble time, and CRs
  start to loose pressure support at overdensities of
  $\delta_\rmn{b}\gtrsim10^4$.

\end{itemize}

Combined with our companion paper on the technical details of the anisotropic CR
transport solver on a moving unstructured mesh \citep{Pakmor2016a}, we have
introduced a new advanced treatment of CR physics in current cosmological
hydrodynamic codes. We expect this will be useful for studying many timely
questions related to CR physics and galaxy formation. In \citet{Pakmor2016b} we
have already studied wind formation in disk galaxies, and in \citet{Simpson2016}
explored the problem of star formation regulation in high-resolution simulations
of the ISM. It will be very interesting to extend these works to further science
applications.

\section*{Acknowledgements}
The authors would like to thank the referee for a constructive report which
helped improving this paper. This work has been supported by the European
Research Council under ERC-CoG grant CRAGSMAN-646955, ERC-StG grant
EXAGAL-308037 and by the Klaus Tschira Foundation. VS and KS acknowledge support
through subproject EXAMAG of the Priority Programme 1648 `Software for Exascale
Computing' of the German Science Foundation.

\bibliography{bibtex/paper}

\begin{thebibliography}{}
\makeatletter
\relax
\def\mn@urlcharsother{\let\do\@makeother \do\$\do\&\do\#\do\^\do\_\do\%\do\~}
\def\mn@doi{\begingroup\mn@urlcharsother \@ifnextchar [ {\mn@doi@}
  {\mn@doi@[]}}
\def\mn@doi@[#1]#2{\def\@tempa{#1}\ifx\@tempa\@empty \href
  {http://dx.doi.org/#2} {doi:#2}\else \href {http://dx.doi.org/#2} {#1}\fi
  \endgroup}
\def\mn@eprint#1#2{\mn@eprint@#1:#2::\@nil}
\def\mn@eprint@arXiv#1{\href {http://arxiv.org/abs/#1} {{\tt arXiv:#1}}}
\def\mn@eprint@dblp#1{\href {http://dblp.uni-trier.de/rec/bibtex/#1.xml}
  {dblp:#1}}
\def\mn@eprint@#1:#2:#3:#4\@nil{\def\@tempa {#1}\def\@tempb {#2}\def\@tempc
  {#3}\ifx \@tempc \@empty \let \@tempc \@tempb \let \@tempb \@tempa \fi \ifx
  \@tempb \@empty \def\@tempb {arXiv}\fi \@ifundefined
  {mn@eprint@\@tempb}{\@tempb:\@tempc}{\expandafter \expandafter \csname
  mn@eprint@\@tempb\endcsname \expandafter{\@tempc}}}

\bibitem[\protect\citeauthoryear{{Achatz}, {Steinacker}  \&
  {Schlickeiser}}{{Achatz} et~al.}{1991}]{1991A&A...250..266A}
{Achatz} U.,  {Steinacker} J.,   {Schlickeiser} R.,  1991, \aap, \href
  {http://adsabs.harvard.edu/abs/1991A%26A...250..266A} {250, 266}

\bibitem[\protect\citeauthoryear{{Ackermann} \& et al.}{{Ackermann} \&
  et~al.}{2013}]{2013Sci...339..807A}
{Ackermann} M.,  et al. 2013, \mn@doi [Science] {10.1126/science.1231160},
  \href {http://adsabs.harvard.edu/abs/2013Sci...339..807A} {339, 807}

\bibitem[\protect\citeauthoryear{{Battaglia}, {Bond}, {Pfrommer}  \&
  {Sievers}}{{Battaglia} et~al.}{2012a}]{2012ApJ...758...74B}
{Battaglia} N.,  {Bond} J.~R.,  {Pfrommer} C.,   {Sievers} J.~L.,  2012a,
  \mn@doi [\apj] {10.1088/0004-637X/758/2/74}, \href
  {http://adsabs.harvard.edu/abs/2012ApJ...758...74B} {758, 74}

\bibitem[\protect\citeauthoryear{{Battaglia}, {Bond}, {Pfrommer}  \&
  {Sievers}}{{Battaglia} et~al.}{2012b}]{2012ApJ...758...75B}
{Battaglia} N.,  {Bond} J.~R.,  {Pfrommer} C.,   {Sievers} J.~L.,  2012b,
  \mn@doi [\apj] {10.1088/0004-637X/758/2/75}, \href
  {http://adsabs.harvard.edu/abs/2012ApJ...758...75B} {758, 75}

\bibitem[\protect\citeauthoryear{{Battaglia}, {Bond}, {Pfrommer}  \&
  {Sievers}}{{Battaglia} et~al.}{2013}]{2013ApJ...777..123B}
{Battaglia} N.,  {Bond} J.~R.,  {Pfrommer} C.,   {Sievers} J.~L.,  2013,
  \mn@doi [\apj] {10.1088/0004-637X/777/2/123}, \href
  {http://adsabs.harvard.edu/abs/2013ApJ...777..123B} {777, 123}

\bibitem[\protect\citeauthoryear{{Bauer} \& {Springel}}{{Bauer} \&
  {Springel}}{2012}]{2012MNRAS.423.2558B}
{Bauer} A.,  {Springel} V.,  2012, \mn@doi [\mnras]
  {10.1111/j.1365-2966.2012.21058.x}, \href
  {http://adsabs.harvard.edu/abs/2012MNRAS.423.2558B} {423, 2558}

\bibitem[\protect\citeauthoryear{{Bell}}{{Bell}}{2004}]{2004MNRAS.353..550B}
{Bell} A.~R.,  2004, \mn@doi [\mnras] {10.1111/j.1365-2966.2004.08097.x}, \href
  {http://adsabs.harvard.edu/abs/2004MNRAS.353..550B} {353, 550}

\bibitem[\protect\citeauthoryear{{Booth}, {Agertz}, {Kravtsov}  \&
  {Gnedin}}{{Booth} et~al.}{2013}]{2013ApJ...777L..16B}
{Booth} C.~M.,  {Agertz} O.,  {Kravtsov} A.~V.,   {Gnedin} N.~Y.,  2013,
  \mn@doi [\apjl] {10.1088/2041-8205/777/1/L16}, \href
  {http://adsabs.harvard.edu/abs/2013ApJ...777L..16B} {777, L16}

\bibitem[\protect\citeauthoryear{{Boulares} \& {Cox}}{{Boulares} \&
  {Cox}}{1990}]{1990ApJ...365..544B}
{Boulares} A.,  {Cox} D.~P.,  1990, \mn@doi [\apj] {10.1086/169509}, \href
  {http://adsabs.harvard.edu/abs/1990ApJ...365..544B} {365, 544}

\bibitem[\protect\citeauthoryear{{Breitschwerdt}, {McKenzie}  \&
  {Voelk}}{{Breitschwerdt} et~al.}{1991}]{1991A&A...245...79B}
{Breitschwerdt} D.,  {McKenzie} J.~F.,   {Voelk} H.~J.,  1991, \aap, \href
  {http://adsabs.harvard.edu/abs/1991A%26A...245...79B} {245, 79}

\bibitem[\protect\citeauthoryear{{Breitschwerdt}, {Dogiel}  \&
  {V{\"o}lk}}{{Breitschwerdt} et~al.}{2002}]{2002A&A...385..216B}
{Breitschwerdt} D.,  {Dogiel} V.~A.,   {V{\"o}lk} H.~J.,  2002, \mn@doi [\aap]
  {10.1051/0004-6361:20020152}, \href
  {http://adsabs.harvard.edu/abs/2002A%26A...385..216B} {385, 216}

\bibitem[\protect\citeauthoryear{{Brunetti} \& {Jones}}{{Brunetti} \&
  {Jones}}{2014}]{2014IJMPD..2330007B}
{Brunetti} G.,  {Jones} T.~W.,  2014, \mn@doi [Int. J. Mod. Phys. D]
  {10.1142/S0218271814300079}, \href
  {http://adsabs.harvard.edu/abs/2014IJMPD..2330007B} {23, 1430007}

\bibitem[\protect\citeauthoryear{{Caprioli} \& {Spitkovsky}}{{Caprioli} \&
  {Spitkovsky}}{2014a}]{2014ApJ...783...91C}
{Caprioli} D.,  {Spitkovsky} A.,  2014a, \mn@doi [\apj]
  {10.1088/0004-637X/783/2/91}, \href
  {http://adsabs.harvard.edu/abs/2014ApJ...783...91C} {783, 91}

\bibitem[\protect\citeauthoryear{{Caprioli} \& {Spitkovsky}}{{Caprioli} \&
  {Spitkovsky}}{2014b}]{2014ApJ...794...46C}
{Caprioli} D.,  {Spitkovsky} A.,  2014b, \mn@doi [\apj]
  {10.1088/0004-637X/794/1/46}, \href
  {http://adsabs.harvard.edu/abs/2014ApJ...794...46C} {794, 46}

\bibitem[\protect\citeauthoryear{{Caprioli}, {Pop}  \& {Spitkovsky}}{{Caprioli}
  et~al.}{2015}]{2015ApJ...798L..28C}
{Caprioli} D.,  {Pop} A.-R.,   {Spitkovsky} A.,  2015, \mn@doi [\apjl]
  {10.1088/2041-8205/798/2/L28}, \href
  {http://adsabs.harvard.edu/abs/2015ApJ...798L..28C} {798, L28}

\bibitem[\protect\citeauthoryear{{Courant} \& {Friedrichs}}{{Courant} \&
  {Friedrichs}}{1948}]{1948sfsw.book.....C}
{Courant} R.,  {Friedrichs} K.~O.,  1948, {Supersonic flow and shock waves}.
Pure and Applied Mathematics, New York: Interscience, 1948

\bibitem[\protect\citeauthoryear{{Di Matteo}, {Springel}  \& {Hernquist}}{{Di
  Matteo} et~al.}{2005}]{DiMatteo2005}
{Di Matteo} T.,  {Springel} V.,   {Hernquist} L.,  2005, \mn@doi [\nat]
  {10.1038/nature03335}, \href
  {http://adsabs.harvard.edu/abs/2005Natur.433..604D} {433, 604}

\bibitem[\protect\citeauthoryear{{Dolag}, {Komatsu}  \& {Sunyaev}}{{Dolag}
  et~al.}{2016}]{2016MNRAS.tmp.1157D}
{Dolag} K.,  {Komatsu} E.,   {Sunyaev} R.,  2016, \mn@doi [\mnras]
  {10.1093/mnras/stw2035}, \href
  {http://adsabs.harvard.edu/abs/2016MNRAS.tmp.1157D} {}

\bibitem[\protect\citeauthoryear{{Donnert}, {Dolag}, {Brunetti}  \&
  {Cassano}}{{Donnert} et~al.}{2013}]{2013MNRAS.429.3564D}
{Donnert} J.,  {Dolag} K.,  {Brunetti} G.,   {Cassano} R.,  2013, \mn@doi
  [\mnras] {10.1093/mnras/sts628}, \href
  {http://adsabs.harvard.edu/abs/2013MNRAS.429.3564D} {429, 3564}

\bibitem[\protect\citeauthoryear{{Dorfi} \& {Breitschwerdt}}{{Dorfi} \&
  {Breitschwerdt}}{2012}]{2012A&A...540A..77D}
{Dorfi} E.~A.,  {Breitschwerdt} D.,  2012, \mn@doi [\aap]
  {10.1051/0004-6361/201118082}, \href
  {http://adsabs.harvard.edu/abs/2012A%26A...540A..77D} {540, A77}

\bibitem[\protect\citeauthoryear{{En{\ss}lin}, {Pfrommer}, {Springel}  \&
  {Jubelgas}}{{En{\ss}lin} et~al.}{2007}]{2007A&A...473...41E}
{En{\ss}lin} T.~A.,  {Pfrommer} C.,  {Springel} V.,   {Jubelgas} M.,  2007,
  \mn@doi [\aap] {10.1051/0004-6361:20065294}, \href
  {http://adsabs.harvard.edu/abs/2007A%26A...473...41E} {473, 41}

\bibitem[\protect\citeauthoryear{{En{\ss}lin}, {Pfrommer}, {Miniati}  \&
  {Subramanian}}{{En{\ss}lin} et~al.}{2011}]{2011A&A...527A..99E}
{En{\ss}lin} T.,  {Pfrommer} C.,  {Miniati} F.,   {Subramanian} K.,  2011,
  \mn@doi [\aap] {10.1051/0004-6361/201015652}, \href
  {http://adsabs.harvard.edu/abs/2011A%26A...527A..99E} {527, A99}

\bibitem[\protect\citeauthoryear{{Everett}, {Zweibel}, {Benjamin}, {McCammon},
  {Rocks}  \& {Gallagher}}{{Everett} et~al.}{2008}]{2008ApJ...674..258E}
{Everett} J.~E.,  {Zweibel} E.~G.,  {Benjamin} R.~A.,  {McCammon} D.,  {Rocks}
  L.,   {Gallagher} III J.~S.,  2008, \mn@doi [\apj] {10.1086/524766}, \href
  {http://adsabs.harvard.edu/abs/2008ApJ...674..258E} {674, 258}

\bibitem[\protect\citeauthoryear{{Everett}, {Schiller}  \& {Zweibel}}{{Everett}
  et~al.}{2010}]{2010ApJ...711...13E}
{Everett} J.~E.,  {Schiller} Q.~G.,   {Zweibel} E.~G.,  2010, \mn@doi [\apj]
  {10.1088/0004-637X/711/1/13}, \href
  {http://adsabs.harvard.edu/abs/2010ApJ...711...13E} {711, 13}

\bibitem[\protect\citeauthoryear{{Farmer} \& {Goldreich}}{{Farmer} \&
  {Goldreich}}{2004}]{2004ApJ...604..671F}
{Farmer} A.~J.,  {Goldreich} P.,  2004, \mn@doi [\apj] {10.1086/382040}, \href
  {http://adsabs.harvard.edu/abs/2004ApJ...604..671F} {604, 671}

\bibitem[\protect\citeauthoryear{{Fujita} \& {Ohira}}{{Fujita} \&
  {Ohira}}{2012}]{2012ApJ...746...53F}
{Fujita} Y.,  {Ohira} Y.,  2012, \mn@doi [\apj] {10.1088/0004-637X/746/1/53},
  \href {http://adsabs.harvard.edu/abs/2012ApJ...746...53F} {746, 53}

\bibitem[\protect\citeauthoryear{{Girichidis} et~al.,}{{Girichidis}
  et~al.}{2016}]{2016ApJ...816L..19G}
{Girichidis} P.,  et~al., 2016, \mn@doi [\apjl] {10.3847/2041-8205/816/2/L19},
  \href {http://adsabs.harvard.edu/abs/2016ApJ...816L..19G} {816, L19}

\bibitem[\protect\citeauthoryear{{Gould}}{{Gould}}{1972}]{1972Physica....58..379G}
{Gould} R.~J.,  1972, Physica, 58, 379

\bibitem[\protect\citeauthoryear{{Guedes}, {Callegari}, {Madau}  \&
  {Mayer}}{{Guedes} et~al.}{2011}]{2011ApJ...742...76G}
{Guedes} J.,  {Callegari} S.,  {Madau} P.,   {Mayer} L.,  2011, \mn@doi [\apj]
  {10.1088/0004-637X/742/2/76}, \href
  {http://adsabs.harvard.edu/abs/2011ApJ...742...76G} {742, 76}

\bibitem[\protect\citeauthoryear{{Guo} \& {Oh}}{{Guo} \&
  {Oh}}{2008}]{2008MNRAS.384..251G}
{Guo} F.,  {Oh} S.~P.,  2008, \mn@doi [\mnras]
  {10.1111/j.1365-2966.2007.12692.x}, \href
  {http://adsabs.harvard.edu/abs/2008MNRAS.384..251G} {384, 251}

\bibitem[\protect\citeauthoryear{{Hall} \& {Sturrock}}{{Hall} \&
  {Sturrock}}{1967}]{1967PhFl...10.2620H}
{Hall} D.~E.,  {Sturrock} P.~A.,  1967, \mn@doi [Physics of Fluids]
  {10.1063/1.1762084}, \href
  {http://adsabs.harvard.edu/abs/1967PhFl...10.2620H} {10, 2620}

\bibitem[\protect\citeauthoryear{{Hanasz}, {Lesch}, {Naab}, {Gawryszczak},
  {Kowalik}  \& {W{\'o}lta{\'n}ski}}{{Hanasz}
  et~al.}{2013}]{2013ApJ...777L..38H}
{Hanasz} M.,  {Lesch} H.,  {Naab} T.,  {Gawryszczak} A.,  {Kowalik} K.,
  {W{\'o}lta{\'n}ski} D.,  2013, \mn@doi [\apjl] {10.1088/2041-8205/777/2/L38},
  \href {http://adsabs.harvard.edu/abs/2013ApJ...777L..38H} {777, L38}

\bibitem[\protect\citeauthoryear{{Helder}, {Vink}, {Bykov}, {Ohira}, {Raymond}
  \& {Terrier}}{{Helder} et~al.}{2012}]{2012SSRv..173..369H}
{Helder} E.~A.,  {Vink} J.,  {Bykov} A.~M.,  {Ohira} Y.,  {Raymond} J.~C.,
  {Terrier} R.,  2012, \mn@doi [\ssr] {10.1007/s11214-012-9919-8}, \href
  {http://adsabs.harvard.edu/abs/2012SSRv..173..369H} {173, 369}

\bibitem[\protect\citeauthoryear{{Henriques}, {White}, {Thomas}, {Angulo},
  {Guo}, {Lemson}, {Springel}  \& {Overzier}}{{Henriques}
  et~al.}{2015}]{Henriques2015}
{Henriques} B.~M.~B.,  {White} S.~D.~M.,  {Thomas} P.~A.,  {Angulo} R.,  {Guo}
  Q.,  {Lemson} G.,  {Springel} V.,   {Overzier} R.,  2015, \mn@doi [\mnras]
  {10.1093/mnras/stv705}, \href
  {http://adsabs.harvard.edu/abs/2015MNRAS.451.2663H} {451, 2663}

\bibitem[\protect\citeauthoryear{{Hong}, {Ryu}, {Kang}  \& {Cen}}{{Hong}
  et~al.}{2014}]{2014ApJ...785..133H}
{Hong} S.~E.,  {Ryu} D.,  {Kang} H.,   {Cen} R.,  2014, \mn@doi [\apj]
  {10.1088/0004-637X/785/2/133}, \href
  {http://adsabs.harvard.edu/abs/2014ApJ...785..133H} {785, 133}

\bibitem[\protect\citeauthoryear{{Hopkins}, {Kere{\v s}}, {O{\~n}orbe},
  {Faucher-Gigu{\`e}re}, {Quataert}, {Murray}  \& {Bullock}}{{Hopkins}
  et~al.}{2014}]{2014MNRAS.445..581H}
{Hopkins} P.~F.,  {Kere{\v s}} D.,  {O{\~n}orbe} J.,  {Faucher-Gigu{\`e}re}
  C.-A.,  {Quataert} E.,  {Murray} N.,   {Bullock} J.~S.,  2014, \mn@doi
  [\mnras] {10.1093/mnras/stu1738}, \href
  {http://adsabs.harvard.edu/abs/2014MNRAS.445..581H} {445, 581}

\bibitem[\protect\citeauthoryear{{Ipavich}}{{Ipavich}}{1975}]{1975ApJ...196..107I}
{Ipavich} F.~M.,  1975, \mn@doi [\apj] {10.1086/153397}, \href
  {http://adsabs.harvard.edu/abs/1975ApJ...196..107I} {196, 107}

\bibitem[\protect\citeauthoryear{{Jacob} \& {Pfrommer}}{{Jacob} \&
  {Pfrommer}}{2016a}]{2016arXiv160906321J}
{Jacob} S.,  {Pfrommer} C.,  2016a, preprint, \href
  {http://adsabs.harvard.edu/abs/2016arXiv160906321J} {} (\mn@eprint {arXiv}
  {1609.06321})

\bibitem[\protect\citeauthoryear{{Jacob} \& {Pfrommer}}{{Jacob} \&
  {Pfrommer}}{2016b}]{2016arXiv160906322J}
{Jacob} S.,  {Pfrommer} C.,  2016b, preprint, \href
  {http://adsabs.harvard.edu/abs/2016arXiv160906322J} {} (\mn@eprint {arXiv}
  {1609.06322})

\bibitem[\protect\citeauthoryear{{Jubelgas}, {Springel}, {En{\ss}lin}  \&
  {Pfrommer}}{{Jubelgas} et~al.}{2008}]{2008A&A...481...33J}
{Jubelgas} M.,  {Springel} V.,  {En{\ss}lin} T.,   {Pfrommer} C.,  2008,
  \mn@doi [\aap] {10.1051/0004-6361:20065295}, \href
  {http://adsabs.harvard.edu/abs/2008A%26A...481...33J} {481, 33}

\bibitem[\protect\citeauthoryear{{Kang}, {Ryu}, {Cen}  \& {Ostriker}}{{Kang}
  et~al.}{2007}]{2007ApJ...669..729K}
{Kang} H.,  {Ryu} D.,  {Cen} R.,   {Ostriker} J.~P.,  2007, \mn@doi [\apj]
  {10.1086/521717}, \href {http://adsabs.harvard.edu/abs/2007ApJ...669..729K}
  {669, 729}

\bibitem[\protect\citeauthoryear{{Kennicutt}}{{Kennicutt}}{1998}]{1998ApJ...498..541K}
{Kennicutt} Jr. R.~C.,  1998, \mn@doi [\apj] {10.1086/305588}, \href
  {http://adsabs.harvard.edu/abs/1998ApJ...498..541K} {498, 541}

\bibitem[\protect\citeauthoryear{{Kravtsov} \& {Borgani}}{{Kravtsov} \&
  {Borgani}}{2012}]{2012ARA&A..50..353K}
{Kravtsov} A.~V.,  {Borgani} S.,  2012, \mn@doi [\araa]
  {10.1146/annurev-astro-081811-125502}, \href
  {http://adsabs.harvard.edu/abs/2012ARA%26A..50..353K} {50, 353}

\bibitem[\protect\citeauthoryear{{Kroupa}}{{Kroupa}}{2001}]{2001MNRAS.322..231K}
{Kroupa} P.,  2001, \mn@doi [\mnras] {10.1046/j.1365-8711.2001.04022.x}, \href
  {http://adsabs.harvard.edu/abs/2001MNRAS.322..231K} {322, 231}

\bibitem[\protect\citeauthoryear{{Krumholz} \& {Thompson}}{{Krumholz} \&
  {Thompson}}{2012}]{2012ApJ...760..155K}
{Krumholz} M.~R.,  {Thompson} T.~A.,  2012, \mn@doi [\apj]
  {10.1088/0004-637X/760/2/155}, \href
  {http://adsabs.harvard.edu/abs/2012ApJ...760..155K} {760, 155}

\bibitem[\protect\citeauthoryear{{Kulsrud} \& {Pearce}}{{Kulsrud} \&
  {Pearce}}{1969}]{1969ApJ...156..445K}
{Kulsrud} R.,  {Pearce} W.~P.,  1969, \mn@doi [\apj] {10.1086/149981}, \href
  {http://adsabs.harvard.edu/abs/1969ApJ...156..445K} {156, 445}

\bibitem[\protect\citeauthoryear{{Lerche}}{{Lerche}}{1967}]{1967ApJ...147..689L}
{Lerche} I.,  1967, \mn@doi [\apj] {10.1086/149045}, \href
  {http://adsabs.harvard.edu/abs/1967ApJ...147..689L} {147, 689}

\bibitem[\protect\citeauthoryear{{Lloyd}}{{Lloyd}}{1982}]{Lloyd}
{Lloyd} S.,  1982, IEEE Trans Inf. Theory, 28, 129

\bibitem[\protect\citeauthoryear{{Loewenstein}, {Zweibel}  \&
  {Begelman}}{{Loewenstein} et~al.}{1991}]{1991ApJ...377..392L}
{Loewenstein} M.,  {Zweibel} E.~G.,   {Begelman} M.~C.,  1991, \mn@doi [\apj]
  {10.1086/170369}, \href {http://adsabs.harvard.edu/abs/1991ApJ...377..392L}
  {377, 392}

\bibitem[\protect\citeauthoryear{{Mannheim} \& {Schlickeiser}}{{Mannheim} \&
  {Schlickeiser}}{1994}]{1994A&A...286..983M}
{Mannheim} K.,  {Schlickeiser} R.,  1994, \aap, 286, 983

\bibitem[\protect\citeauthoryear{{Marinacci}, {Pakmor}  \&
  {Springel}}{{Marinacci} et~al.}{2014}]{2014MNRAS.437.1750M}
{Marinacci} F.,  {Pakmor} R.,   {Springel} V.,  2014, \mn@doi [\mnras]
  {10.1093/mnras/stt2003}, \href
  {http://adsabs.harvard.edu/abs/2014MNRAS.437.1750M} {437, 1750}

\bibitem[\protect\citeauthoryear{{McCarthy}, {Le Brun}, {Schaye}  \&
  {Holder}}{{McCarthy} et~al.}{2014}]{2014MNRAS.440.3645M}
{McCarthy} I.~G.,  {Le Brun} A.~M.~C.,  {Schaye} J.,   {Holder} G.~P.,  2014,
  \mn@doi [\mnras] {10.1093/mnras/stu543}, \href
  {http://adsabs.harvard.edu/abs/2014MNRAS.440.3645M} {440, 3645}

\bibitem[\protect\citeauthoryear{{McCarthy}, {Schaye}, {Bird}  \& {Le
  Brun}}{{McCarthy} et~al.}{2016}]{2016arXiv160302702M}
{McCarthy} I.~G.,  {Schaye} J.,  {Bird} S.,   {Le Brun} A.~M.~C.,  2016,
  preprint, \href {http://adsabs.harvard.edu/abs/2016arXiv160302702M} {}
  (\mn@eprint {arXiv} {1603.02702})

\bibitem[\protect\citeauthoryear{{McKenzie} \& {Voelk}}{{McKenzie} \&
  {Voelk}}{1982}]{1982A&A...116..191M}
{McKenzie} J.~F.,  {Voelk} H.~J.,  1982, \aap, \href
  {http://adsabs.harvard.edu/abs/1982A%26A...116..191M} {116, 191}

\bibitem[\protect\citeauthoryear{{McNamara} \& {Nulsen}}{{McNamara} \&
  {Nulsen}}{2007}]{2007ARA&A..45..117M}
{McNamara} B.~R.,  {Nulsen} P.~E.~J.,  2007, \mn@doi [\araa]
  {10.1146/annurev.astro.45.051806.110625}, \href
  {http://adsabs.harvard.edu/abs/2007ARA%26A..45..117M} {45, 117}

\bibitem[\protect\citeauthoryear{{McNamara} \& {Nulsen}}{{McNamara} \&
  {Nulsen}}{2012}]{2012NJPh...14e5023M}
{McNamara} B.~R.,  {Nulsen} P.~E.~J.,  2012, \mn@doi [New Journal of Physics]
  {10.1088/1367-2630/14/5/055023}, \href
  {http://adsabs.harvard.edu/abs/2012NJPh...14e5023M} {14, 055023}

\bibitem[\protect\citeauthoryear{{Miniati}, {Ryu}, {Kang}  \&
  {Jones}}{{Miniati} et~al.}{2001a}]{2001ApJ...559...59M}
{Miniati} F.,  {Ryu} D.,  {Kang} H.,   {Jones} T.~W.,  2001a, \mn@doi [\apj]
  {10.1086/322375}, \href {http://adsabs.harvard.edu/abs/2001ApJ...559...59M}
  {559, 59}

\bibitem[\protect\citeauthoryear{{Miniati}, {Jones}, {Kang}  \&
  {Ryu}}{{Miniati} et~al.}{2001b}]{2001ApJ...562..233M}
{Miniati} F.,  {Jones} T.~W.,  {Kang} H.,   {Ryu} D.,  2001b, \mn@doi [\apj]
  {10.1086/323434}, \href {http://adsabs.harvard.edu/abs/2001ApJ...562..233M}
  {562, 233}

\bibitem[\protect\citeauthoryear{{Miyoshi} \& {Kusano}}{{Miyoshi} \&
  {Kusano}}{2005}]{2005JCoPh.208..315M}
{Miyoshi} T.,  {Kusano} K.,  2005, \mn@doi [J. Comput. Phys.]
  {10.1016/j.jcp.2005.02.017}, \href
  {http://adsabs.harvard.edu/abs/2005JCoPh.208..315M} {208, 315}

\bibitem[\protect\citeauthoryear{{Morlino} \& {Caprioli}}{{Morlino} \&
  {Caprioli}}{2012}]{2012A&A...538A..81M}
{Morlino} G.,  {Caprioli} D.,  2012, \mn@doi [\aap]
  {10.1051/0004-6361/201117855}, \href
  {http://adsabs.harvard.edu/abs/2012A%26A...538A..81M} {538, A81}

\bibitem[\protect\citeauthoryear{{Murray}, {Quataert}  \& {Thompson}}{{Murray}
  et~al.}{2005}]{2005ApJ...618..569M}
{Murray} N.,  {Quataert} E.,   {Thompson} T.~A.,  2005, \mn@doi [\apj]
  {10.1086/426067}, \href {http://adsabs.harvard.edu/abs/2005ApJ...618..569M}
  {618, 569}

\bibitem[\protect\citeauthoryear{{Navarro}, {Frenk}  \& {White}}{{Navarro}
  et~al.}{1997}]{1997ApJ...490..493N}
{Navarro} J.~F.,  {Frenk} C.~S.,   {White} S.~D.~M.,  1997, \apj, \href
  {http://adsabs.harvard.edu/abs/1997ApJ...490..493N} {490, 493}

\bibitem[\protect\citeauthoryear{{Oppenheimer} \& {Dav{\'e}}}{{Oppenheimer} \&
  {Dav{\'e}}}{2006}]{Oppenheimer2006}
{Oppenheimer} B.~D.,  {Dav{\'e}} R.,  2006, \mn@doi [\mnras]
  {10.1111/j.1365-2966.2006.10989.x}, \href
  {http://adsabs.harvard.edu/abs/2006MNRAS.373.1265O} {373, 1265}

\bibitem[\protect\citeauthoryear{{Pakmor} \& {Springel}}{{Pakmor} \&
  {Springel}}{2013}]{2013MNRAS.432..176P}
{Pakmor} R.,  {Springel} V.,  2013, \mn@doi [\mnras] {10.1093/mnras/stt428},
  \href {http://adsabs.harvard.edu/abs/2013MNRAS.432..176P} {432, 176}

\bibitem[\protect\citeauthoryear{{Pakmor}, {Bauer}  \& {Springel}}{{Pakmor}
  et~al.}{2011}]{2011MNRAS.418.1392P}
{Pakmor} R.,  {Bauer} A.,   {Springel} V.,  2011, \mn@doi [\mnras]
  {10.1111/j.1365-2966.2011.19591.x}, \href
  {http://adsabs.harvard.edu/abs/2011MNRAS.418.1392P} {418, 1392}

\bibitem[\protect\citeauthoryear{{Pakmor}, {Springel}, {Bauer}, {Mocz},
  {Munoz}, {Ohlmann}, {Schaal}  \& {Zhu}}{{Pakmor}
  et~al.}{2016a}]{2016MNRAS.455.1134P}
{Pakmor} R.,  {Springel} V.,  {Bauer} A.,  {Mocz} P.,  {Munoz} D.~J.,
  {Ohlmann} S.~T.,  {Schaal} K.,   {Zhu} C.,  2016a, \mn@doi [\mnras]
  {10.1093/mnras/stv2380}, \href
  {http://adsabs.harvard.edu/abs/2016MNRAS.455.1134P} {455, 1134}

\bibitem[\protect\citeauthoryear{{Pakmor}, {Pfrommer}, {Simpson}, {Kannan}  \&
  {Springel}}{{Pakmor} et~al.}{2016b}]{Pakmor2016a}
{Pakmor} R.,  {Pfrommer} C.,  {Simpson} C.~M.,  {Kannan} R.,   {Springel} V.,
  2016b, \mn@doi [\mnras] {10.1093/mnras/stw1761}, \href
  {http://adsabs.harvard.edu/abs/2016MNRAS.462.2603P} {462, 2603}

\bibitem[\protect\citeauthoryear{{Pakmor}, {Pfrommer}, {Simpson}  \&
  {Springel}}{{Pakmor} et~al.}{2016c}]{Pakmor2016b}
{Pakmor} R.,  {Pfrommer} C.,  {Simpson} C.~M.,   {Springel} V.,  2016c, \mn@doi
  [\apjl] {10.3847/2041-8205/824/2/L30}, \href
  {http://adsabs.harvard.edu/abs/2016ApJ...824L..30P} {824, L30}

\bibitem[\protect\citeauthoryear{{Peterson} \& {Fabian}}{{Peterson} \&
  {Fabian}}{2006}]{2006PhR...427....1P}
{Peterson} J.~R.,  {Fabian} A.~C.,  2006, \mn@doi [\physrep]
  {10.1016/j.physrep.2005.12.007}, \href
  {http://adsabs.harvard.edu/abs/2006PhR...427....1P} {427, 1}

\bibitem[\protect\citeauthoryear{{Pfrommer}}{{Pfrommer}}{2008}]{2008MNRAS.385.1242P}
{Pfrommer} C.,  2008, \mn@doi [\mnras] {10.1111/j.1365-2966.2008.12957.x},
  \href {http://adsabs.harvard.edu/abs/2008MNRAS.385.1242P} {385, 1242}

\bibitem[\protect\citeauthoryear{{Pfrommer}}{{Pfrommer}}{2013}]{2013ApJ...779...10P}
{Pfrommer} C.,  2013, \mn@doi [\apj] {10.1088/0004-637X/779/1/10}, \href
  {http://adsabs.harvard.edu/abs/2013ApJ...779...10P} {779, 10}

\bibitem[\protect\citeauthoryear{{Pfrommer} \& {En{\ss}lin}}{{Pfrommer} \&
  {En{\ss}lin}}{2004}]{2004A&A...413...17P}
{Pfrommer} C.,  {En{\ss}lin} T.~A.,  2004, \mn@doi [\aap]
  {10.1051/0004-6361:20031464}, \href
  {http://adsabs.harvard.edu/abs/2004A%26A...413...17P} {413, 17}

\bibitem[\protect\citeauthoryear{{Pfrommer}, {Springel}, {En{\ss}lin}  \&
  {Jubelgas}}{{Pfrommer} et~al.}{2006}]{2006MNRAS.367..113P}
{Pfrommer} C.,  {Springel} V.,  {En{\ss}lin} T.~A.,   {Jubelgas} M.,  2006,
  \mn@doi [\mnras] {10.1111/j.1365-2966.2005.09953.x}, \href
  {http://adsabs.harvard.edu/abs/2006MNRAS.367..113P} {367, 113}

\bibitem[\protect\citeauthoryear{{Pfrommer}, {En{\ss}lin}, {Springel},
  {Jubelgas}  \& {Dolag}}{{Pfrommer} et~al.}{2007}]{2007MNRAS.378..385P}
{Pfrommer} C.,  {En{\ss}lin} T.~A.,  {Springel} V.,  {Jubelgas} M.,   {Dolag}
  K.,  2007, \mn@doi [\mnras] {10.1111/j.1365-2966.2007.11732.x}, \href
  {http://adsabs.harvard.edu/abs/2007MNRAS.378..385P} {378, 385}

\bibitem[\protect\citeauthoryear{{Pfrommer}, {En{\ss}lin}  \&
  {Springel}}{{Pfrommer} et~al.}{2008}]{2008MNRAS.385.1211P}
{Pfrommer} C.,  {En{\ss}lin} T.~A.,   {Springel} V.,  2008, \mn@doi [\mnras]
  {10.1111/j.1365-2966.2008.12956.x}, \href
  {http://adsabs.harvard.edu/abs/2008MNRAS.385.1211P} {385, 1211}

\bibitem[\protect\citeauthoryear{{Pinzke} \& {Pfrommer}}{{Pinzke} \&
  {Pfrommer}}{2010}]{2010MNRAS.409..449P}
{Pinzke} A.,  {Pfrommer} C.,  2010, \mn@doi [\mnras]
  {10.1111/j.1365-2966.2010.17328.x}, \href
  {http://adsabs.harvard.edu/abs/2010MNRAS.409..449P} {409, 449}

\bibitem[\protect\citeauthoryear{{Pinzke}, {Oh}  \& {Pfrommer}}{{Pinzke}
  et~al.}{2013}]{2013MNRAS.435.1061P}
{Pinzke} A.,  {Oh} S.~P.,   {Pfrommer} C.,  2013, \mn@doi [\mnras]
  {10.1093/mnras/stt1308}, \href
  {http://adsabs.harvard.edu/abs/2013MNRAS.435.1061P} {435, 1061}

\bibitem[\protect\citeauthoryear{{Pinzke}, {Oh}  \& {Pfrommer}}{{Pinzke}
  et~al.}{2015}]{2015arXiv150307870P}
{Pinzke} A.,  {Oh} S.~P.,   {Pfrommer} C.,  2015, preprint, \href
  {http://adsabs.harvard.edu/abs/2015arXiv150307870P} {} (\mn@eprint {arXiv}
  {1503.07870})

\bibitem[\protect\citeauthoryear{{Planck Collaboration} et~al.,}{{Planck
  Collaboration} et~al.}{2016}]{2016A&A...594A..13P}
{Planck Collaboration} et~al., 2016, \mn@doi [\aap]
  {10.1051/0004-6361/201525830}, \href
  {http://adsabs.harvard.edu/abs/2016A%26A...594A..13P} {594, A13}

\bibitem[\protect\citeauthoryear{{Planelles} \& {Quilis}}{{Planelles} \&
  {Quilis}}{2013}]{2013MNRAS.428.1643P}
{Planelles} S.,  {Quilis} V.,  2013, \mn@doi [\mnras] {10.1093/mnras/sts142},
  \href {http://adsabs.harvard.edu/abs/2013MNRAS.428.1643P} {428, 1643}

\bibitem[\protect\citeauthoryear{{Powell}, {Roe}, {Linde}, {Gombosi}  \& {De
  Zeeuw}}{{Powell} et~al.}{1999}]{1999JCoPh.154..284P}
{Powell} K.~G.,  {Roe} P.~L.,  {Linde} T.~J.,  {Gombosi} T.~I.,   {De Zeeuw}
  D.~L.,  1999, \mn@doi [J. Comput. Phys.] {10.1006/jcph.1999.6299}, \href
  {http://adsabs.harvard.edu/abs/1999JCoPh.154..284P} {154, 284}

\bibitem[\protect\citeauthoryear{{Ptuskin}, {Voelk}, {Zirakashvili}  \&
  {Breitschwerdt}}{{Ptuskin} et~al.}{1997}]{1997A&A...321..434P}
{Ptuskin} V.~S.,  {Voelk} H.~J.,  {Zirakashvili} V.~N.,   {Breitschwerdt} D.,
  1997, \aap, \href {http://adsabs.harvard.edu/abs/1997A%26A...321..434P} {321,
  434}

\bibitem[\protect\citeauthoryear{{Puchwein} \& {Springel}}{{Puchwein} \&
  {Springel}}{2013}]{Puchwein2013}
{Puchwein} E.,  {Springel} V.,  2013, \mn@doi [\mnras] {10.1093/mnras/sts243},
  \href {http://adsabs.harvard.edu/abs/2013MNRAS.428.2966P} {428, 2966}

\bibitem[\protect\citeauthoryear{{Recchia}, {Blasi}  \& {Morlino}}{{Recchia}
  et~al.}{2016}]{2016MNRAS.462.4227R}
{Recchia} S.,  {Blasi} P.,   {Morlino} G.,  2016, \mn@doi [\mnras]
  {10.1093/mnras/stw1966}, \href
  {http://adsabs.harvard.edu/abs/2016MNRAS.462.4227R} {462, 4227}

\bibitem[\protect\citeauthoryear{{Rephaeli}}{{Rephaeli}}{1979}]{1979ApJ...227..364R}
{Rephaeli} Y.,  1979, \mn@doi [\apj] {10.1086/156740}, \href
  {http://adsabs.harvard.edu/abs/1979ApJ...227..364R} {227, 364}

\bibitem[\protect\citeauthoryear{{Rodrigues}, {Sarson}, {Shukurov}, {Bushby}
  \& {Fletcher}}{{Rodrigues} et~al.}{2016}]{2016ApJ...816....2R}
{Rodrigues} L.~F.~S.,  {Sarson} G.~R.,  {Shukurov} A.,  {Bushby} P.~J.,
  {Fletcher} A.,  2016, \mn@doi [\apj] {10.3847/0004-637X/816/1/2}, \href
  {http://adsabs.harvard.edu/abs/2016ApJ...816....2R} {816, 2}

\bibitem[\protect\citeauthoryear{{Rosdahl}, {Schaye}, {Teyssier}  \&
  {Agertz}}{{Rosdahl} et~al.}{2015}]{2015MNRAS.451...34R}
{Rosdahl} J.,  {Schaye} J.,  {Teyssier} R.,   {Agertz} O.,  2015, \mn@doi
  [\mnras] {10.1093/mnras/stv937}, \href
  {http://adsabs.harvard.edu/abs/2015MNRAS.451...34R} {451, 34}

\bibitem[\protect\citeauthoryear{{Ruszkowski}, {Yang}  \&
  {Zweibel}}{{Ruszkowski} et~al.}{2016}]{2016arXiv160204856R}
{Ruszkowski} M.,  {Yang} H.-Y.~K.,   {Zweibel} E.,  2016, preprint, \href
  {http://adsabs.harvard.edu/abs/2016arXiv160204856R} {} (\mn@eprint {arXiv}
  {1602.04856})

\bibitem[\protect\citeauthoryear{{Ryu}, {Kang}, {Hallman}  \& {Jones}}{{Ryu}
  et~al.}{2003}]{2003ApJ...593..599R}
{Ryu} D.,  {Kang} H.,  {Hallman} E.,   {Jones} T.~W.,  2003, \mn@doi [\apj]
  {10.1086/376723}, \href {http://adsabs.harvard.edu/abs/2003ApJ...593..599R}
  {593, 599}

\bibitem[\protect\citeauthoryear{{Salem} \& {Bryan}}{{Salem} \&
  {Bryan}}{2014}]{2014MNRAS.437.3312S}
{Salem} M.,  {Bryan} G.~L.,  2014, \mn@doi [\mnras] {10.1093/mnras/stt2121},
  \href {http://adsabs.harvard.edu/abs/2014MNRAS.437.3312S} {437, 3312}

\bibitem[\protect\citeauthoryear{{Salem}, {Bryan}  \& {Hummels}}{{Salem}
  et~al.}{2014}]{2014ApJ...797L..18S}
{Salem} M.,  {Bryan} G.~L.,   {Hummels} C.,  2014, \mn@doi [\apjl]
  {10.1088/2041-8205/797/2/L18}, \href
  {http://adsabs.harvard.edu/abs/2014ApJ...797L..18S} {797, L18}

\bibitem[\protect\citeauthoryear{{Samui}, {Subramanian}  \& {Srianand}}{{Samui}
  et~al.}{2010}]{2010MNRAS.402.2778S}
{Samui} S.,  {Subramanian} K.,   {Srianand} R.,  2010, \mn@doi [\mnras]
  {10.1111/j.1365-2966.2009.16099.x}, \href
  {http://adsabs.harvard.edu/abs/2010MNRAS.402.2778S} {402, 2778}

\bibitem[\protect\citeauthoryear{{Sarazin}}{{Sarazin}}{1999}]{1999ApJ...520..529S}
{Sarazin} C.~L.,  1999, \mn@doi [\apj] {10.1086/307501}, \href
  {http://adsabs.harvard.edu/abs/1999ApJ...520..529S} {520, 529}

\bibitem[\protect\citeauthoryear{{Schaal} \& {Springel}}{{Schaal} \&
  {Springel}}{2015}]{2015MNRAS.446.3992S}
{Schaal} K.,  {Springel} V.,  2015, \mn@doi [\mnras] {10.1093/mnras/stu2386},
  \href {http://adsabs.harvard.edu/abs/2015MNRAS.446.3992S} {446, 3992}

\bibitem[\protect\citeauthoryear{{Schaal} et~al.,}{{Schaal}
  et~al.}{2016}]{2016MNRAS.461.4441S}
{Schaal} K.,  et~al., 2016, \mn@doi [\mnras] {10.1093/mnras/stw1587}, \href
  {http://adsabs.harvard.edu/abs/2016MNRAS.461.4441S} {461, 4441}

\bibitem[\protect\citeauthoryear{{Schaye} \& {Dalla Vecchia}}{{Schaye} \&
  {Dalla Vecchia}}{2008}]{Schaye2008}
{Schaye} J.,  {Dalla Vecchia} C.,  2008, \mn@doi [\mnras]
  {10.1111/j.1365-2966.2007.12639.x}, \href
  {http://adsabs.harvard.edu/abs/2008MNRAS.383.1210S} {383, 1210}

\bibitem[\protect\citeauthoryear{{Schaye} et~al.,}{{Schaye}
  et~al.}{2010}]{2010MNRAS.402.1536S}
{Schaye} J.,  et~al., 2010, \mn@doi [\mnras]
  {10.1111/j.1365-2966.2009.16029.x}, \href
  {http://adsabs.harvard.edu/abs/2010MNRAS.402.1536S} {402, 1536}

\bibitem[\protect\citeauthoryear{{Schaye} et~al.,}{{Schaye}
  et~al.}{2015}]{Schaye2015Eagle}
{Schaye} J.,  et~al., 2015, \mn@doi [\mnras] {10.1093/mnras/stu2058}, \href
  {http://adsabs.harvard.edu/abs/2015MNRAS.446..521S} {446, 521}

\bibitem[\protect\citeauthoryear{{Schlickeiser}}{{Schlickeiser}}{2002}]{2002cra..book.....S}
{Schlickeiser} R.,  2002, {Cosmic Ray Astrophysics}

\bibitem[\protect\citeauthoryear{{Sedov}}{{Sedov}}{1959}]{1959sdmm.book.....S}
{Sedov} L.~I.,  1959, {Similarity and Dimensional Methods in Mechanics}

\bibitem[\protect\citeauthoryear{{Sijacki}, {Pfrommer}, {Springel}  \&
  {En{\ss}lin}}{{Sijacki} et~al.}{2008}]{2008MNRAS.387.1403S}
{Sijacki} D.,  {Pfrommer} C.,  {Springel} V.,   {En{\ss}lin} T.~A.,  2008,
  \mn@doi [\mnras] {10.1111/j.1365-2966.2008.13310.x}, \href
  {http://adsabs.harvard.edu/abs/2008MNRAS.387.1403S} {387, 1403}

\bibitem[\protect\citeauthoryear{{Sijacki}, {Vogelsberger}, {Kere{\v s}},
  {Springel}  \& {Hernquist}}{{Sijacki} et~al.}{2012}]{2012MNRAS.424.2999S}
{Sijacki} D.,  {Vogelsberger} M.,  {Kere{\v s}} D.,  {Springel} V.,
  {Hernquist} L.,  2012, \mn@doi [\mnras] {10.1111/j.1365-2966.2012.21466.x},
  \href {http://adsabs.harvard.edu/abs/2012MNRAS.424.2999S} {424, 2999}

\bibitem[\protect\citeauthoryear{{Simpson}, {Pakmor}, {Marinacci}, {Pfrommer},
  {Springel}, {Glover}, {Clark}  \& {Smith}}{{Simpson}
  et~al.}{2016}]{Simpson2016}
{Simpson} C.~M.,  {Pakmor} R.,  {Marinacci} F.,  {Pfrommer} C.,  {Springel} V.,
   {Glover} S.~C.~O.,  {Clark} P.~C.,   {Smith} R.~J.,  2016, \mn@doi [\apjl]
  {10.3847/2041-8205/827/2/L29}, \href
  {http://adsabs.harvard.edu/abs/2016ApJ...827L..29S} {827, L29}

\bibitem[\protect\citeauthoryear{{Skilling}}{{Skilling}}{1971}]{1971ApJ...170..265S}
{Skilling} J.,  1971, \mn@doi [\apj] {10.1086/151210}, \href
  {http://adsabs.harvard.edu/abs/1971ApJ...170..265S} {170, 265}

\bibitem[\protect\citeauthoryear{{Skilling}}{{Skilling}}{1975}]{1975MNRAS.172..557S}
{Skilling} J.,  1975, \mn@doi [\mnras] {10.1093/mnras/172.3.557}, \href
  {http://adsabs.harvard.edu/abs/1975MNRAS.172..557S} {172, 557}

\bibitem[\protect\citeauthoryear{{Skillman}, {O'Shea}, {Hallman}, {Burns}  \&
  {Norman}}{{Skillman} et~al.}{2008}]{2008ApJ...689.1063S}
{Skillman} S.~W.,  {O'Shea} B.~W.,  {Hallman} E.~J.,  {Burns} J.~O.,   {Norman}
  M.~L.,  2008, \mn@doi [\apj] {10.1086/592496}, \href
  {http://adsabs.harvard.edu/abs/2008ApJ...689.1063S} {689, 1063}

\bibitem[\protect\citeauthoryear{{Skinner} \& {Ostriker}}{{Skinner} \&
  {Ostriker}}{2015}]{2015ApJ...809..187S}
{Skinner} M.~A.,  {Ostriker} E.~C.,  2015, \mn@doi [\apj]
  {10.1088/0004-637X/809/2/187}, \href
  {http://adsabs.harvard.edu/abs/2015ApJ...809..187S} {809, 187}

\bibitem[\protect\citeauthoryear{{Socrates}, {Davis}  \&
  {Ramirez-Ruiz}}{{Socrates} et~al.}{2008}]{2008ApJ...687..202S}
{Socrates} A.,  {Davis} S.~W.,   {Ramirez-Ruiz} E.,  2008, \mn@doi [\apj]
  {10.1086/590046}, \href {http://adsabs.harvard.edu/abs/2008ApJ...687..202S}
  {687, 202}

\bibitem[\protect\citeauthoryear{{Springel}}{{Springel}}{2010}]{2010MNRAS.401..791S}
{Springel} V.,  2010, \mn@doi [\mnras] {10.1111/j.1365-2966.2009.15715.x},
  \href {http://adsabs.harvard.edu/abs/2010MNRAS.401..791S} {401, 791}

\bibitem[\protect\citeauthoryear{{Springel} \& {Hernquist}}{{Springel} \&
  {Hernquist}}{2003}]{2003MNRAS.339..289S}
{Springel} V.,  {Hernquist} L.,  2003, \mn@doi [\mnras]
  {10.1046/j.1365-8711.2003.06206.x}, \href
  {http://esoads.eso.org/cgi-bin/nph-bib_query?bibcode=2003MNRAS.339..289S&db_key=AST}
  {339, 289}

\bibitem[\protect\citeauthoryear{{Springel}, {Di Matteo}  \&
  {Hernquist}}{{Springel} et~al.}{2005}]{Springel2005}
{Springel} V.,  {Di Matteo} T.,   {Hernquist} L.,  2005, \mn@doi [\mnras]
  {10.1111/j.1365-2966.2005.09238.x}, \href
  {http://adsabs.harvard.edu/abs/2005MNRAS.361..776S} {361, 776}

\bibitem[\protect\citeauthoryear{{Stage}, {Allen}, {Houck}  \& {Davis}}{{Stage}
  et~al.}{2006}]{2006NatPh...2..614S}
{Stage} M.~D.,  {Allen} G.~E.,  {Houck} J.~C.,   {Davis} J.~E.,  2006, \mn@doi
  [Nature Physics] {10.1038/nphys391}, \href
  {http://adsabs.harvard.edu/abs/2006NatPh...2..614S} {2, 614}

\bibitem[\protect\citeauthoryear{{Takahashi}, {Yamada}  \&
  {Yamada}}{{Takahashi} et~al.}{2014}]{2014JPlPh..80..255T}
{Takahashi} K.,  {Yamada} S.,   {Yamada} 2014, \mn@doi [Journal of Plasma
  Physics] {10.1017/S0022377813001268}, \href
  {http://adsabs.harvard.edu/abs/2014JPlPh..80..255T} {80, 255}

\bibitem[\protect\citeauthoryear{{Thompson}, {Quataert}  \&
  {Murray}}{{Thompson} et~al.}{2005}]{2005ApJ...630..167T}
{Thompson} T.~A.,  {Quataert} E.,   {Murray} N.,  2005, \mn@doi [\apj]
  {10.1086/431923}, \href {http://adsabs.harvard.edu/abs/2005ApJ...630..167T}
  {630, 167}

\bibitem[\protect\citeauthoryear{{T{\"u}llmann}, {Dettmar}, {Soida}, {Urbanik}
  \& {Rossa}}{{T{\"u}llmann} et~al.}{2000}]{2000A&A...364L..36T}
{T{\"u}llmann} R.,  {Dettmar} R.-J.,  {Soida} M.,  {Urbanik} M.,   {Rossa} J.,
  2000, \aap, \href {http://adsabs.harvard.edu/abs/2000A%26A...364L..36T} {364,
  L36}

\bibitem[\protect\citeauthoryear{{Uhlig}, {Pfrommer}, {Sharma}, {Nath},
  {En{\ss}lin}  \& {Springel}}{{Uhlig} et~al.}{2012}]{2012MNRAS.423.2374U}
{Uhlig} M.,  {Pfrommer} C.,  {Sharma} M.,  {Nath} B.~B.,  {En{\ss}lin} T.~A.,
  {Springel} V.,  2012, \mn@doi [\mnras] {10.1111/j.1365-2966.2012.21045.x},
  \href {http://adsabs.harvard.edu/abs/2012MNRAS.423.2374U} {423, 2374}

\bibitem[\protect\citeauthoryear{{Vazza}, {Brunetti}  \& {Gheller}}{{Vazza}
  et~al.}{2009}]{2009MNRAS.395.1333V}
{Vazza} F.,  {Brunetti} G.,   {Gheller} C.,  2009, \mn@doi [\mnras]
  {10.1111/j.1365-2966.2009.14691.x}, \href
  {http://adsabs.harvard.edu/abs/2009MNRAS.395.1333V} {395, 1333}

\bibitem[\protect\citeauthoryear{{Vazza}, {Dolag}, {Ryu}, {Brunetti},
  {Gheller}, {Kang}  \& {Pfrommer}}{{Vazza} et~al.}{2011}]{2011MNRAS.418..960V}
{Vazza} F.,  {Dolag} K.,  {Ryu} D.,  {Brunetti} G.,  {Gheller} C.,  {Kang} H.,
   {Pfrommer} C.,  2011, \mn@doi [\mnras] {10.1111/j.1365-2966.2011.19546.x},
  \href {http://adsabs.harvard.edu/abs/2011MNRAS.418..960V} {418, 960}

\bibitem[\protect\citeauthoryear{{Vazza}, {Br{\"u}ggen}, {Gheller}  \&
  {Brunetti}}{{Vazza} et~al.}{2012}]{2012MNRAS.421.3375V}
{Vazza} F.,  {Br{\"u}ggen} M.,  {Gheller} C.,   {Brunetti} G.,  2012, \mn@doi
  [\mnras] {10.1111/j.1365-2966.2012.20562.x}, \href
  {http://adsabs.harvard.edu/abs/2012MNRAS.421.3375V} {421, 3375}

\bibitem[\protect\citeauthoryear{{Vazza}, {Gheller}  \& {Br{\"u}ggen}}{{Vazza}
  et~al.}{2014}]{2014MNRAS.439.2662V}
{Vazza} F.,  {Gheller} C.,   {Br{\"u}ggen} M.,  2014, \mn@doi [\mnras]
  {10.1093/mnras/stu126}, \href
  {http://adsabs.harvard.edu/abs/2014MNRAS.439.2662V} {439, 2662}

\bibitem[\protect\citeauthoryear{{Vogelsberger}, {Sijacki}, {Kere{\v s}},
  {Springel}  \& {Hernquist}}{{Vogelsberger}
  et~al.}{2012}]{2012MNRAS.425.3024V}
{Vogelsberger} M.,  {Sijacki} D.,  {Kere{\v s}} D.,  {Springel} V.,
  {Hernquist} L.,  2012, \mn@doi [\mnras] {10.1111/j.1365-2966.2012.21590.x},
  \href {http://adsabs.harvard.edu/abs/2012MNRAS.425.3024V} {425, 3024}

\bibitem[\protect\citeauthoryear{{Vogelsberger} et~al.,}{{Vogelsberger}
  et~al.}{2014}]{2014MNRAS.444.1518V}
{Vogelsberger} M.,  et~al., 2014, \mn@doi [\mnras] {10.1093/mnras/stu1536},
  \href {http://adsabs.harvard.edu/abs/2014MNRAS.444.1518V} {444, 1518}

\bibitem[\protect\citeauthoryear{{Wiener}, {Oh}  \& {Guo}}{{Wiener}
  et~al.}{2013}]{2013MNRAS.434.2209W}
{Wiener} J.,  {Oh} S.~P.,   {Guo} F.,  2013, \mn@doi [\mnras]
  {10.1093/mnras/stt1163}, \href
  {http://adsabs.harvard.edu/abs/2013MNRAS.434.2209W} {434, 2209}

\bibitem[\protect\citeauthoryear{{Wiener}, {Pfrommer}  \& {Oh}}{{Wiener}
  et~al.}{2016}]{2016arXiv160802585W}
{Wiener} J.,  {Pfrommer} C.,   {Oh} S.~P.,  2016, preprint, \href
  {http://adsabs.harvard.edu/abs/2016arXiv160802585W} {} (\mn@eprint {arXiv}
  {1608.02585})

\bibitem[\protect\citeauthoryear{{Zhuravleva} et~al.,}{{Zhuravleva}
  et~al.}{2014}]{2014Natur.515...85Z}
{Zhuravleva} I.,  et~al., 2014, \mn@doi [\nat] {10.1038/nature13830}, \href
  {http://esoads.eso.org/abs/2014Natur.515...85Z} {515, 85}

\bibitem[\protect\citeauthoryear{{Zirakashvili}, {Breitschwerdt}, {Ptuskin}  \&
  {Voelk}}{{Zirakashvili} et~al.}{1996}]{1996A&A...311..113Z}
{Zirakashvili} V.~N.,  {Breitschwerdt} D.,  {Ptuskin} V.~S.,   {Voelk} H.~J.,
  1996, \aap, \href {http://adsabs.harvard.edu/abs/1996A%26A...311..113Z} {311,
  113}

\makeatother
\end{thebibliography}
\bibliographystyle{mnras}

\onecolumn
\appendix

\section{Cosmic ray hydrodynamics}
\label{sec:CRhydro}

In this Appendix, we derive the energy equation for CRs in a magnetised plasma
\citep{1982A&A...116..191M, 2008MNRAS.384..251G}, additionally augmented with an
expression for momentum diffusion as a result of second-order Fermi
acceleration. When CRs are streaming along the local magnetic field at a speed
faster than the Alfv\'en speed, they resonantly excite Alfv\'en waves at the
gyroscale by the CR streaming instability \citep{1967ApJ...147..689L,
  1969ApJ...156..445K}. These waves scatter the CRs in pitch angle and attempt
to confine them to the frame comoving with the Alfv\'en waves. Here, we only
consider forward Alfv\'en waves that propagate nearly parallel to the
unperturbed background magnetic field, in the direction of the streaming CRs
\citep[backward Alfv\'en waves are damped; see][]{1967ApJ...147..689L,
  1969ApJ...156..445K}. Depending on the damping rate of the forward Alfv\'en
waves due to turbulent damping or non-linear Landau damping in an ionised plasma
\citep{2004ApJ...604..671F, 1969ApJ...156..445K}, and additional ion-neutral
damping in a sufficiently neutral plasma, this confinement can be incomplete,
leading to a diffusive motion relative to the forward Alfv\'en wave frame
\citep{2013MNRAS.434.2209W}.

Defining a dimensionless momentum of a particle, $\vec{p} = \vec{P}_\p/(m c)$ and its
magnitude $p=|\vec{p}|$, we start with the relativistic Vlasov equation for the
3D CR distribution function $f_\p$ and derive a Fokker-Planck
equation for the transport of CRs \citep{1971ApJ...170..265S,
  1975MNRAS.172..557S, 2002cra..book.....S},
\begin{equation}
  \label{eq:fp1}
  \pp{f_\p}{t} + (\bvel + \bvel_{\rmn{st}}) \bcdot \bnabla f_\p
  = \bnabla\bcdot \left[\kappa_\p \bb\, \left(\bb\bcdot\bnabla f_\p\right)\right]
  + \frac{1}{3} p\pp{f_\p}{p}\,\bnabla\bcdot(\bvel + \bvel_{\rmn{st}})
  + \frac{1}{p^2}\pp{}{p}\left[p^2\Gamma_\p\,\pp{f_\p}{p}\right]
  + Q_\p\,,
\end{equation}
where $f_\p=f_\p(\vec{x}, p, t)$ is the isotropic momentum part of the CR phase
space distribution function. This equation has been derived in the quasi-linear
approximation that assumes small-amplitude electro-magnetic fluctuations. It
requires incoherent mode coupling of the fluctuating electromagnetic fields
described as the superposition of individual plasma wave modes.  It is only
valid on timescales long compared to the pitch angle scattering relaxation time
$\tau\sim\mathcal{O}(D_{\mu\mu}^{-1})$ where
$D_{\mu\mu}=D_{\mu\mu}(\vec{x},p,\mu)$ is the Fokker-Planck coefficient
representing the frequency of pitch angle scattering of CRs by hydromagnetic
waves. Here $\mu\equiv \vec{b}\bcdot\vec{p}/p$ denotes the pitch-angle cosine, and
$\vec{b}=\vec{B}/|\vec{B}|$ is a unit vector along the local magnetic
field. 

Under these conditions, the particles can locally reach near-equilibrium, which
results in a small anisotropy of the distribution function, i.e.~$\delta f_\p
\ll f_\p$.  Note that equation~(\ref{eq:fp1}) employs a mixed coordinate frame
in which the configuration-space coordinates ($\vec{x}$) are measured in the
laboratory system and the momentum-space coordinates
$\vec{p}=\left(p\sqrt{1-\mu^2} \cos\varphi,
p\sqrt{1-\mu^2}\sin\varphi,p\mu\right)$ are defined with respect to the rest
frame of the streaming CRs, i.e., in the frame comoving with the velocity
$\bvel+\bvel_{\rmn{st}}$.  In this equation, $\bvel$ is the mean velocity of the
thermal background plasma, $\bvel_{\rmn{st}}=-\bvel_{\rmn{A}}\,
\rmn{sgn}(\bB\bcdot\bnabla f_\p)$ is the streaming velocity of CRs,
$\bvel_{\rmn{A}} = \vec{B}/\sqrt{4\upi\rho}$ is the local Alfv\'en velocity (in
the cgs system of units), and $\bnabla\equiv \partial/\partial \vec{x}$.  The
spatial diffusion coefficient $\kappa_\p$ (in units of cm$^2$ s$^{-1}$) and the
momentum diffusion rate $\Gamma_\p$ (in units of s$^{-1}$) are given by
\begin{equation}
\label{eq:kappa}
\kappa_\p(\vec{x},p)
=\frac{\beta^2c^2}{8}\,\int_{-1}^{+1}\frac{(1-\mu^2)^2}{D_{\mu\mu}}\,\d \mu,
\qquad\mbox{and}
\qquad
\Gamma_\p=
\frac{1}{2}\int_{-1}^{+1}
\left[D_{pp}-
  \frac{D_{\mu p}^2}{D_{\mu\mu}}\right]\d\mu,
\end{equation}
where $\beta=p/\gamma$ is the dimensionless CR particle speed and $\gamma=
\sqrt{1+p^2}$ is its Lorentz factor. Here $D_{pp}=D_{pp}(\vec{x},p,\mu)$ and
$D_{\mu p}=D_{\mu p}(\vec{x},p,\mu)$ are the Fokker-Planck coefficients
representing the ensemble averages of the rate of change of the particles'
momenta $p$ with themselves and with the rate of change of the particles' pitch
angles $\mu$, evaluated along the first-order corrections to the particle orbits
\citep{1967PhFl...10.2620H, 1991A&A...250..266A},
\begin{equation}
  \label{eq:FokkerPlanck}
  D_{\mu\mu}=\mathcal{R}\int_0^\infty \left\bra\dot{\mu}(t)\dot{\mu}^*(t+\tau)\right\ket\d\tau,
  \qquad
  D_{\mu p}=\mathcal{R}\int_0^\infty \left\bra\dot{\mu}(t)\dot{p}^*(t+\tau)\right\ket\d\tau,
  \qquad
  D_{pp}=\mathcal{R}\int_0^\infty \left\bra\dot{p}(t)\dot{p}^*(t+\tau)\right\ket\d\tau.
\end{equation}
Here, the asterisk denotes the complex conjugate, $\mathcal{R}$ denotes the real
part of the integral, and the equation of motion is
\begin{equation}
  \label{eq:EoM}
  \dot{\vec{p}}=Ze\left[\delta\vec{E}
    + \frac{1}{m c\gamma} \vec{p}\btimes\left(\vec{B}_0+\delta\vec{B}\right)\right], 
\end{equation}
where $Ze$ denotes the charge of the particle, and $(\delta\vec{E},
\delta\vec{B})$ denote the fluctuations of the electromagnetic field with
respect to the mean magnetic field $\vec{B}_0$.
The term on the left-hand side of equation~(\ref{eq:fp1}) accounts for advective
transport of the CR distribution function with the Alfv\'en wave frame relative
to the laboratory rest frame, while the terms on the right-hand side represent,
from left to right: diffusive transport along magnetic field lines,
diffusive shock (first-order Fermi) acceleration, second-order Fermi
acceleration (which is equivalent to momentum-space diffusion), and sources and
sinks for the distribution function (generally denoted by $Q_\p$).  

To derive the evolution equation for the CR number and energy density,
we define three thermodynamic quantities as moments of $f_\p$, namely CR
number density $n_\CR$, CR pressure $P_\CR$, and CR energy density
$\eps_\CR$:
\begin{eqnarray}
  \label{eq:ncr}
  n_{\CR} &=& 4\upi\int_0^\infty p^2\,f_\p(p)\,\d p
  = \frac{4\upi\, C}{\alpha-3}\, q^{3-\alpha},\\
  P_\CR &=& \frac{4\upi\,m c^2}{3}\,\int_0^\infty \beta p^3\, f_\p(p)\,\d p
  =\frac{4\upi\,C\,m c^2}{6} \, 
  \B_{\frac{1}{1+q^2}} \left( \frac{\alpha-4}{2},\frac{5-\alpha}{2} \right), \\
  \eps_\CR &=&4\upi \int_0^\infty p^2 \,E_\p(p)\,f_\p(p) \,\d p
  =\frac{4\upi\,C\, m c^2}{\alpha-3} \,
  \left[\frac{1}{2}\, \B_{\frac{1}{1+q^2}}
    \left(\frac{\alpha-4}{2},\frac{5-\alpha}{2}\right)
    + q^{3-\alpha}
    \left(\sqrt{1+q^2}-1 \right)
    \right] . \label{eq:ecr}
\end{eqnarray}
Here $\B_x(a,b)$ denotes the incomplete beta function (assuming $\alpha>4$) and
$E_\p(p)$ is the kinetic energy of a CR particle with momentum $p$,
\begin{equation}
E_\p(p) = \left(\sqrt{1+p^2} -1\right)\, m c^2.
\end{equation}
For the explicit forms of equations~(\ref{eq:ncr}) to (\ref{eq:ecr}), we adopted a
power-law CR distribution function,
\begin{equation}
  \label{eq:f}
  f_\p(\vec{x}, p, t) \equiv \frac{\d N}{\d^3 p\,\d V}
  = C \, p^{-\alpha}\,\theta(p-q),
\end{equation}
with low-momentum cutoff $q$, normalisation $C$ and 3D spectral index
$\alpha$.\footnote{Note that it is not required to specify the form of the CR
  distribution function for the derivation of the CR energy equation. However,
  we will employ such a simplified form for the distribution function of the
  injected CR population when we derive the CR cooling rates in
  Sect.~\ref{sec:loss}.}

Integration of equation~(\ref{eq:fp1}) over all particle momenta yields the
evolution equation for the CR number density $n_\CR$:
\begin{equation}
  \label{eq:evolution_ncr}
  \pp{n_\CR}{t}
  +\bnabla\bcdot\vec{F}_n = \bar{Q}_n,
  \quad\mbox{where}\quad
  \vec{F}_n= (\bvel+\bvel_{\rmn{st}})\,n_\CR
  - \kappa_n \vec{b}\,\left(\vec{b}\bcdot\bnabla n_\CR\right).
\end{equation}
Here, $\kappa_n$ is the momentum-space averaged spatial diffusion coefficient and
$\bar{Q}_n$ is the net source of CRs, given by
\begin{equation}
  \label{eq:kappa_n}
  \kappa_n =
  \frac{\int_0^\infty p^2 \kappa_\p(\vec{x},p)\,(\vec{b}\bcdot\bnabla f_\p)\,\d p}
       {\int_0^\infty p^2 \,(\vec{b}\bcdot\bnabla f_\p)\,\d p},
   \quad\mbox{and}\quad
  \bar{Q}_n = 
  4\upi\int_0^\infty p^2 Q_\p(\vec{x},p)\,\d p.
\end{equation}
Note that the second-order Fermi acceleration term (with the rate coefficient
$\Gamma_\p$) drops from this equation because this process conserves CR particle
number. For this to hold mathematically, we have to require that the combination
$\Gamma_\p\partial f_\p/\partial p$ vanishes at infinity in momentum space.

Multiplication of equation (\ref{eq:fp1}) by $E_\p(p)$ and integration over all
particle momenta results in an evolution equation for the CR energy density,
\begin{eqnarray}
  \label{eq:evolution_ecr}
  \pp{\eps_\CR}{t} + \bnabla\bcdot \vec{F}_\eps
  = (\bvel+\bvel_{\rmn{st}})\bcdot\bnabla P_\CR
  + \Gamma_{\rmn{acc}} + \bar{Q}_\eps,
  \quad\mbox{where}\quad
  \vec{F}_\eps =
  (\bvel+\bvel_{\rmn{st}})\,(P_\CR+\eps_\CR)
  - \kappa_\eps \vec{b}\,\left(\vec{b}\bcdot\bnabla\eps_\CR\right).
\end{eqnarray}
Here $\kappa_\eps$ is the kinetic energy-weighted spatial diffusion coefficient,
$\bar{Q}_\eps$ is the net source of mean kinetic energy density of CRs, and
$\Gamma_{\rmn{acc}}$ is the gain rate of energy density due to second-order
Fermi acceleration in units of erg cm$^{-3}$ s$^{-1}$:
\begin{eqnarray}
  \label{eq:kappa_cr}
  \kappa_\eps &=&
  \frac{\int_0^\infty p^2 E_\p(p) \kappa_\p(\vec{x},p)\,
    (\vec{b}\bcdot\bnabla f_\p)\,\d p}
       {\int_0^\infty p^2 E_\p(p)\,(\vec{b}\bcdot\bnabla f_\p)\,\d p},\\
  \bar{Q}_\eps &=& 
  4\upi\int_0^\infty p^2 E_\p(p) Q_\p(\vec{x},p)\,\d p,\mbox{ and}\\
  \Gamma_{\rmn{acc}} &=& -4\upi m c^2\,\int_0^\infty
  \beta p^2 \Gamma_\p\pp{f_\p}{p}\,\d p > 0
  \quad\mbox{for}\quad
  \pp{f_\p}{p}<0.
\end{eqnarray}

The flux function $\vec{F}_\eps$ of equation~(\ref{eq:evolution_ecr}) represents
the advective transport of CR enthalpy density ($h_\CR=P_\CR+\eps_\CR$) with the
total velocity ($\bvel + \bvel_{\rmn{st}}$) as well as the anisotropic diffusive
transport of CR energy density into and out of a given volume element. The first
term on the right-hand side describes the energy-loss rate of CRs due to the
volume work of the CR pressure gradient on the background plasma
($\bvel\bcdot\bnabla P_\CR$) and the generation of Alfv\'en waves
($\bvel_{\rmn{st}}\bcdot\bnabla P_\CR$).  $\Gamma_{\rmn{acc}}$ accounts for
energy gain due to second-order Fermi acceleration (which is only a positive
gain process if $\partial f_\p/\partial p<0$) and
$\bar{Q}_\eps=\Gamma_\CR+\Lambda_\CR$ represents various gain and loss processes
for the CR energy density.  Equation~(\ref{eq:evolution_ecr}) is mathematically
equivalent to the fourth row of the matrix equation~(\ref{eq:conservation})
employed in our code (where we subsumed $\Gamma_{\rmn{acc}}$ into $\Gamma_\CR$
for clarity of the notation). To first approximation, we will adopt a constant
spatial diffusion coefficient so that $\kappa_\p = \kappa_\eps$. Future work
will employ a momentum dependence of $\kappa_\p$ so that the energy-weighted
spatial diffusion coefficient automatically acquires a spatial dependence
through the gradient of the distribution function ($\bnabla f_\p$) in
equation~(\ref{eq:kappa_cr}). Omission of this spatial dependence will lead to
results that are not self-consistent.

\section{Riemann shock-tube problem with cosmic ray acceleration}
\label{sec:Riemann}

Exact solvers of the Riemann shock-tube problem are of eminent
importance for understanding the hydrodynamic behaviour of a fluid and
for validating numerical implementations of approximate Riemann
solvers. While exact solutions to the problem have been put forward
for a single polytropic fluid \citep{1948sfsw.book.....C}, for a MHD
fluid \citep{2014JPlPh..80..255T}, and a two-component fluid composed
of CRs and thermal gas \citep{2006MNRAS.367..113P}, such an exact
solution of the Riemann problem with CR acceleration at shocks is
still lacking.

Collisionless shocks in astrophysical plasmas are able to accelerate
thermal ions through the process of diffusive shock acceleration. The
presence of freshly injected CRs modifies the classical
Rankine-Hugoniot jump conditions due to the softer equation of state
of CRs, which leads to a more compressible composite gas in the
post-shock regime and thus an enhanced density jump compared to the
classical case of a purely thermal gas. In reality, CRs are diffusing
multiple times across the shock front and develop a precursor in the
upstream that adiabatically heats the incoming fluid before it
encounters the discontinuity at the subshock. While it is possible for
the highest-energy CRs to escape upstream of the shock, the majority
-- and in particular the pressure-carrying CRs with energies
$E\sim m c^2$ -- are swept downstream of the shock.  Since we are
interested in scales much larger than the diffusion length, which is
given by the spatial extent of the CR precursor that we do not aim to
resolve, we represent the shock region by a discontinuity and assume
that the freshly accelerated CRs are injected into the downstream
region of the shock.

The mathematical complexity of the solution differs depending on the
presence of CRs in the initial conditions. Hence, in this section we
first present the exact solution for the Riemann problem in the case
of a polytropic gas (i.e., $P_\th = (\gamma-1)\eps_\th$) experiencing
a collisionless shock that is sufficiently strong to accelerate
CRs. Then, in Appendix~\ref{sec:Riemann+CRs} we consider the case of a
gas composed of a pre-existing population of CRs and thermal gas and
allowing for the acceleration of CRs. The freshly injected CRs obey an
equation of state $P_\inj = (\gamma_\inj - 1) \eps_\inj$, where
$\gamma_\inj = 4/3$ for an ultra-relativistic CR population that can
be accelerated at a strong shock. The injected energy density into
CRs, $\eps_\inj$, is a constant fraction of the total dissipated
energy density at the shock (which is equal to the generated internal
energy density corrected for adiabatic compression),
\begin{equation}
  \label{eq:epsinj}
  \eps_\inj = \zeta \eps_\rmn{diss} = \zeta (\eps_2 - \eps_1
  x_\rmn{s}^\gamma),
\end{equation}
where the compression ratio at the shock is denoted by $x_\rmn{s} =
\rho_2/\rho_1$, the total post-shock energy density is $\eps_2=\eps_\inj +
\eps_{\th2}$, and $\zeta$ is the effective energy injection efficiency
after correcting for the fraction of low-energy CRs that is immediately
re-thermalized by Coulomb interactions with thermal protons.  Here, $\eps_1$
and $\eps_2$ indicate the total energy densities in the upstream and
downstream regime of the shock, respectively.

In the following, we summarise the steps which lead to the solution of the
Riemann problem, for completeness. Without loss of generality, we assume an
initial state with higher pressure in the left half-space. At time $t>0$, the
evolving solution is characterised by five regions of gas with different
hydrodynamical states which are numbered in ascending order from the right. From
the left to right, these regions are separated by the head and the tail of the
leftwards propagating rarefaction wave, and the rightwards propagating contact
discontinuity and the shock. A Galilean transformation of the Rankine-Hugoniot
shock jump conditions from the shock to the laboratory rest system leads to the
generalised Rankine-Hugoniot conditions of mass, momentum, and energy
conservation at a shock,
\begin{eqnarray}
  \label{eq:generalRH}
  \vel_\rmn{s}[\rho] &=& [\rho \vel], \nonumber\\
  \vel_\rmn{s}[\rho \vel] &=& [\rho \vel^2 + P], \\
  \vel_\rmn{s}\left[\rho \frac{\vel^2}{2} + \eps\right] &=& 
  \left[\left(\rho \frac{\vel^2}{2} + \eps + P\right)\vel\right]. \nonumber
\end{eqnarray}
Here $\vel_\rmn{s}$ and $\vel$ denote the shock and the mean gas velocity
measured in the laboratory rest system and we defined the abbreviation $[F] \equiv
F_i - F_j$ for the jump of some quantity $F$ across the shock.  Note that we
assume the pressure of freshly injected CRs to be only nonzero in regime 2 in
between the shock and the contact discontinuity. The leftwards propagating
rarefaction wave is characterised by an isentropic change of state, $\d s = 0$
($s$ is the specific entropy), that conserves the Riemann invariant $\Gamma^+$:
\begin{equation}
  \label{eq:Riemann}
  \Gamma^+ = \vel + \int_0^\rho \frac{c_{\rmn{sound}}(\rho')}{\rho'}\d \rho'
   = \vel + \frac{2\, c_{\rmn{sound}}(\rho)}{\gamma - 1} = \rmn{const}.
\end{equation}
In the last step, we assumed a polytropic equation of state of the thermal gas,
$P = A \rho^\gamma$, where $A=\rmn{const.}$ for an isentropic change of
state. Appropriately combining these equations, the solution reads as follows:
\begin{equation}
  \label{eq:rho}
  \rho(x, t) = 
  \left\{ \begin{array}{ll}
      \rho_5, & x \le -c_5 t, \\
      \rho_5 \left[-\eta^2 \frac{\dps x}{\dps c_5 t} + (1 - \eta^2)\right]^{2/(\gamma-1)}, 
      & -c_5 t < x \le -\vel_\rmn{t} t, \\
      \rho_3, & -\vel_\rmn{t} t < x \le \vel_2 t, \\
      \rho_2, & \vel_2 t < x \le \vel_\rmn{s} t, \\
      \rho_1, & x > \vel_\rmn{s} t, \\
    \end{array} \right.
\end{equation}
\begin{equation}
  \label{eq:P}
  P(x, t) = 
  \left\{ \begin{array}{ll}
      P_5, & x \le -c_5 t, \\
      P_5 \left[-\eta^2 \frac{\dps x}{\dps c_5 t} + (1 - \eta^2)\right]^{2\gamma/(\gamma-1)}, 
      & -c_5 t < x \le -\vel_\rmn{t} t, \\
      P_2 = P_3, & -\vel_\rmn{t} t < x \le \vel_\rmn{s} t, \\
      P_1, & x > \vel_\rmn{s} t, \\
    \end{array} \right.
\end{equation}
\begin{equation}
  \label{eq:vel}
  \vel(x, t) = 
  \left\{ \begin{array}{ll}
      0, & x \le -c_5 t, \\
      (1 - \eta^2)\left( \frac{\dps x}{\dps t} + c_5\right),
      & -c_5 t < x \le -\vel_\rmn{t} t, \\
      \vel_2 = \vel_3, & -\vel_\rmn{t} t < x \le \vel_\rmn{s} t, \\
      0, & x > \vel_\rmn{s} t. \\
    \end{array} \right.
\end{equation}
Here $\eta^2 = (\gamma - 1) / (\gamma + 1)$, $c_1 = \sqrt{\gamma P_1 / \rho_1}$,
and $c_5 = \sqrt{\gamma P_5 / \rho_5}$ are the speeds of sound in the
unperturbed state to the right and left, respectively, $\vel_\rmn{s}$ is the
shock speed and $\vel_\rmn{t}$ is the speed of propagation of the rarefaction
wave's tail in the laboratory system. The total post-shock pressure
$P_2=P_{\th2} + P_\inj$ is obtained by solving (numerically) the non-linear
equation for the compression ratio $x_\s$, which is derived from the generalised
Rankine-Hugoniot conditions over the shock while ensuring the conservation of
the Riemann invariant that connects the states 5 and 3 according to
equation~(\ref{eq:Riemann}):
\begin{equation}
  \label{eq:NLeq}
  \left[\frac{P_2(x_\rmn{s})}{P_1} - 1\right]
  \frac{\A(x_\rmn{s})}{1+\A(x_\rmn{s})}
  - \frac{2\gamma}{(\gamma - 1)^2}\frac{c_5^2}{c_1^2} 
  \left\{1 - \left[\frac{P_2(x_\rmn{s})}{P_5}\right]^{(\gamma-1)/(2\gamma)}\right\}^2 = 0.
\end{equation}
To derive this equation, we introduce the jump of the thermal pressure across
the shock, $y_\s$, the ratio of CR-to-thermal energy flux generated at the
shock, $\xi = \zeta / (1 - \zeta)$, and the Atwood number, $\A$, and find
\begin{eqnarray}
  \label{eq:ys}
  y_\s(x_\s) &=&
  \frac{P_{\th2}(x_\s)}{P_1}
  = \frac{\xi\,[x_\s(\gamma_\inj - 1) - (\gamma_\inj + 1)]\,x_\s^\gamma
           - x_\s(\gamma+1) + (\gamma - 1)}
         {\xi\,[x_\s(\gamma_\inj - 1) - (\gamma_\inj + 1)]\ph{\,x_\s^\gamma\ }
           + x_\s(\gamma-1) - (\gamma + 1)},\\
  \label{eq:Atwood}
  \A(x_\s) &\equiv&
  \frac{\rho_2-\rho_1}{\rho_2+\rho_1} =
  \frac{P_{\th2} (\gamma_\inj-1)\ph{\gamma} + P_\inj (\gamma-1) \ph{\gamma_\inj} - P_1(\gamma_\inj-1)\ph{\gamma}}
       {P_{\th2} (\gamma_\inj-1)\gamma      + P_\inj (\gamma-1) \gamma_\inj      + P_1(\gamma_\inj-1)\gamma     }
   =
   \frac{y_\s(x_\s)\ph{\gamma} + \xi\,\left[y_\s(x_\s) - x_\s^\gamma\right]\ph{\gamma_\inj} - 1}
        {y_\s(x_\s)\gamma      + \xi\,\left[y_\s(x_\s) - x_\s^\gamma\right]\gamma_\inj      + \gamma},\\
  \label{eq:P2}
  P_2(x_\rmn{s}) &\equiv& P_{\th2}(x_\s) + P_\inj(x_\rmn{s}) =
  \left\{y_\s(x_\s) + \frac{\gamma_\inj-1}{\gamma-1}\xi\,
  \left[y_\s(x_\s) - x_\s^\gamma\right]\right\} P_1, \\
  \label{eq:Pinj}
  P_\inj(x_\rmn{s}) &\equiv& 
  \frac{\gamma_\inj-1}{\gamma-1}\xi\,
  \left[y_\s(x_\s) - x_\s^\gamma\right] P_1.
\end{eqnarray}
In the limiting case of no CR injection ($\xi = 0$), it can be shown
straightforwardly that equation (\ref{eq:NLeq}) reduces to equation (A6) in
\citet{2006MNRAS.367..113P} for the classical case of a polytropic fluid without
CR acceleration.  The density on the left of the contact discontinuity is
$\rho_3 = \rho_5 [P_2(x_\s) / P_5]^{1 / \gamma}$, since the gas is adiabatically
connected to the left. The post-shock density is simply given by $\rho_2 = x_\s
\rho_1$.  The velocity of the post-shock gas, $\vel_2$, is obtained by combining
the rarefaction wave equation, $x / t = \vel - c$, and the Riemann invariant
$\Gamma^+$:
\begin{equation}
  \label{eq:v3}
  \vel_2 = \vel_3 = \frac{2 c_5}{(\gamma - 1)}
  \left[1 - \left(\frac{P_2(x_\s)}{P_5}\right)^{(\gamma - 1) / (2\gamma)}\right].
\end{equation}
Mass conservation across the shock yields $\vel_\rmn{s}$, and the speed of
propagation of the rarefaction wave's tail, $\vel_\rmn{t}$, is derived with the
aid of (\ref{eq:vel}),
\begin{eqnarray}
  \label{eq:vs}
  \vel_\rmn{s} = \frac{\vel_\rmn{2}}{1 - \rho_1 / \rho_2}
  \quad\mbox{and}\quad
  \vel_\rmn{t} = c_5 - \frac{\vel_2}{1 - \eta^2}.
\end{eqnarray}

\section{Riemann shock-tube problem for a composite of cosmic rays and thermal
  gas with cosmic ray acceleration} 
\label{sec:Riemann+CRs}

\subsection{Derivation}

A composite gas consisting of CRs and thermal particles does not obey a
polytropic equation of state with a constant adiabatic index; only the
sup-populations (thermal gas and CRs) fulfil such a relation separately. In the
following, we summarise the key considerations that lead to the exact solution
of the Riemann shock-tube problem for a composite of thermal gas and CRs (that
are adiabatically compressed at the shock) while allowing for CR acceleration at
the shock. Accounting for CR shock acceleration yields a composite gas in the
post-shock region that is more compressible and experiences an enhanced density
jump in comparison to the case without CR acceleration
\citep{2006MNRAS.367..113P}. To proceed, we adopt the following three
approximations: (i) as before, we only consider scales much larger than the CR
diffusion length so that we can represent the region surrounding the shock by a
discontinuous jump of thermodynamic quantities, (ii) we assume that the
pre-existing CR population is adiabatically compressed over the shock and keep
the adiabatic index of this CR population, $\gamma_\CR$, constant over the
shock-tube, and (iii) we assume that the freshly injected CR population obeys
the equation of state, $P_\inj = (\gamma_\inj - 1) \eps_\inj$, where
$\gamma_\inj = 4/3$ for an ultra-relativistic CR population that can be
accelerated at a strong shock.  Note that the pressure of freshly injected CRs
is only nonzero in the post-shock regime between the shock and the contact
discontinuity. Taking $\gamma_\CR=\rmn{const}.$ is justified as long as the
pre-existing CR pressure is not dominated by trans-relativistic CRs of low
energy.

As before, we adopt the convention that the high-pressure state in the initial
condition is on the left-hand side. The evolving solution for time $t>0$ is
characterised by five regions of gas with different hydrodynamical states which
are numbered in ascending order from the right. Starting at the low-pressure
state to the right, we encounter the pre-shock region (1), the post-shock region
(2), the region trapped in between the contact discontinuity and the rarefaction
wave (3), the rarefaction wave itself (4), and the unperturbed high-pressure
region to the left (5).  We use the notation $P_2 = P_{\inj} + P_{\CR2} +
P_{\th2}$, $\eps_2 = \eps_{\inj} + \eps_{\CR2} + \eps_{\th2}$, $P_3 = P_{\CR3} +
P_{\th3}$, and $\eps_3 = \eps_{\CR3} + \eps_{\th3}$ for the total composite
pressures and energy densities in the respective regions. The exact solution of
the initial value problem requires to determine the time evolution of 14 unknown
quantities in the regions (2) and (3): $\rho_2$, $\vel_2$, $P_{\th2}$,
$P_{\CR2}$, $P_{\inj}$, $\eps_{\th2}$, $\eps_{\CR2}$, $\eps_{\inj}$, and
$\rho_3$, $\vel_3$, $P_{\th3}$, $P_{\CR3}$, $\eps_{\th3}$, $\eps_{\CR3}$ (the
behaviour of the rarefaction wave directly follows from these).  The regions (2)
and (3) are separated by a contact discontinuity, which implies a vanishing mass
flux across it and thus, $\vel_2 = \vel_3$ and $P_2 = P_3$. The thermal gas
obeys a polytropic equation of state, i.e.~$\eps_{\th\, i} = P_{\th\, i} /
(\gamma_\th - 1)$ for $i \in \{2,3\}$. This reduces the dimensionality of our
problem to 10 unknowns. According to our assumption (iii), the freshly injected
CR population obeys a polytropic equation of state with adiabatic index
$\gamma_{\inj}$ and its energy density, $\eps_\inj$, is a constant fraction of
the total dissipated energy density at the shock (i.e., the generated internal
energy density corrected for adiabatic compression),
\begin{equation}
  \label{eq:epsinj2}
  \eps_{\inj} = \zeta \eps_\rmn{diss} =
  \zeta \left(\eps_\inj + \eps_{\th2} - \eps_{\th1} x_\rmn{s}^{\gamma_\th}\right),
\end{equation}
where the compression ratio at the shock is denoted by $x_\rmn{s} =
\rho_2/\rho_1$, and $\zeta<1$ is the effective energy injection efficiency
after correcting for the fraction of low-energy CRs that is immediately
re-thermalized by Coulomb interactions with thermal protons.  This reduces the
dimensionality by 2 unknowns. Thermal gas and pre-existing CRs are adiabatically
expanded over the rarefaction wave and, in our approximation, the pre-existing
CRs are adiabatically compressed at the shock, yielding the following relations,
\begin{equation}
  \label{eq:CRadiabatic}
  \begin{array}{lcl lcl lcl}
    P_{\th3} & = & 
    P_{\th5} \left(\frac{\dps\rho_3}{\dps\rho_5}\right)^{\gamma_\th},
    \quad & & & 
    \\ \rule{0cm}{0.6cm}
    P_{\CR3} & = & 
    P_{\CR5} \left(\frac{\dps\rho_3}{\dps\rho_5}\right)^{\gamma_\CR},
    \quad &
    \eps_{\CR3} & = & 
    \eps_{\CR5} \left(\frac{\dps\rho_3}{\dps\rho_5}\right)^{\gamma_\CR},
    \\ \rule{0cm}{0.6cm}
    P_{\CR2} & = & 
    P_{\CR1} \left(\frac{\dps\rho_2}{\dps\rho_1}\right)^{\gamma_\CR},
    \quad &
    \eps_{\CR2} & = & 
    \eps_{\CR1} \left(\frac{\dps\rho_2}{\dps\rho_1}\right)^{\gamma_\CR},
  \end{array}
\end{equation}
which further reduces the dimensionality by 5 unknowns.  Thus, solving this
system requires three more linearly independent equations, two of which are
obtained by considering the generalised Rankine-Hugoniot conditions
(\ref{eq:generalRH}). The last equation is given by the Riemann invariant
$\Gamma^+$. Using the effective speed of sound, $c_{\rmn{sound}} = \sqrt{\gamma_{\eff} P /
  \rho}$, we obtain
\begin{equation}
  \label{eq:RiemannCR}
  \Gamma^+ = \vel + \int_0^\rho \frac{c_{\rmn{sound}}(\rho')}{\rho'}\d \rho'
  = \vel + I(\rho) = \rmn{const.}
  \quad\mbox{with}\quad
  I(\rho) = \int_0^\rho 
  \sqrt{\tA_\CR x^{\gamma_\CR - 3} + \tA_\th x^{\gamma_\th - 3}} \d x.
\end{equation}
Here, we introduced the abbreviations $\tA_i = \gamma_i A_i$ where $i \in
\{\rmn{th},\rmn{CR}\}$ and $A_{i}^{} = P_{i5}^{}\,\rho_5^{-\gamma_i}=
P_{i3}^{}\,\rho_3^{-\gamma_i}$ denotes the adiabatic function that is conserved
across the rarefaction wave. Defining the difference of the adiabatic indexes of
the two populations, $\Delta\gamma \equiv \gamma_\th - \gamma_\CR$, we obtain
the solution to the integral $I(\rho)$,
\begin{equation}
  \label{eq:RIsolution}
  I(\rho) = \frac{\sqrt{\tA_\CR}}{\Delta\gamma}
  \left(\frac{\tA_\CR}{\tA_\th}\right)^{(\gamma_\CR - 1)/(2 \Delta\gamma)}
  \B_{x(\rho)} \left(\frac{\gamma_\CR - 1}{2\Delta\gamma},
             \frac{1 - \gamma_\th}{2\Delta\gamma}\right)
  \quad\mbox{with}\quad
  x(\rho) = \frac{\tA_\th\, \rho^{\gamma_\th}}
  {\tA_\CR\, \rho^{\gamma_\CR} + \tA_\th\, \rho^{\gamma_\th}},
\end{equation}
where $\B_x(a,b)$ denotes the incomplete beta function. While the second
argument of the incomplete beta function is always negative, the expression for
$I(\rho)$ is well defined as long as we consider a non-vanishing CR pressure
which is characterised by $\tA_\CR > 0$ and $\Delta\gamma > 0$. For the
remaining case $\tA_\CR = 0$, the integral can be solved in closed form,
yielding $I(\rho) = 2 c_{\rmn{sound}}(\rho) / (\gamma_\th-1)$. Note that the
solution of the rarefaction wave fan remains conceptually the same in comparison
to the case without CR acceleration \citep{2006MNRAS.367..113P}, albeit the wave
solution connects to a different $P_3$ and $\rho_3$, which respond to the softer
equation of state in the post-shock regime as a result of CR acceleration.

\subsection{Solution of the Riemann problem}

The densities to the left and right of the contact discontinuity, $\rho_3$ and
$\rho_2$, are obtained by matching the possible post-shock states (pressure and
density in regime 2) with the possible rarefaction-wave states (regime 3)
while simultaneously obeying the conservation laws over the rarefaction wave and
the shock. In practice, we have to (numerically) solve the following non-linear
system of equations:
\begin{equation}
  \label{eq:sys}
  \begin{array}{lclcl}
    f_1(x_\s,x_\rmn{r}) & \equiv &
    [P_2(x_\rmn{r}) - P_1]\, (x_\rmn{s} - 1) - 
    \rho_1 x_\rmn{s} \left[ I(\rho_5) - I(x_\rmn{r}\rho_5) \right]^2 & = & 0,
    \\ \rule{0cm}{0.4cm}
    f_2(x_\rmn{s},x_\rmn{r}) &\equiv &
    [P_2(x_\rmn{r}) + P_1]\, (x_\rmn{s} - 1) + 
    2 [x_\rmn{s} \eps_1 - \eps_2(x_\rmn{s},x_\rmn{r})] & = & 0.
  \end{array}
\end{equation}
Here, we expressed the system of equations in terms of dimensionless density
ratios: the shock compression ratio $x_\rmn{s} \equiv \rho_2 / \rho_1$ and the
rarefaction wave ratio $x_\rmn{r} \equiv \rho_3 / \rho_5$.  The implicit
functional dependencies on $x_\rmn{s}$ and $x_\rmn{r}$ read as follows,
\begin{eqnarray}
  \label{eq:P2CR}
  P_2(x_\rmn{r}) &=& P_3(x_\rmn{r}) = P_{\CR5} x_\rmn{r}^{\gamma_\CR} + 
  P_{\th5} x_\rmn{r}^{\gamma_\th}, \\
  \label{eq:ethCR}
  \eps_{\th,\rmn{ad}}(x_\s) &=& \eps_{\th1}x_\s^{\gamma_\th}, \\
  \label{eq:e2CR}
  \eps_2(x_\rmn{s},x_\rmn{r}) &=&
  \left[(1-\zeta)\frac{\gamma_\th-1}{\gamma_\inj-1}+\zeta\right]^{-1}
  \left[\frac{P_2(x_\rmn{r})}{\gamma_\inj-1} + \xi \eps_{\th,\rmn{ad}}(x_\s)
    -\frac{\gamma_\CR-1}{\gamma_\inj-1}\eps_{\CR2}(x_\s)\right]
    + \eps_{\CR2}(x_\s) - \xi \eps_{\th,\rmn{ad}}(x_\s),
\end{eqnarray}
where $\xi = \zeta / (1 - \zeta)$.  The post-shock pressure is obtained by
inserting the root $x_r$ into equation~(\ref{eq:P2CR}). The post-shock velocity
$\vel_2 = \vel_3$ and the shock speed $\vel_\rmn{s}$ are obtained from the
generalised Rankine-Hugoniot relations,
\begin{eqnarray}
  \label{eq:v3CR}
  \vel_2 &=& 
  \sqrt{[P_2(x_\rmn{r}) - P_1]\, \frac{\rho_2 - \rho_1}{\rho_2 \, \rho_1}}, \\
  \vel_\rmn{s} &=& \frac{\rho_2 \vel_2}{\rho_2 - \rho_1}.
\end{eqnarray}
Using these results, we find the solution to the generalised Riemann problem for
a composite of pre-existing CRs and thermal gas that experiences CR shock
acceleration:
\begin{equation}
  \label{eq:rhoCR}
  \begin{array}{lll}
  \rho(x, t) & = &
  \left\{ \begin{array}{ll}
      \rho_5, & x \le -c_5 t, \\
      \rho(x,t), & -c_5 t < x \le -\vel_\rmn{t} t, \\
      \rho_3, & -\vel_\rmn{t} t < x \le \vel_2 t, \\
      \rho_2, & \vel_2 t < x \le \vel_\rmn{s} t, \\
      \rho_1, & x > \vel_\rmn{s} t, \\
    \end{array} \right. \\
  \label{eq:PCR}
  [P,P_\th,P_{\CR,\rmn{tot}}](x, t) & = &
  \left\{ \begin{array}{llll}
      [P_5, & P_{\th5}, & P_{\CR5}], & x \le -c_5 t, \\
      {[A_\CR\, \rho(x,t)^{\gamma_\CR} + A_\th\, \rho(x,t)^{\gamma_\th}},
        & A_\th\, \rho(x,t)^{\gamma_\th},
        & A_\CR\, \rho(x,t)^{\gamma_\CR}],
      & -c_5 t < x \le -\vel_\rmn{t} t, \\
      {[P_3},     & P_{\th3}, & P_{\CR3}], & -\vel_\rmn{t} t < x \le \vel_2 t, \\
      {[P_2=P_3}, & P_{\th2}, & P_{\CR2}+P_\inj], & \vel_2 t < x \le \vel_\rmn{s} t, \\
      {[P_1},     & P_{\th1}, & P_{\CR1}], & x > \vel_\rmn{s} t, \\
    \end{array} \right. \\
  \label{eq:velCR}
  \vel(x, t) & = &
  \left\{ \begin{array}{ll}
      0, & x \le -c_5 t, \\
      \frac{\dps x}{\dps t} + 
      \sqrt{\tA_\CR\, \rho(x,t)^{\gamma_\CR-1} + \tA_\th\, \rho(x,t)^{\gamma_\th-1}},
      & -c_5 t < x \le -\vel_\rmn{t} t, \\
      \vel_2 = \vel_3, & -\vel_\rmn{t} t < x \le \vel_\rmn{s} t, \\
      0, & x > \vel_\rmn{s} t. \\
    \end{array} \right.
    \end{array}
\end{equation}
Here, $P_{\CR,\rmn{tot}}$ denotes the total CR pressure in a given regime, $c_5
= \sqrt{\gamma_{\eff5} P_5 / \rho_5}$ is the effective speed of sound,
$\vel_\rmn{t}$ is the velocity of the rarefaction wave's tail, and
$\vel_\rmn{s}$ is the shock velocity.  Matching the rarefaction wave equation to
the density of the post-contact discontinuity (regime 3) yields $\vel_\rmn{t}$:
\begin{equation}
  \label{eq:vtCR}
  \vel_\rmn{t} = I(\rho_3) - I(\rho_5) +
  \sqrt{\tA_\CR\, \rho_3^{\gamma_\CR-1} + \tA_\th\, \rho_3^{\gamma_\th-1}}.
\end{equation}
The stratified density within the rarefaction wave (regime 4) is obtained by
(numerically) solving the non-linear equation for a given characteristic
$(x,t)$, which is derived from the rarefaction wave equation,
\begin{equation}
  \label{eq:rfCR}
  I[\rho(x,t)] - I(\rho_5) + \frac{\dps x}{\dps t} + 
  \sqrt{\tA_\CR\, \rho(x,t)^{\gamma_\CR-1} + \tA_\th\, \rho(x,t)^{\gamma_\th-1}}
  = 0.
\end{equation}
Finally, the partial pressures left to the contact discontinuity can be obtain
from (\ref{eq:CRadiabatic}) while the partial pressure quantities in the
post-shock region are obtained by the following relations,
\begin{eqnarray}
  \label{eq:PsolCR}
  P_{\th2} (x_\s,x_{\rmn{r}}) &=& \frac{1}{1+\zeta}
  \left[ P_2(x_\rmn{r}) + \xi \frac{\gamma_\inj-1}{\gamma_\th-1} P_{\th1} x_\s^{\gamma_\th}
    - P_{\CR1} x_\s^{\gamma_\CR}\right], \\
  P_\inj (x_\s) &=& \xi \frac{\gamma_\inj-1}{\gamma_\th-1}
  \left[P_{\th2} (x_\s,x_{\rmn{r}}) - P_{\th1} x_\s^{\gamma_\th} \right].
\end{eqnarray}
In the limiting case of no CR injection ($\xi = 0$), it can be shown
straightforwardly that the system of equations (\ref{eq:sys}) reduces to
equation (B4) in \citet{2006MNRAS.367..113P} for the Riemann problem of a
composite fluid without CR acceleration.

\bsp

\label{lastpage}

\end{document}